\documentclass[12pt]{article}

\usepackage[margin=1in]{geometry}
\usepackage{amsthm,amsmath,amsfonts,amssymb}
\usepackage[colorlinks,citecolor=blue,urlcolor=blue]{hyperref}
\usepackage{graphicx}
\usepackage[numbers,sort&compress]{natbib}
\usepackage{caption}
\usepackage{algorithm}
\usepackage{algpseudocode}
\algrenewcommand\algorithmicrequire{\textbf{Input:}}
\algrenewcommand\algorithmicensure{\textbf{Output:}}

\theoremstyle{plain}

\newtheorem{theorem}{Theorem}[section]
\newtheorem{lemma}[theorem]{Lemma}
\theoremstyle{remark}
\newtheorem{definition}[theorem]{Definition}
\newtheorem{condition}[theorem]{Condition}

\theoremstyle{plain}
\newtheorem{proposition}[theorem]{Proposition}

\theoremstyle{remark}
\newtheorem{remark}{Remark}

\newcommand{\x}{x}
\newcommand{\xn}[1]{\x_{#1}}
\newcommand{\xln}[1]{\x_{#1}}
\newcommand{\bx}{X}
\newcommand{\bxh}{\hat{X}}
\newcommand{\bxhln}[1]{\hat{X}_{#1}}
\newcommand{\bxln}[1]{\bx_{#1}}

\newcommand{\xpn}[1]{\xp_{#1} }
\newcommand{\xs}{x^*}

\newcommand{\xsn}[1]{\xs_{#1}}
\newcommand{\bxt}{\tilde{\bx}}

\newcommand{\xt}{\tilde{x}}
\newcommand{\xtn}[1]{\xt_{#1}}
\newcommand{\bxp}{\bx '}

\newcommand{\bxls}{\bx _{*}}

\newcommand{\bxsp}{\bxls'}

\newcommand{\bxs}{{\bx_*}}

\newcommand{\y}{y}
\newcommand{\yn}[1]{y_{#1}}
\newcommand{\yln}[1]{\y_{#1}}
\newcommand{\by}{Y}
\newcommand{\byh}{\hat{Y}}
\newcommand{\byhun}[1]{\hat{Y}^{(#1)}}
\newcommand{\byhln}[1]{\hat{Y}_{#1}}
\newcommand{\byln}[1]{\by_{#1}}

\newcommand{\ys}{y^*}
\newcommand{\yls}{y_*}
\newcommand{\ysn}[1]{\ys_{#1}}
\newcommand{\byt}{\tilde{\by}}
\newcommand{\bytln}[1]{\byt_{#1}}
\newcommand{\yt}{\tilde{y}}
\newcommand{\ytn}[1]{\yt_{#1}}
\newcommand{\byp}{\by '}

\newcommand{\byls}{\by _{*}}

\newcommand{\bys}{\by_*}

\newcommand{\z}{z}

\newcommand{\bz}{Z}
\newcommand{\bzh}{\hat{\bz}}

\newcommand{\zsn}[1]{\z^*_{#1}}

\newcommand{\bzp}{\bz'}
\newcommand{\bzs}{\bz^*}

\newcommand{\n}{n}

\newcommand{\bn}{N}
\newcommand{\np}{\n'}
\newcommand{\bnp}{N'}

\newcommand{\m}{m}

\newcommand{\mmp}{\m'}
\newcommand{\bmp}{M'}
\newcommand{\bmsn}[1]{\mathcal{S}(#1)}

\newcommand{\bo}{O}

\newcommand{\bmo}{\mathcal{O}}
\newcommand{\p}{p}

\newcommand{\pln}[1]{p_{#1}}

\newcommand{\pmin}[2]{p_{\min}(\bx,\by)}
\newcommand{\pmax}[2]{p_{\max}(\bx,\by)}
\newcommand{\kk}{k}
\newcommand{\kkp}{k'}

\newcommand{\rrln}[1]{r_{#1}}

\newcommand{\mr}{\mathcal{R}}
\newcommand{\mrln}[1]{\mr_{#1}}

\newcommand{\w}{w}

\newcommand{\sa}{a}

\newcommand{\baa}{A}

\newcommand{\sbb}{b}

\newcommand{\bbb}{B}

\newcommand{\scc}{c}

\newcommand{\bcc}{C}

\newcommand{\bmcn}[1]{\mathcal{C}(#1)}
\newcommand{\sdd}{d}

\newcommand{\iconstant}{i}
\newcommand{\jconstant}{j}

\newcommand{\lconstant}{l}
\newcommand{\lln}[1]{\lconstant_{#1}}

\newcommand{\st}{t}
\newcommand{\tconstant}{t}
\newcommand{\bt}{T}

\newcommand{\prob}[1]{\text{Pr}\left(#1\right)} % probability
\newcommand{\Prob}[1]{\text{Pr}\left(#1\right)}   % probablity
\newcommand{\rank}[2]{R(#1, #2)} % rank
\newcommand{\indicator}[1]{\text{I}\left(#1\right)} % indicator function
\newcommand{\WMW}[2]{T({#1},{#2})}  % Wilcoxon-Mann-Whitney
\newcommand{\WMWmin}[2]{T_{\text{min}}({#1},{#2})}  % Wilcoxon-Mann-Whitney
\newcommand{\WMWmax}[2]{T_{\text{max}}({#1},{#2})}  % Wilcoxon-Mann-Whitney
\newcommand{\ABT}[2]{T(#1,#2)} % Ansari-Brdley
\newcommand{\card}[1]{\left|#1\right|} %cardinality 
\newcommand{\sgn}[1]{\text{sgn}(#1)} %% sign function
\newcommand{\func}[2]{f_{#1}(#2)} % f: function
\newcommand{\mysetr}{\mathcal{R}}
\newcommand{\mysetrln}[1]{\mathcal{R}_{#1}}
\newcommand{\mysetx}{\mathcal{X}}

\newcommand{\mysety}{\mathcal{\by}}

\newcommand{\mysetz}{\mathcal{\bz} }

\newcommand{\nh}{H_0}

\newcommand{\dnm}[2]{\mathcal{D}(#1,#2)}

\newcommand{\case}[1]{(\text{#1})}

\newcommand{\Lepage}{L}
\newcommand{\xx}{x}
\newcommand{\xp}{\xx '}
\newcommand{\yy}{y}

\newcommand{\misss}{s}

\begin{document}

\title{Scale two-sample testing with arbitrarily missing data}

\author{Yijin Zeng \and Niall M. Adams \and Dean A. Bodenham}

\date{{\normalsize Department of Mathematics, Imperial College London,} \\
	{\normalsize South Kensington Campus, London SW7 2AZ, U.K.} \\
	{\normalsize yijin.zeng20@imperial.ac.uk, 
		\quad n.adams@imperial.ac.uk, 
		\quad dean.bodenham@imperial.ac.uk}
}

\maketitle

\begin{abstract}
This work proposes a novel rank-based scale two-sample testing method for univariate, distinct data when a subset of the data may be missing. 
Our approach is based on mathematically tight bounds of the Ansari-Bradley test statistic in the presence of missing data, and rejects the null hypothesis if the test statistic is significant regardless of the missing values. This proposed scale testing method is then combined with the location testing method proposed by \citet{zeng2024two} using the Holm-Bonferroni correction for location-scale testing. We show that our methods control the Type I error regardless of the values of the missing data.  Simulation results demonstrate that our methods have good statistical power, typically when less than 10\% of the data are missing, while other missing data methods, such as case deletion or imputation methods, fail to control the Type I error when the data are missing not at random. We illustrate the proposed location-scale testing method on hepatitis C virus dataset where a subset of values is unobserved. 
\end{abstract}

\section{Introduction}

Two-sample hypothesis testing is often problematic 
in the presence of missing data. Standard hypothesis testing methods 
generally do not take missing data into account, and the testing results
cannot be determined directly when a subset of data is missing.
In certain cases, it is possible to make reasonable assumptions about the missing data, and the testing results might be taken as valid. For example, with clinical trials it frequently occurs that participants drop out before the trial concludes, resulting in missing data for those participants. Several analyses of clinical trials data 
\citep{feldman1993effects, marguliesetal2016} have dealt 
with missing values by assuming them having the worst-possible ranks, 
or first assuming them to have the worst-possible 
values and then converting these values to ranks, as
\citet{lachin1999} sought to formalise. 

However, such assumptions of missing data are usually empirical and untestable using the observed data alone. In many real-world scenarios, the missing data mechanisms are unclear or too complicated to be analyzed \citep{bakrisetal2015}.
The missing data assumptions are often made primarily to 
facilitate reaching testing conclusions without any accompanying 
analysis \citep{mathur2023m}, and are based 
mainly on its simplicity to use \citep{staudt2022sensitivity}. 
Such practices can be invalid, as they risk increasing the  
Type I error of the testing results beyond a pre-specified significance level. For evaluating the credibility of these testing results, one common
approach is to consider the proportion of the missing data: 
some authors suggest that less than $5$\% missing data
\citep{Schafer1999Multiple, heymans2022handling} could be inconsequential to the
results, some increase the proportion to 10\% \citep{bennett2001can},
while others argue that the proportion of missing data should not be used as an indication~\citep{madley2019proportion}.

The concern that the missing data assumptions could lead to false testing
results often leads researchers to perform a sensitivity analysis 
\citep{little1996intent, goldberg2021data, thabane2013tutorial}, where a set of missing data assumptions is performed to evaluate whether a consensus about the testing result can be established. If several different missing data assumptions lead to a consistent testing result, then it is taken as the final result with more confidence \citep{bakrisetal2015}.

Recently, \citet{zeng2024two} proposed a new framework for performing 
two sample testing with controlled Type I error
in the presence of missing data, without the need for making any 
missing data assumptions.  Their approach
takes all possible missing data into account, and rejects the null hypothesis only when all possible missing data lead to a consistent significant result.
For considering all possible missing data, they do not consider
the values of the missing data directly, but evaluate all possible \emph{ranks}
of the missing data. Specifically, they consider a rank-based
test statistic, known as the Wilcoxon-Mann-Whitney
test statistic \citep{mann1947test}, and derive the tight lower and 
upper bounds of this test statistic with missing data. Then, 
the corresponding $p$-values in the presence of missing data are derived, and
if the maximum possible $p$-value is smaller than a pre-specified 
significance level, a significant testing result will be declared. It is demonstrated that this approach controls the Type I error regardless of missing data, while also having good statistical power when the proportion of missing data is around 10\% to 20\%. Since this approach is based on the Wilcoxon-Mann-Whitney test statistic, it can only be used for the location testing problem.

% motivation
However, in practice, a change in location
is often accompanied by a change in variability \citep{marozzi2013nonparametric, zhang2009variability, neuhauser2001adaptive}. Moreover, detecting heteroscedasticity between two samples is often important, as revealing differences in population variability may carry practical significance. For instance, in clinical trials, the presence of heteroscedasticity can indicate a treatment effect \citep{murakami2025two}.
\citet{zhu2019intra} report that the intra-individual variability of total cholesterol could be a risk factor of cardiovascular mortality, irrespective of mean TC level.
This work is therefore motivated to allow valid 
scale and location-scale two sample testing in the presence of missing data, without relying on missingness assumptions.

Following a similar approach to \cite{zeng2024two}, this work considers the scale testing problem in the presence of missing data based on the Ansari-Bradley test \citep{ansari1960rank} when data are
univariate and distinct. Similar to the Wilcoxon-Mann-Whitney
test, the Ansari-Bradley test is also based on rank, but 
is used for testing the scale difference, rather than the location difference.
We consider all possible Ansari-Bradley test statistics by deriving the tight
lower and upper  bounds of this test statistic. Based on these
bounds, we construct the bounds of the corresponding
$p$-values and propose to reject the null hypothesis when
all possible $p$-values are smaller than a significant level $\alpha$.
We prove this method controls the Type I error regardless of 
the values of the missing data. 
The proposed scale testing method is then combined with the location testing method proposed by~\citet{zeng2024two}
using the Holm-Bonferroni correction for the location-scale testing problem,
where one wishes to detect the differences in both the location and scale.
Simulation results demonstrate our scale and location-scale testing methods have good statistical power, typically when less than $10\%$ of the data is missing, while other common missing data methods, such as case deletion and imputation methods, fail to control the Type I error. We also illustrate the proposed location-scale testing method on hepatitis C virus dataset where a subset of values is unobserved. 

Our proposed testing methods can also be interpreted as a sensitivity analysis that considers all possible assumptions on the missing data. As mentioned earlier, our methods take all possible imputations into account by providing bounds of $p$-values. Hence, 
if a significant result is obtained by our methods, then this
result is consistent among all possible missing data assumptions, since every possible imputation leads to a significant result. If, however, our methods fail to reject the null hypothesis, there are two possible outcomes: either all possible missing
data assumptions lead to insignificant results when the lower bound of the $p$-values is greater than $\alpha$, or a subset of assumptions leads to significant results when the lower bound of the $p$-value is less than or equal to $\alpha$, 
but the upper bound is larger than $\alpha$.
Hence our methods provide a theoretically rigorous
answer to whether different approaches to handling missing data lead to different conclusions.

\section{Background}
\subsection{Location and scale  two-sample testing problems}
\label{sec:locationscaletest}

To formally define the location, scale and location-scale two-sample testing problems,
let $\bxln{1}, \ldots, \bxln{\n}$ and $\byln{1}, \ldots, \byln{\m}$ be independent random variables following continuous cumulative distribution functions $F(\xx)$ and $G(\y)$, respectively. Assume the differences of the two functions are determined
only by the location and the scale parameters $a$ and $b$ such that 
\begin{align*}
	G(y) = F(by + a),~\text{where } a > 0.
\end{align*}

Suppose the scale parameter $b = 1$, the problem of testing whether 
the location parameter $a = 0$ is known as the location test. 
A popular testing method for this problem is the Wilcoxon-Mann-Whitney test \citep{mann1947test}. On the other hand, suppose the location
parameter $a = 0$, the problem of testing whether the scale 
parameter $b = 1$ is known as the scale test. As mentioned previously, 
the Ansari-Bradley test \citep{ansari1960rank} is often used for the scale 
testing problem. Finally, the problem of testing whether $a = 1$ and $b = 0$ is the location-scale testing problem.
By calculating the sum of squared standardized Wilcoxon-Mann-Whitney test statistic and the Ansari–Bradley test statistic, \citet{lepage1971combination} introduces a location-scale testing method known as the Lepage test. 

The location and scale testing problems described above are closely related to the mean and variance two-sample testing problems, where one wishes to test whether the mean values and variances of random
variables from $F$ and $G$ are different. When the 
data are normally distributed, i.e. $F$ and $G$ are cumulative 
distribution of normal distributions, it can be shown that the location
testing problem is equivalent to the mean value testing problem,
and the scale testing problem is the variance testing problem.

Apart from the Wilcoxon-Mann-Whitney test,
Ansari-Bradley test, and the Lepage test mentioned above, 
other testing methods for location, scale and location-scale
testing problems are also available. When the data are normally
distributed, student's $t$-test \citep{student1908probable} and
F-test \citep{fisher1970statistical} are proven to be the most
powerful testing methods for the location and scale testing problems, respectively. A non-parametric testing method that does
not assume the data distribution for the location-scale
testing problem is known as the Cucconi test \citep{cucconi1968nuovo}.

In this paper, we focus mainly on the Ansari-Bradley test
statistic for the scale testing problem. We consider an
extension of this test in the presence of missing data
by deriving the tight lower and upper bounds of the Ansari-Bradley test statistic. This proposed method is then combined with the location
testing method proposed by \citet{zeng2024two} based on
the Wilcoxon-Mann-Whitney test for the location-scale testing
problem.

\subsection{The Ansari-Bradley test} \label{ansaribradley}

As mentioned in Section~\ref{sec:locationscaletest}, the Ansari-Bradley test is a non-parametric scale testing method based on rank. The concept of rank is central in this work, hence we define it carefully. 
To start, let $\bx = \{\xn{1}, \ldots, \xn{\n}\}$ and $\by = \{\yn{1}, \ldots, \yn{\m}\}$
be two sets of distinct real-valued samples of size $\n$ and $\m$, respectively.
Let us denote $\bz = \bx \cup \by$, and let $\indicator{A}$ be the indicator function such that if the statement $A$ is correct,  $\indicator{A} = 1$, otherwise $\indicator{A} = 0$. Subsequently, the rank is defined formally as follows.

\begin{definition} \label{def:rank}
	Suppose $\bz$ is a set of $\bn$ distinct real-value observations. 
	Then for any $\xx \in \mathbb{R}$, the rank of $\xx$ in $\bz$ is defined as 
	\begin{align*}
		\rank{\xx}{\bz} = \sum_{\z \in \bz} \indicator{\z \le \xx }.
	\end{align*}
	where $I$ is the indicator function such that 
	$\indicator{A} = 1$ if  $A$ is true; otherwise $\indicator{A} = 0$.
\end{definition}
Using the above definition of rank, 
the Ansari-Bradley test statistic can be defined as
\begin{definition} \label{def:ABT}
	Suppose $\bx = \{\xn{1}, \ldots, \xn{\n}\}$ and $\by = \{\yn{1}, \ldots, \yn{\m}\}$ are two sets of distinct real values. The Ansari-Bradley test statistic $\ABT{\bx}{\by}$ is defined as
	\begin{align*}
		\ABT{\bx}{\by} = \sum_{i = 1}^{\n} \left|\rank{\xn{\iconstant}}{\bx \cup \by} - \frac{1}{2}(\n + \m +1)\right|.	
	\end{align*}
\end{definition}
Note that the above definition of the Ansari-Bradley test statistic is slightly different from that used by other authors \citep{ansari1960rank, lepage1971combination}.
However, our definition differs from theirs only by a constant 
depending on the sample sizes $\n$ and $\m$. Definition \ref{def:ABT} is used in this work mainly for the purpose of simplifying notation.

For performing the Ansari-Bradley test, the distribution
of the Ansari-Bradley test statistic under the null hypothesis is often required. \citet{ansari1960rank} provide a recursion formula and a frequency generating function for calculating the exact distribution of $\ABT{\bx}{\by}$ for small sample sizes, e.g. $\n, \m \le 10$.
They also show that $\ABT{\bx}{\by}$ asymptotically follows a  normal distribution with mean
\begin{align} \label{eqn:mean}
	\mu =  \left\{ \begin{array}{lc}
		\n\bn/4, & \mbox{when}~\bn~\mbox{ is even,} \\ 
		\n(\bn^2-1)/(4\bn),  & \mbox{when}~\bn~\mbox{is odd,}
	\end{array}\right.
\end{align}
and variance
\begin{align} \label{eqn:var}
	\sigma^2 =  \left\{ \begin{array}{lc}
		\n\m(\bn^2-4)/\{48(\bn-1)\}, & \mbox{when}~\bn~\mbox{is even,}  \\ 
		\n\m(\bn+1)(\bn^2+3)/(48\bn^2),  & \mbox{when}~\bn~\mbox{is odd.}
	\end{array}\right.
\end{align}
The performance of the normal approximation is studied by
\cite{fahoome2000review}. Their results suggest that
this approximation can be applied when both $\n,\m$ are larger than $15$ for a significance level $\alpha = 0.05$,  while for $\alpha = 0.01$, 
both $\n, \m$ should be larger than $29$. \citet{R:2010} apply 
the normal approximation when both $\n,\m \ge 50$.

\subsection{Missingness mechanisms}

A framework for missing data is introduced by \citet{rubin1976inference}, where the missingness mechanisms are divided into three types, namely missing completely at random, missing at random, and missing not at random. Following \cite{rubin1976inference}, it is more convenient to consider vectors of data, rather than sets of data. Consider a vector of univariate real-valued samples $z = (z_1,z_2,\cdots,z_N)$. Let $\iota = (\iota_1,\cdots,\iota_N)$ be an indicator function of $z$ such that $\iota_i$ taking value 1 if $z_i$ is missing and 0 if $z_i$ is observed. The core idea of \cite{rubin1976inference} is to admit $\iota$ as a probabilistic phenomenon.

Suppose $z$ is a realized value of a random vector $Z$. Let $g_{\theta}(\iota | z)$ denote the probability of $I = \iota$ given $Z = z$, where $\theta$ denotes any unknown parameters of the distribution.  Then, the missingness mechanism is missing completely at random if
\begin{align} \label{MCAR}
	g_{\theta}({\iota} | z) = g_{\theta}(\iota | \widetilde{z}), \forall \iota, z, \widetilde{z}.
\end{align}
In such cases, the missingness mechanism $I$ is independent of the values of samples, and ignoring missing data before testing may be justified. 

Denote $z'$ as a sub-vector of $z$ including all observed samples in $z$, i.e. including all $z_i$ such that $\iota_i = 0$. Then, the missingness mechanism is missing at random if
\begin{align} \label{MAR}
	g_{\theta}({\iota} | z) = g_{\theta}(\iota | \widetilde{z}), \forall \iota, z, \widetilde{z} \text{ such that } z' = \widetilde{z}'. 
\end{align}
In such cases, the missingness mechanism $I$ is independent of the values of the missing values, and some imputing practices before testing may be justified. 

If neither Equation (\ref{MCAR}) nor (\ref{MAR}) holds, the missingness mechanism is missing not at random. In this case, ignoring missing data or imputing the missing values could be invalid, since they risk increasing the Type I error beyond a pre-specified significance level $\alpha$. Missing not at random is difficult to deal with properly, since it usually requires one to explicitly specify the distribution for the missingness \citep{schafer2002missing}.

\citet{rubin1976inference} divides the missingness mechanisms
into three different cases according to whether the distribution 
of the missingness mechanism depends on the missing values or observed values. However, the types of missingness mechanisms 
are often difficult to determine in practice; \citet{van2012flexible}
suggests that it is impossible to test missing not at random versus
missing at random using the observed data alone. To the best
of our knowledge, there is no testing method for deciding
whether the data are missing completely at random for univariate data.

We end this section by emphasizing that the proposed methods
in this work do not make any assumptions of the missing data. The 
unique contribution of this paper is the construction of 
scale and location-scale
testing methods with controlled Type I error without the need for
making any missing data assumptions.

\subsection{Related work}

As mentioned earlier, our proposed scale testing method
follows a similar approach to that used by \citet{zeng2024two} for location testing, where the tight bounds of the test statistics with missing data are construed, and later are applied for deciding whether all possible
test statistics are significant. 

To the best of our knowledge, there is no existing literature that discusses the scale testing, or variance testing problem in the missing data directly. A common practice is to either ignore missing values or impute them with a single value, such as the mean or median of the observed data \citep{schafer2002missing}. When both the underlying data distribution and the missingness mechanism are known, likelihood-based methods such as multiple imputation \citep{rubin1976inference} or the expectation–maximization algorithm \citep{dempster1977maximum} can be employed.

Various authors have suggested modifications and refinements to
rank-based tests to accommodate missing values, typically under the assumption that the missingness mechanism is known.
\citet{cheung2005exact} and \citet{lee1997} adapt
the Wilcoxon-Mann-Whitney test when the data are
assumed to be missing at random. On the other hand, \citet{lachin1999}
considers the Wilcoxon-Mann-Whitney test under a special
missing not at random case, where the missing data are assumed to have the 
largest values. In the context of independence testing, \citet{zeng2025exact, Alvo1995RankCM} modify rank-based 
statistical association measurements to accommodate
missing data, although this topic is different from the scale and location-scale testing problem considered in this paper.

\section{Main Results} \label{mainresults}

This section presents the main results of the paper. In Section~\ref{bounds}, we derive the tight lower and upper bounds of the Ansari–Bradley test statistic, with full technical details provided in the Supplementary Material. Building on these results, Section~\ref{ABtestwithmissingdata} introduces a scale testing method that controls the Type I error in the presence of missing data, without making assumptions about the missingness mechanism.

\subsection{Bounding the Ansari-Bradley test statistic with missing data} \label{bounds}

We now consider the bounds of the Ansari-Bradley test statistic $\ABT{\bx}{\by}$ in the presence of missing data, when all values are
univariate and assumed to be distinct. This section presents only
the main results regarding the bounds, while
all the technical details can be found in the Supplementary Material.
Our proof starts by considering the case when only a single
value is missing, and is later generalized to arbitrary
missingness patterns without making any missing data assumptions,
as we will show below.

We start by proving the following Proposition~\ref{prop:3}, which considers the minimum possible Ansari-Bradley test statistic $\ABT{\bx}{\by}$ when a \emph{single} value $\x \in \bx$ is unobserved.

\begin{proposition} \label{prop:3}
	Suppose $\bx = \{\xn{1}, \ldots, \xn{\n}\}$ 
	and $\by = \{\yn{1}, \ldots, \yn{\m}\}$ are samples of distinct real values.  Denote $\bn =\n + \m$. Suppose $\x \in \bx$ 
	is a value in $\bx$ and denote $\bxp = \bx \setminus \{\x\}$. Consider a real value
	$\xs$ that is distinct to all values in $\bxp \cup \by$, and denote 
	$\bxs = \{\xs\} \cup \bxp$. Then, if 
	\begin{align*}
		\left|\rank{\xs}{\bxs \cup \by} - \frac{1}{2}(\bn + 1)\right| =  \left\{ \begin{array}{lll}
			0, & \text{when} & \bn~\text{is odd,}  \\ 1/2, & \mbox{when} & \bn~\text{is even,}
		\end{array}\right.
	\end{align*}
	we have $\ABT{\bx}{\by} \ge \ABT{\bxs}{\by}$.
\end{proposition}	

\begin{proof}
	The proof is included in Section~\ref{append:ab:proofofsinglemissngdata}
	of the Supplementary Material.
\end{proof}

Proposition~\ref{prop:3} considers the minimum possible
Ansari-Bradley test statistic when only a single value
in $\bx$ is missing. To extend this result to the case
where \emph{multiple} values in $\bx$ can be missing, we first
make the following definitions, which classify all possible configurations
of the total sample sizes $\bn = \n + \m$
and the number of missing values $\n - \np$ 
based on their parity.

\begin{definition}\label{def:conditions}
	Let $\bn \ge \n \ge \np$ be positive integers. We define the 
	following four cases 
	$\bmcn{1}, \bmcn{2}, \bmcn{3}$ and $\bmcn{4}$ 
	according to the parity of $\bn$ and $\n - \np$:
	\begin{align*}
		&\bmcn{1}:~\text{$\bn$ is odd but $\n - \np$ is even},\\
		&\bmcn{2}:~\text{$\bn$ is odd and $\n - \np$ is odd},\\
		&\bmcn{3}:~\text{$\bn$ is even but $\n - \np$ is odd},\\
		&\bmcn{4}:~\text{$\bn$ is even and $\n - \np$ is even}.
	\end{align*}
\end{definition}

Using this definition, we can then state the following theorem,
which presents the tight  lower bound of the Ansari-Bradley test statistic when a subset of values in $\bx$ is missing, while all values in $\by$ are observed.

\begin{theorem} \label{theorem:1}
	Suppose $\bx = \{\xn{1}, \ldots, \xn{\n}\}$ and $\by = \{\yn{1}, \ldots, \yn{\m}\}$ are 
	samples of distinct real values, and $\bxp \subset \bx$ is a subset of $\bx$ 
	with sample size $|\bx| = \np$, which are observed. Then, the minimum possible Ansari-Bradley test statistic, across all possible values of
	missing data, is given as follows:
	\begin{align*}
		\min_{\bx \setminus \bxp \in \mathbb{R}^{\n - \np}}\ABT{\bx}{\by} = \ABT{\bxs}{\by} 
		= \left\{ \begin{array}{lcl}
			\ABT{\bxp}{\by} + {(\n^2 - \np^2)}/{4} &~\text{if}
			&\bmcn{1} \text{ or } \bmcn{4} \text{ holds}, \\ 
			\ABT{\bxp}{\by} + {(\n^2 - \np^2 - 1)}/{4} &~\text{if}
			&\bmcn{2} \text{ holds},\\ 
			\ABT{\bxp}{\by} + {(\n^2 - \np^2 + 1)}/{4} &~\text{if}
			&\bmcn{3} \text{ holds}.
		\end{array}\right.
	\end{align*}
\end{theorem}

\begin{proof}
	The proof is included in Section~\ref{appendix:lowebound:ab:x} of the 
	Supplementary Material.
\end{proof}

Theorem~\ref{theorem:1} provides the minimum possible
Ansari-Bradley test statistic when multiple values in
$\bx$ can be missing, while all values in $\by$ are observed.
We now study a similar case where multiple values in $\by$ can be
missing, but in $\bx$ are all observed. For presenting
the main result under such case, we first need to make the following
definition.

\begin{definition} \label{def:func}
	Suppose $\bx = \{\xn{1}, \ldots, \xn{\n}\}$ and $\byp = \{\yn{1}, \ldots, \yn{\mmp}\}$ are samples of distinct real values, and $\m \in \mathbb{\bn}$ is a positive integer such that $\m \ge \mmp$. Denote $\bzp = \bx \cup \byp$, $\bnp = (\n + \mmp + 1)/2$, and $\bmp = (\m-\mmp)/2$. For any 
	given $\kk  \in \{0, \ldots, \m - \mmp\}$, denote $\sa = \min\left\{0, \kk - \bmp \right\}$, $\sbb = \max \left\{0, \kk - \bmp \right\}$.
	For any $\iconstant \in \{1,\ldots, \n\}$, let $A_{\iconstant} = \indicator{\rank{\xn{\iconstant}}{ \bzp} < \bnp + \sa},
	B_{\iconstant} =  \indicator{\bnp + \sa \le \rank{\xn{\iconstant}}{ \bzp} \le \bnp + \sbb}$,	and $C_{\iconstant} = \indicator{\rank{\xn{\iconstant}}{\bzp} >  \bnp + \sbb }$.
	Then, we define
	\begin{align*}
		\func{\bx, \byp, \m}{\kk} = (\kk - \bmp) \sum_{\iconstant = 1}^{\n} (A_{\iconstant} - C_{\iconstant} ) + \sgn{\kk - \bmp} \sum_{\iconstant = 1}^{\n} B_{\iconstant} \left(2\bnp + \kk - \bmp - 2\rank{\xn{\iconstant}}{\bzp} \right),
	\end{align*}
	where $\sgn{\x} = 1$ if $\x \ge 0$, otherwise $\sgn{\x} = -1$.
\end{definition}

The function $\func{\bx, \byp, \m}{\kk}$ is important
for presenting the following result, which 
provides the minimum possible Ansari-Bradley test statistic when all values in $\bx$ are observed but not all values in $\by$ are observed,

\begin{theorem} \label{theorem:2}
	Suppose $\bx = \{\xn{1}, \ldots, \xn{\n}\}$ and $\by = \{\yn{1}, \ldots, \yn{\m}\}$ are samples of distinct real values and $\byp \subset \by$ is a subset of $\by$ with sample size $\card{\byp} = \mmp$, which are observed. Then, the minimum possible Ansari-Bradley test statistic, across all unobserved values, is given as follows:
	\begin{align*}
		\min_{\by \setminus \byp \in \mathbb{R}^{\m - \mmp}}\ABT{\bx}{\by} = \ABT{\bx}{\byp} + \min_{\kk \in \{0,\ldots,\m - \mmp \}} \func{\bx, \byp, \m}{\kk} ,
	\end{align*}
	where $\func{\bx, \byp, \m}{\kk}$ is defined in Definition~\ref{def:func}.
\end{theorem}
\begin{proof}
	The proof is included in Section~\ref{appendix:lowebound:ab:y}
	of the Supplementary Material.
\end{proof}

Finally, by combining Theorem \ref{theorem:1} and \ref{theorem:2}, 
we present our main result in this section.

\begin{theorem} \label{theorem:3}
	Suppose $\bx = \{\xn{1}, \ldots, \xn{\n}\}$ and $\by = \{\yn{1}, \ldots, \yn{\m}\}$ are samples of distinct real values. Let $\bxp \subset \bx$ and $\byp \subset \by$ be observed subsets with sizes $|\bxp| = \np$ and $|\byp| = \mmp$. Define $\sa = \max \left\{{(\m -\np -2\mmp + 1)}/{2}, 0\right\}$, $\sbb =  \min \left\{{(\m + \np + 1)}/{2},  \m -\mmp \right\}$,  $\scc =  \max \left\{{(\m -\np -2\mmp)}/{2}, 0\right\}$ and $\sdd =  \min \left\{{(\m + \np)}/{2},  \m -\mmp \right\}$. Then, 
	the minimum possible Ansari-Bradley test statistic over all
	possible missing values is:
	\begin{align*}
		\min_{\substack{\bx \setminus \bxp \in \mathbb{R}^{\n - \np}, \\ \by \setminus \byp \in \mathbb{R}^{\m - \mmp} }} \ABT{\bx}{\by} =  \ABT{\bxp}{\byp} + \left\{ \begin{array}{lcl}
			\min\limits_{\kk \in \{\sa, \ldots, \sbb\}} \func{\bxp, \byp, \m}{\kk} + {(\n^2 - \np^2)}/{4} &\text{~if}~\bmcn{1}, \\ 
			\min\limits_{\kk \in \{\scc, \ldots, \sdd\}} \func{\bxp, \byp, \m}{\kk} + {(\n^2 - \np^2 - 1)}/{4} &\text{~if}~\bmcn{2},\\ 
			\min\limits_{\kk \in \{\sa, \ldots, \sbb\}} \func{\bxp, \byp, \m}{\kk} + {(\n^2 - \np^2 + 1)}/{4} &\text{~if}~\bmcn{3},\\
			\min\limits_{\kk \in \{\scc, \ldots, \sdd\}} \func{\bxp, \byp, \m}{\kk} + {(\n^2 - \np^2)}/{4} &\text{~if}~\bmcn{4}, \\ 
		\end{array}\right.
	\end{align*}
	and the maximum possible Ansari-Bradley test statistic is:
	\begin{align*}
		\max_{\substack{\bx \setminus \bxp \in \mathbb{R}^{\n - \np}, \\ \by \setminus \byp \in \mathbb{R}^{\m - \mmp} }} \ABT{\bx}{\by}=  - \min_{\substack{\bx \setminus \bxp \in \mathbb{R}^{\n - \np}, \\ \by \setminus \byp \in \mathbb{R}^{\m - \mmp} }}  \ABT{\by}{\bx}   + \left\{ \begin{array}{ll}
			{\bn^2}/{4} 
			& \bn~\text{is even,} \\ 
			{(\bn^2 - 1)}/{4}
			& \bn~\text{is odd.} 
		\end{array}\right.
	\end{align*}
\end{theorem}

\begin{proof}
	The proof is included in Section~\ref{appendix:general}
	in the Supplementary Material.
\end{proof}

\begin{remark}
	The computational complexity of applying Theorem~\ref{theorem:3}
	for the lower and upper bounds of the Ansari-Bradley test statistic is analyzed as follows. 
	To compute the lower bound of the Ansari–Bradley test statistic, 
	$\ABT{\bxp}{\byp}$ can be computed in linear time once the combined 
	set of observed values $\bxp \cup \byp$ has been ranked, 
	which requires $\bmo((\np + \mmp)\log(\np + \mmp))$ time. 
	Given the ranked values, the computation of $ \func{\bxp, \byp, \m}{\kk}$
	for a fixed $\kk$ requires a cost of $\bo(\np)$. Since $\sbb - \sa, \sdd - \scc \le \m - \mmp$, evaluating the minimum of $ \func{\bxp, \byp, \m}{\kk}$ over the ranges $\kk \in \{\sa,\ldots,\sbb\}$ or $\kk \in \{\scc,\ldots,\sdd\}$ yields a worst-case cost of $\bmo\{(\m - \mmp)\np\}$. Therefore, Theorem~\ref{theorem:3} enables the computation of the lower bound of $\ABT{\bx}{\by}$ with total computational complexity
	\begin{align*}
		\bmo\{(\np+\mmp)\log(\np+\mmp) +  (\m - \mmp)\np \} = \bmo(\bn^2),
	\end{align*}
	where $\bn = \n + \m$ is the total sample size of $\bx$ and $\by$.
	The upper bound  of $\ABT{\bx}{\by}$ requires the same computational complexity $\bmo(\bn^2)$, by a similar argument.
\end{remark}

\subsection{Determine statistical significance without imputation} \label{ABtestwithmissingdata}

This section develops a two-sample scale testing method in the presence of missing data using the bounds of the Ansari-Bradley 
test statistic derived in Section~\ref{bounds}. 
The key idea of our testing method is to reject the null hypothesis 
when all possible test statistics are significant. 
We first provide the conditions for deciding
whether all possible test statistics are significant.
Subsequently, we derive the bounds of the $p$-values of the test
statistic, and show that all possible test statistics
are significant if and only if the maximum possible $p$-value
is less than or equal to the significance level $\alpha$.

\subsubsection{Rejection conditions for deciding significance}
\label{rejectconditions}

To start, recall that under the null hypothesis, 
the Ansari-Bradley test statistic 
$\ABT{\bx}{\by}$ asymptotically follows a normal distribution 
with mean $\mu$ and variance $\sigma^2$ defined in Equation~\eqref{eqn:mean}
and \eqref{eqn:var}, respectively.
Hence, when the sample sizes $\n, \m$ are large enough for normal approximation, e.g. both $\n, \m \ge 50$, the Ansari-Bradley test rejects
the null hypothesis if 
\begin{align}
	&(\ABT{\bx}{\by} - \mu)/ \sigma  \ge \Phi^{-1}(1 - \alpha/2) \label{eqn:sec:wmw:1}\\
	\text{or }&(\ABT{\bx}{\by} - \mu)/ \sigma  \le \Phi^{-1}(\alpha/2) \label{eqn:sec:wmw:2} 
\end{align}
for any given significance level $0 < \alpha < 1$, where $\Phi$ denotes the 
cumulative distribution function of a standard normal distribution.
Denote the rejection region $\mrln{\alpha}$ with respect to $\alpha$ as
\begin{align*}
	\mrln{\alpha} = (-\infty,  \Phi^{-1}(\alpha/2)) \cup (\Phi^{-1}(1-\alpha/2), \infty).
\end{align*}
Then it is equivalent to reject the null hypothesis
if the Ansari-Bradley test statistic 
$\ABT{\bx}{\by}$  falls into the rejection region,
i.e.
\begin{align*}
	(\ABT{\bx}{\by} - \mu)/ \sigma  \in 	\mrln{\alpha} .
\end{align*}

Suppose only a subset of data $\bxp \subset \bx, \byp \subset \by$
are observed, where either $\bxp \neq \bx$ or $\byp \neq \by$. Then 
$\ABT{\bx}{\by}$ can not be computed directly,
and one cannot assess whether 
Inequality $\eqref{eqn:sec:wmw:1}$
or $\eqref{eqn:sec:wmw:2}$ holds.
However, following Theorem~\ref{theorem:3}, we can 
still compute the minimum and maximum possible Ansari-Bradley
test statistics $\WMWmin{\bx}{\by}$, and $\WMWmax{\bx}{\by}$.

Base on these bounds, we propose to reject the null hypothesis
when all possible Ansari-Bradley test
statistics fall into the rejection region $\mrln{\alpha}$.
This is done by rejecting the null hypothesis
when either one of the following two conditions is true.
\begin{condition} \label{condition:wmw:1}
	\begin{align*}
		\frac{\WMWmin{\bx}{\by} - \mu}{\sigma}  \ge \Phi^{-1}\left(1 - \frac{\alpha}{2}\right)\text{ and } \frac{\WMWmax{\bx}{\by}  - \mu}{\sigma}  \ge \Phi^{-1}\left(1 - \frac{\alpha}{2}\right).
	\end{align*}
\end{condition}
\begin{condition} \label{condition:wmw:2}
	\begin{align*}
		\frac{\WMWmin{\bx}{\by} - \mu}{\sigma}  \le \Phi^{-1}\left( \frac{\alpha}{2}\right)\text{ and } \frac{\WMWmax{\bx}{\by}  - \mu}{\sigma}  \le \Phi^{-1}\left( \frac{\alpha}{2}\right).
	\end{align*}
\end{condition}

We now show that when Conditions~\ref{condition:wmw:1}
or \ref{condition:wmw:2} hold, all possible
Ansari-Bradley test statistics fall into 
the rejection region $\mrln{\alpha}$.
Let $\bt$ be any possible Ansari-Bradley 
test statistic across all possible values of missing data.
Since $\WMWmax{\bx}{\by} \ge \bt \ge \WMWmin{\bx}{\by}$, 
and $\Phi^{-1}(\x)$ is a monotonic increasing function,
\begin{align*}
	&\text{Condition~\ref{condition:wmw:1}} \implies \bt \ge \Phi^{-1}\left(1 - \frac{\alpha}{2}\right),\\
	&\text{Condition~\ref{condition:wmw:2}} \implies \bt \le \Phi^{-1}\left(\frac{\alpha}{2}\right).
\end{align*}
In other words, if either Condition~\ref{condition:wmw:1} or 
Condition~\ref{condition:wmw:2}
is true, any possible Ansari-Bradley
test statistic $\bt$ must fall into the rejection
region $\mrln{\alpha}$.

In particular, the Ansari-Bradley
test statistic $\WMW{\bx}{\by}$, which can be computed if the missing
data were fully observed, must also fall into the rejection region, i.e.,
\begin{align*}
	\text{Condition~\ref{condition:wmw:1} or Condition~\ref{condition:wmw:2}  are true} \implies \WMW{\bx}{\by} \in \mrln{\alpha}.
\end{align*}
One important conclusion is that
if the null hypothesis is rejected when
either Condition~\ref{condition:wmw:1} or 
\ref{condition:wmw:2} hold, our approach 
controls the Type I error regardless of the 
number and values of missing data, since we have
\begin{align*}
	\Prob{\text{Condition~\ref{condition:wmw:1} or Condition~\ref{condition:wmw:2}  are true} | \nh} \le \prob{ \WMW{\bx}{\by} \in \mrln{\alpha}|\nh} = \alpha.
\end{align*}
This result is central to our method. Hence we state it more formally as
\begin{theorem}
	Suppose the Type I error of the  Ansari-Bradley test
	is equal to the significance level $\alpha$ when
	the normal approximation is used. The Type I error of the proposed method, which rejects the null hypothesis when either Conditions~\ref{condition:wmw:1} or \ref{condition:wmw:2} are true, 
	is also no greater than the significance
	level $\alpha$, regardless of the number and values of the missing data.
\end{theorem}

\subsubsection{Computing a $p$-value of the proposed method}
\label{wmwm:pvalue}

In Section~\ref{rejectconditions}, we consider
conditions for rejecting all possible test statistics.
We now consider the bounds of $p$-values of the 
Ansari-Bradley test statistic, and show that
all possible test statistics are significant if and only if
the maximum possible $p$-value is less than or equal to the
significance level $\alpha$.

When the normal approximation is used,
the $p$-value of the Ansari-Bradley test
is defined as
\begin{align} \label{eqn:wmwm:pvalue}
	\p(\WMW{\bx}{\by}) = 2\min \left\{ \Phi\left( \frac{\WMW{\bx}{\by} - \mu}{\sigma} \right), 
	1 -  \Phi\left( \frac{\WMW{\bx}{\by} - \mu}{\sigma} \right)\right\}.
\end{align}

In the presence of missing data, while $\p(\WMW{\bx}{\by})$
can not be computed directly, one can compute
the corresponding $p$-values of the minimum and maximum
Ansari-Bradley test statistic as
\begin{align*}
	\p(\WMWmin{\bx}{\by}) = 2\min \left\{ \Phi\left( \frac{\WMWmin{\bx}{\by} - \mu}{\sigma} \right), 1 -  \Phi\left( \frac{\WMWmin{\bx}{\by} - \mu}{\sigma} \right)\right\},\\
	\text{and~}\p(\WMWmax{\bx}{\by}) = 2\min \left\{ \Phi\left( \frac{\WMWmax{\bx}{\by} - \mu}{\sigma} \right), 1 -  \Phi\left( \frac{\WMWmax{\bx}{\by} - \mu}{\sigma} \right)\right\}.
\end{align*}
One caveat here is to consider $\p(\WMWmin{\bx}{\by})$ and
$\p(\WMWmax{\bx}{\by})$ as the minimum and the maximum
possible $p$-value of the Ansari-Bradley test directly.  
In fact, we have the following result regarding the bounds of
the $p$-value of the Ansari-Bradley test in 
the presence of missing data.

\begin{proposition} \label{prop:wmw:boundspvalue}
	Suppose that $\bx = \{\xln{1}, \cdots, \xln{\n}\}$ and 
	$\by = \{\yln{1}, \cdots, \yln{\m}\}$ are samples of distinct, 
	real-valued observations. Suppose that $\bxp \subset \bx$ is 
	a subset of $\np$ values in $\bx$, and suppose that $\byp \subset \by$ 
	is a subset of $\mmp$ values in $\by$. Defining $\bz = \bx \cup \by$ 
	and $\bzp = \bxp \cup \byp$ and supposing only $\bzp$ is known, then
	the $p$-value of the Ansari-Bradley test $\p(\WMW{\bx}{\by})$
	is bounded such that
	\begin{align*}
		\p(\WMW{\bx}{\by})  &\ge \pmin{\bx}{\by} = \min\{	\p(\WMWmin{\bx}{\by}), \p(\WMWmax{\bx}{\by})  \}, \\
		\p(\WMW{\bx}{\by})  &\le \pmax{\bx}{\by} = \left\{ \begin{array}{cl}
			1,~~~\text{ if } (\WMWmin{\bx}{\by} - \mu)(\WMWmax{\bx}{\by} - \mu) \le 0,\\
			\max\{	\p(\WMWmin{\bx}{\by}), \p(\WMWmax{\bx}{\by})  \},~~~\text{otherwise}.
		\end{array}\right.
	\end{align*}
\end{proposition}

\begin{proof}
	The proof is included in Section~\ref{supp:bounds:p-value}
	in the Supplementary Material.
\end{proof}

Proposition~\ref{prop:wmw:boundspvalue} provides
all possible $p$-values of the Ansari-Bradley
test when the data are not entirely observed.
An equivalent approach of rejecting the null hypothesis 
when Condition~\ref{condition:wmw:1} or \ref{condition:wmw:2} holds
is to reject the null hypothesis
when the maximum possible $p$-value
is smaller than or equal to the significance level $\alpha$,
as demonstrated by the following result.

\begin{proposition} \label{prop:wmw:pvalueequivalent}
	Suppose that $\bx = \{\xln{1}, \cdots, \xln{\n}\}$ and 
	$\by = \{\yln{1}, \cdots, \yln{\m}\}$ are 
	partially observed samples of distinct, 
	real-valued observations. 
	Then for any given significance level
	$\alpha < 1$, 
	\begin{align*}
		\text{ Condition~\ref{condition:wmw:1} or \ref{condition:wmw:2} hold} \iff \pmax{\bx}{\by} \le \alpha,
	\end{align*}
	where $\pmax{\bx}{\by}$ is defined in Proposition~\ref{prop:wmw:boundspvalue}.
	Hence, the $p$-value of the proposed method is the maximum possible $p$-value $\pmax{\bx}{\by}$ over all possible values of missing data.
\end{proposition}
\begin{proof}
	The proof is included in Section~\ref{supp:bounds:p-value}
	in the Supplementary Material.
\end{proof}

\section{Simulations for the scale test} \label{chap4:sec:numeric}

This section performs numerical simulations for
investigating the Type I error and statistical power of  
the proposed scale testing method in the presence of missing data. 
The performance of our method is compared 
with the Ansari-Bradly test when the missing data have been imputed using either mean imputation or hot deck imputation, or when the missing data are ignored.

\subsection{As the proportion of missing data varies}
\label{chap4:sec:numeric:proportion}

We first consider experiments comparing the performance
of the proposed method and other methods with varying 
proportion of missing data, while
sample sizes, missingness mechanisms, 
and the distributions of the data, are fixed for each experiment.

\subsubsection{Case 1: Missing completely at random data}

The first experiment considers the case where data are missing completely at 
random. Observations in $\bx$ are sampled independently from a 
$\mathrm{N}(0,1)$ distribution, while observations in $\by$ are sampled 
independently from a $\mathrm{N}(0,1)$ distribution to evaluate the Type I
error, and from a $\mathrm{N}(0, \sigma^2)$ distribution to evaluate the statistical 
power with $\sigma = 3$. 
A proportion $\misss \in [0, 0.3]$ of the observations
$\bx=\{\xx_1, \dots, \xx_{\n}\}$ are selected completely at random to be marked
as missing. The same proportion $\misss$ of $\by=\{\yy_1, \dots, \yy_{\m}\}$
are selected completely at random to be marked as missing. 
Sample sizes $\n = \m = 100$ are considered.

\begin{figure}
	\includegraphics[width=\textwidth]{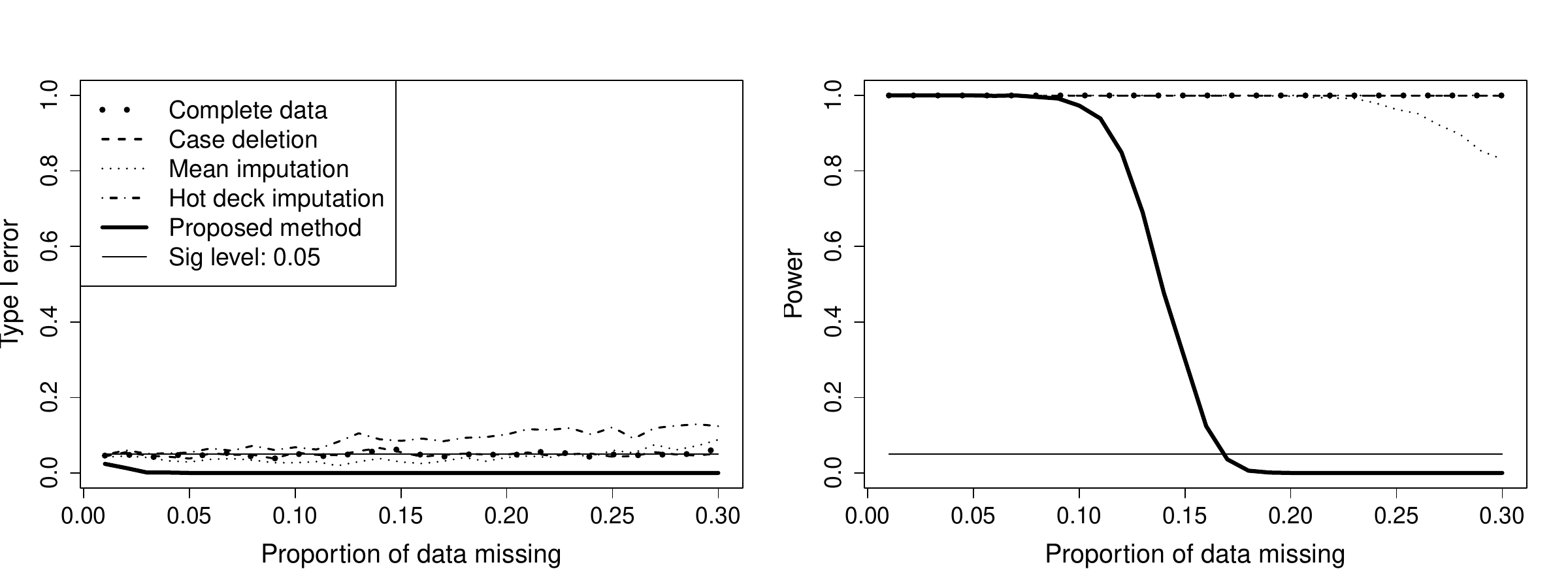}
	\caption{The Type I error and statistical power of the proposed method and 
		the standard Ansari-Bradley test after the missing data is either known or
		has been imputed or ignored as the proportion of missing data increases.
		The data is missing completely at random.
		(Left) Type I error: $\mathrm{N}(0,1)$ vs $\mathrm{N}(0,1)$; 
		(Right) Power: $\mathrm{N}(0,1)$ vs $\mathrm{N}(0,\sigma^2)$,
		with the scale parameter $\sigma = 3$. For both figures, 
		a significance threshold of $\alpha=0.05$ has been used and the total
		sample sizes are $\n=100$, $\m=100$, and $1000$ trials were used.}
	\label{fig:AB:mcar1}
\end{figure}

Figure~\ref{fig:AB:mcar1} shows that the Type I error is not controlled by
either hot deck imputation or mean imputation, although it is controlled
for the proposed method and for the case deletion case when the missing data 
are ignored. On the other hand, all methods have good power, except that
the power for the proposed method decreases significantly when more
than $10\%$ of the data is missing.

For this experiment, $\n=\m=100$, but Figure~\ref{fig:ab:mcar2} in Section~\ref{addtionalexp:ab:1} of the Supplementary Material considers an
experiment with different
sample sizes $\n, \m = 500$, which shows the increasing power
of the proposed method.
Furthermore, if there is a larger difference between the two samples for the
power experiment, for example if the scale parameter $\sigma$ of the
second sample is 5 rather than 3, then the proposed
method can still have good power for over $15\%$ of the data is missing; see
Figure~\ref{fig:ab:mcar3} in Section~\ref{addtionalexp:ab:1} of the Supplementary Material.

%%- - - - - - - - - - - - - - - - - - - - - - - - - - - - - - - - - - - - -%%

\subsubsection{Case 2: Missing not at random data}

The second experiment is the same as the first, except that 
in this experiment the data 
are missing not at random. The missingness mechanism is as follows: if $\misss$
is the proportion of observations to be missing, then for any observation
$\xx_i \in \{\xx_1, \dots, \xx_{\n}\}$, the probability of $\xx_i$ being 
missing is 
\begin{align}
	\prob{\textrm{$\xx_i$ is missing}} = 
	\begin{cases}
		\min(1, \misss \n / \sum_{j=1}^{\n} \indicator{|\xx_j| < 1}), &\quad\text{if $|\xx_i| < 1$}, \\
		\max(0, \misss \n / \sum_{j=1}^{\n} \indicator{|\xx_j| < 1} -1), &\quad\text{otherwise}, \\
	\end{cases}
	\label{eqn:AB:missmechonex}
\end{align}
where $\indicator{A}$ is the indicator function for the event $A$.
In other words, when there are more values in $\bx$ with 
their absolute values smaller than one
than the number of missing values in $\bx$, i.e.
$\misss \n / \sum_{j=1}^{\n} \indicator{|\xx_j| < 1} < 1$,
only values with 
absolute values smaller than one in $\bx$ can be missing. Otherwise
all values with 
absolute values smaller than one in $\bx$ will be missing, and 
values with 
absolute values larger
than one will be randomly missing.

For $\yy_i \in \{\yy_1, \dots, \yy_{\m}\}$, a similar missingness mechanism
\begin{align}
	\prob{\textrm{$\yy_i$ is missing}} = 
	\begin{cases}
		\min(1, \misss \m / \sum_{j=1}^{\m} \indicator{|\yy_j| > 1}), &\quad\text{if $|\yy_i| > 1$}, \\
		\max(0, \misss \m / \sum_{j=1}^{\m} \indicator{|\yy_j| > 1} -1), &\quad\text{otherwise}, \\
	\end{cases}
	\label{eqn:AB:missmechoney}
\end{align}
is used. When there are more values in $\by$ with 
its absolute values larger than one
than the number of missing values in $\by$, i.e.
$\misss \m / \sum_{j=1}^{\m} \indicator{|\yy_j| > 1} < 1$,
only values with absolute values lager than one in $\by$ can be missing. Otherwise
all values with 
absolute values larger than one in $\by$ will be missing, and 
values with absolute values smaller than one will be randomly missing.
We choose missingness mechanisms for 
$\bx$ and $\by$ as Equation~\eqref{eqn:AB:missmechonex}
and \eqref{eqn:AB:missmechoney} in order to create
different missing patterns for $\bx$ and $\by$.

\begin{figure}
	\includegraphics[width=\textwidth]{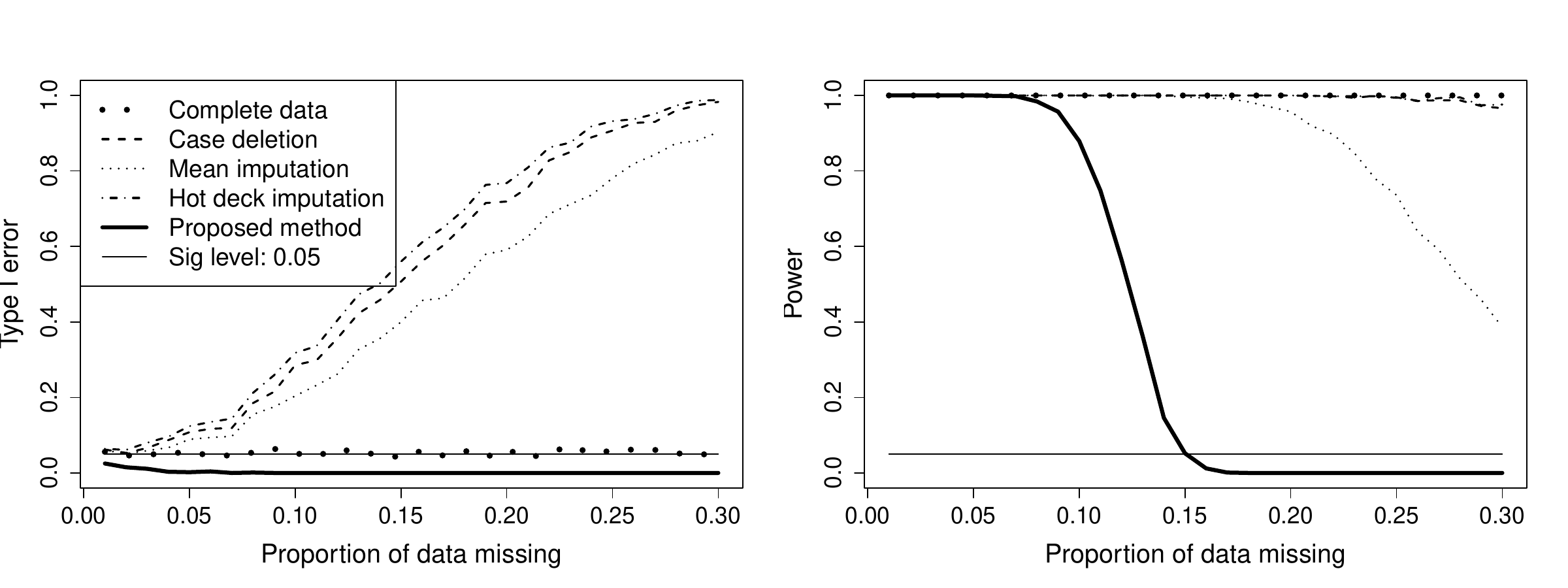}
	\caption{The Type I error and statistical power of the proposed method and 
		the standard Ansari-Bradley test after the missing data is either known or
		has been imputed or ignored as the proportion of missing data increases.
		The data is missing not at random, according to the mechanism in 
		Equation~\eqref{eqn:AB:missmechonex}, and \eqref{eqn:AB:missmechoney}.
		(Left) Type I error: $\mathrm{N}(0,1)$ vs $\mathrm{N}(0,1)$; 
		(Right) Power: $\mathrm{N}(0,1)$ vs $\mathrm{N}(0,\sigma)$,
		with scale parameter $\sigma = 3$. For both figures, 
		a significance threshold of $\alpha=0.05$ has been used and the total
		sample sizes are $\n=100$, $\m=100$, and $1000$ trials were used.}
	\label{fig:AB:mnar1}
\end{figure}

Figure~\ref{fig:AB:mnar1} shows that when the data from both samples
are missing not at random and follow the above missingness mechanisms, then the Type I error is not controlled 
for the imputation methods nor the case deletion method.
This relatively simple example illustrates the potential peril of not taking 
missing data into account.
On the other hand, the proposed method controls the Type I error rate for this case.
The statistical power of these methods appears similar to that for the missing 
completely at random case in Figure~\ref{fig:AB:mcar1}; all methods have good power,  although the proposed method's power decreases as the proportion of missing data increases beyond $10\%$ of the total.

\subsection{As the sample size varies}
\label{chap4:sec:numeric:scalesamplesize}

We still consider the data to be missing not at random 
following the mechanisms specified in 
Equation~\eqref{eqn:AB:missmechonex}, and \eqref{eqn:AB:missmechoney},
but now as the sample sizes $\n, \m$ increase, with a fixed proportion
of missing data $\misss = 0.1$.

\begin{figure}
	\includegraphics[width=\textwidth]{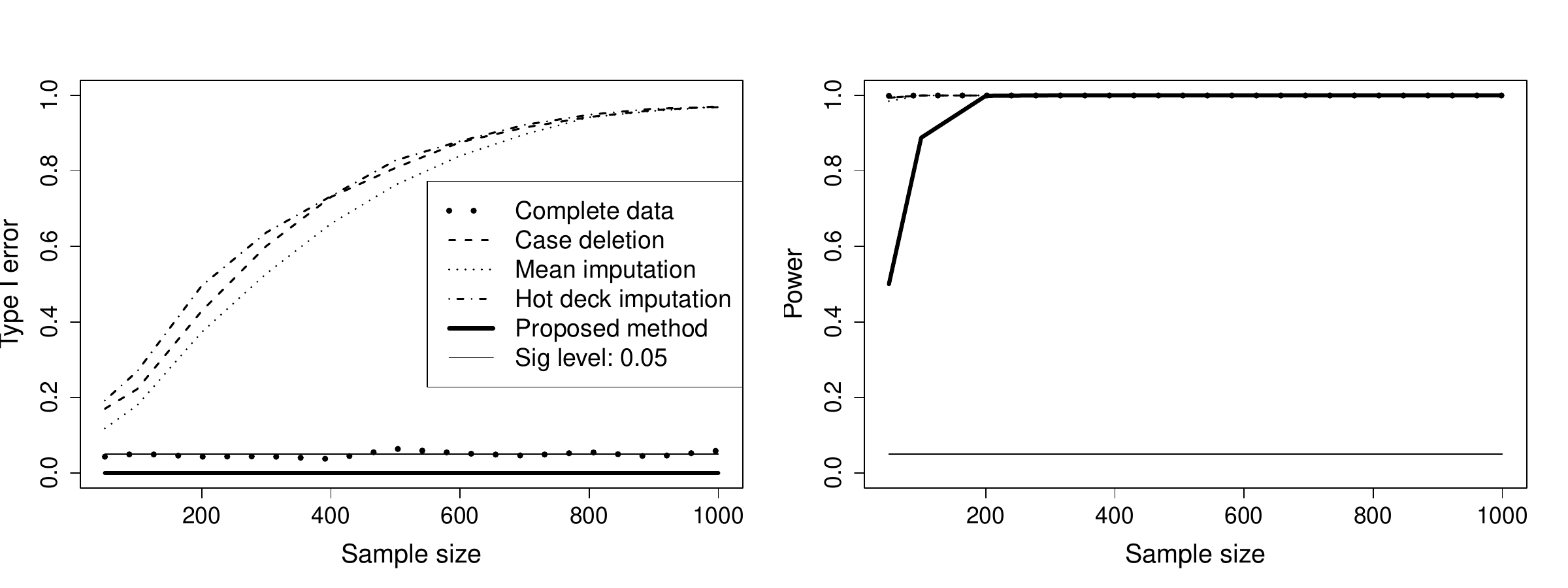}
	\caption{The Type I error and statistical power of the proposed method and 
		the standard Ansari-Bradley test after the missing data is either known or
		has been imputed or ignored as the sample sizes increase.
		The data is missing not at random, according to the mechanism in 
		Equation~\eqref{eqn:AB:missmechonex}, and \eqref{eqn:AB:missmechoney}.
		(Left) Type I error: $\mathrm{N}(0,1)$ vs $\mathrm{N}(0,1)$; 
		(Right) Power: $\mathrm{N}(0,1)$ vs $\mathrm{N}(0,\sigma^2)$ with the scale parameter $\sigma = 3$. For both figures, 
		a significance threshold of $\alpha=0.05$ has been used and the proportion
		of missing data for both samples is $\misss = 0.1$, and $1000$ trials were used.}
	\label{fig:AB:mnar:samplesize1}
\end{figure}

Figure~\ref{fig:AB:mnar:samplesize1} further highlights the potential
peril of not taking the missing data into account: the Type I error of case deletion,
mean imputation and hot deck imputation methods appear
to be asymptotically converge to 1 as the sample size increases
with a fixed proportion of missing data $s = 0.1$ in this experiment.
In contrast, the proposed method
controls the Type I error across all sample sizes considered. All methods
appear to have good power. 
The power of the proposed method increases 
with sample sizes. When the sample size is 200, the power of the proposed method
is close to 1.

This section evaluates the Type I error and statistical
power of the proposed scale testing 
method in the presence of missing data.
The method is compared with the Ansari-Bradley
test when the missing data have been imputed using 
mean or hot deck imputation, or when the data are ignored,
or when the data are known, i.e. without missing data.

The simulation results show that
the proposed method controls the Type I error, while imputation methods
and case deletion fail to control the Type I error, especially when the data
are missing not at random. The problem of the imputation and case deletion methods
appear to be more serious with larger sample sizes.
Figure~\ref{fig:AB:mnar:samplesize1} 
shows that the Type I error for these methods 
asymptotically converges to 
$1$ as the sample size increases.

The power of the proposed method
drops significantly when more than $10\%$
of the data are missing, as we have shown in 
Figure~\ref{fig:AB:mcar1} and \ref{fig:AB:mnar1}.
This is because our method rejects the null hypothesis
only when all possible test statistics are significant,
across all possible values of imputations. When 
more data are missing, there are more possible different combinations
of imputations. Hence, the upper and lower bounds
of the Ansari-Bradley test statistics increase, and decrease,
respectively, resulting in a higher $p$-value for the proposed method.

The power of the proposed method is also related to the 
sample size and the scale parameter $\sigma$.
Greater power of the proposed method can be achieved,
when the sample sizes $\n, \m$ increase, and/or when
larger scale parameter $\sigma$ of the distribution 
generating $\by$ are provided,
as demonstrated by experiments in Section~\ref{chap4:sec:numeric:scalesamplesize}
and Section~\ref{chap4:sec:numeric:mean}.

Additional experiments for evaluating
the proposed method are included in the
Supplementary Material.
The simulations which consider increasing 
scale parameters of the second sample $\by$ are provided in Section~\ref{chap4:sec:numeric:mean}. 
In this section, we consider the case when the data
are normally distributed. However, the Ansari-Bradley test
and the proposed method can also be applied for 
non-normal distributed data. Section~\ref{addtionalexp:ab:3} in
Supplementary Material considers the data to be Gamma-distributed,
rather than normally distributed. In addition, Section~\ref{addtionalexp:ab:4}
considers a different missing not at random mechanism that is different from the mechanism specified by Equation~\eqref{eqn:AB:missmechonex} and \eqref{eqn:AB:missmechoney}, while also presenting similar results as shown in this section.

\section{Location-scale test with missing data} \label{combiningABTandWMWwithmissing}

In Section~\ref{mainresults}, we propose
a scale testing method in the presence of 
missing data, based on the Ansari-Bradley
test. Based on this, we now introduce a location-scale 
testing method in the presence of missing data by
combining the proposed scale testing 
method with the location testing method proposed by \citet{zeng2024two}, using the Holm–Bonferroni method \citep{holm1979simple}.

As discussed in Section~\ref{sec:locationscaletest}, the location-scale testing problem is a combination of both the 
location test and the scale test problems: the goal is to 
determine whether the distributions of $\bx$ and $\by$ have equal location and scale, or if either location or scale differs significantly.

\subsection{Extension to a location-scale testing method} \label{subsec:holmbonferroni}

For constructing the proposed location-scale testing method,
we first review the Holm-Bonferroni method, which is a common procedure used to control the Type I error when performing multiple hypothesis testing. Specifically,
suppose we are testing $\kk$ hypotheses with $p$-values
$\pln{1},\ldots,\pln{\kk}$, the Holm-Bonferroni procedure 
for a significance level $\alpha \in (0,1)$ 
is as follows: 
\begin{itemize}	
	\item Sort the $p$-values in ascending order:
	\begin{align*}
		\pln{(1)} &\leq \dots \leq \pln{(\kk)}.
	\end{align*}
	Let $H_{(1)}, \dots, H_{(\kk)}$ be the corresponding null hypotheses.
	
	\item For each $i \in \{1, \dots, \kk\}$, compare \(p_{(i)}\) with the threshold:
	\begin{align*}
		\pln{(i)} &\leq \frac{\alpha}{\kk - i + 1}.
	\end{align*}
	
	\item Find the smallest index $t$ such that
	\begin{align*}
		\pln{(t)} &> \frac{\alpha}{\kk - t + 1}.
	\end{align*}
	
	\item Reject all null hypotheses $H_{(1)}, \ldots, H_{(t-1)}$
	and do not reject $H_{(t)}, \ldots, H_{(\kk)}$.
\end{itemize}

Then the probability of rejecting at least one true $H_{(1)}, \dots, H_{(\kk)}$,
that is, making at least one Type I error, 
is smaller than the significance level $\alpha$.

Applying the Holm-Bonferroni method, a location-scale 
testing method based on the Wilcoxon-Mann-Whitney test
and the Ansari-Bradley test can be constructed as follows:

\begin{itemize}
	\item Compute the $p$-values for the Wilcoxon-Mann-Whitney test and the Ansari-Bradley test.
	\item Sort the $p$-values in ascending order:
	\begin{align*}
		\pln{(1)} \leq \pln{(2)}.
	\end{align*}
	\item Compare $\pln{(1)}$ with the thresholds ${\alpha}/{2}$.
	If the inequality $\pln{(1)} \leq \alpha/2$ holds, we reject the null hypothesis, otherwise we do not reject the null hypothesis.
\end{itemize}

As the goal for the location-scale testing problem
is to determine if there is location
or scale difference, we reject the null hypothesis
so long as the null hypothesis for one of the location
test or the scale test is rejected. This means we only
need to consider if $\pln{(1)} \leq \alpha/2$ according to the Holm-Bonferroni method: if 
$\pln{(1)} \leq \alpha/2$, then at least one hypothesis is rejected, 
regardless the value of $\pln{(2)}$. However, 
if $\pln{(1)} > \alpha/2$, then both hypotheses are not rejected.

Further, since the null hypothesis is rejected when $\pln{(1)} \le \alpha/2$, we can define the $p$-value of this location-scale testing method as $\pln{\text{Holm}} = \min \{1, 2 \pln{(1)}\}$, where $\pln{(1)}$ is the minimum of the $p$-values for the Wilcoxon-Mann-Whitney test and the Ansari-Bradley test. More formally,
we define

\begin{definition} \label{pvalue:holm}
	Suppose $\bx$ and $\by$ are samples of distinct real values. Denote
	the $p$-values of the Wilcoxon-Mann-Whitney test and the Ansari-Bradley test over $\bx$ and $\by$ as $\pln{\text{WMW}}$
	and $\pln{\text{ABT}}$, respectively. Then, we define the $p$-value
	of the combined Wilcoxon-Mann-Whitney test and Ansari-Bradley test
	for location-scale testing using the Holm-Bonferroni method as
	\begin{align*}
		\pln{\text{Holm}} = \min \{1, 2 \pln{(1)}\}, \text{~where~}\pln{(1)} = \min \{\pln{\text{WMW}}, \pln{\text{ABT}}\}.
	\end{align*}
\end{definition}
Following the Definition~\ref{pvalue:holm}, the null hypothesis of the location-scale test based on the Holm-Bonferroni method is rejected when the $p$-value $\pln{\text{Holm}}$ is smaller than a significance level $\alpha$.

For constructing a valid location-scale testing method
in the presence of missing data, we propose to bound 
the $p$-value $\pln{\text{Holm}}$ using the bounds of $p$-value of the Ansari-Bradley test from Proposition~\ref{prop:wmw:boundspvalue},
and the bounds of $p$-value of the Wilcoxon-Mann-Whitney test provided by \cite{zeng2024two}. Specifically, we have the following result:

\begin{proposition} \label{prop:location-scale}
	Suppose $\bx$ and $\by$ are samples of distinct real-valued observations, with only a subset of their values observed.
	Let $\pln{\text{ABT}, \min}$ and $\pln{\text{ABT}, \max}$ denote the minimum and maximum $p$-values of the Ansari–Bradley test, and let $\pln{\text{WMW}, \min}$ and $\pln{\text{WMW}, \max}$ denote those of the Wilcoxon–Mann–Whitney test. Then, the $p$-value of the combined location–scale test based on these two tests, using the Holm–Bonferroni procedure, is bounded as follows
	\begin{align*}
		\min\{1, 2\min\{\pln{\text{ABT}, \min}, \pln{\text{WMW}, \min}\}\} \le \pln{\text{Holm}} \le \min\{1, 2\min\{\pln{\text{ABT}, \max}, \pln{\text{WMW}, \max}\}\}.
	\end{align*}
\end{proposition}
\begin{proof}
	To start, recall that the $p$-value of the combined location–scale test base on the Wilcoxon–Mann–Whitney test and the Ansari–Bradley test is defined in Definition~\ref{pvalue:holm} as 
	\begin{align*}
		\pln{\text{Holm}} = \min \{1, 2 \pln{(1)}\}, \text{~where~}\pln{(1)} = \min \{\pln{\text{WMW}}, \pln{\text{ABT}}\}.
	\end{align*}
	Since
	\begin{align*}
		\pln{\text{WMW}, \min} \le \pln{\text{WMW}}, \text{ and } \pln{\text{ABT}, \min} \le \pln{\text{ABT}},
	\end{align*}
	we have 
	\begin{align*}
		&2\min\{\pln{\text{ABT}, \min}, \pln{\text{WMW}, \min}\} \le 2 \min \{\pln{\text{WMW}}, \pln{\text{ABT}}\} =  2 \pln{(1)},\\
		\implies & \min\{1, 2\min\{\pln{\text{ABT}, \min}, \pln{\text{WMW}, \min}\}\} \le \pln{\text{Holm}}.
	\end{align*}
	Similarly, since 
	\begin{align*}
	\pln{\text{WMW}} \le \pln{\text{WMW}, \max}, \text{ and } \pln{\text{ABT}} \le \pln{\text{ABT}, \max},
	\end{align*}
	we have
	\begin{align*}
		&2 \pln{(1)} = 2 \min \{\pln{\text{WMW}}, \pln{\text{ABT}}\} \le 2\min\{\pln{\text{ABT}, \max}, \pln{\text{WMW}, \max}\} \\
		\implies & \pln{\text{Holm}} \le \min\{1, 2\min\{\pln{\text{ABT}, \max}, \pln{\text{WMW}, \max}\}\}.
	\end{align*}
	This completes our proof.
\end{proof}

Following Proposition~\ref{prop:location-scale}, the proposed location-scale testing method based on the Holm-Bonferroni method
rejects the null hypothesis when 
\begin{align*}
	\min\{1, 2\min\{\pln{\text{ABT}, \max}, \pln{\text{WMW}, \max}\}\} \le \alpha.
\end{align*}
Then, since the $p$-value $\pln{\text{Holm}}$ controls the Type I error, the proposed method also controls the type I error, without assuming missingness mechanisms.

\subsection{Evaluation of the location-scale test without missing data}

Section~\ref{subsec:holmbonferroni} proposed a location-scale
testing with missing data method by combining the Wilcoxon-Mann-Whitney test 
and the Ansari-Bradley test using the Holm-Bonferroni
method. However, as we mentioned previously in Section~\ref{sec:locationscaletest}, 
a more common method for the location-scale testing problem
is known as the Lepage test. The Lepage test statistic is defined as the 
sum of the squared Wilcoxon-Mann-Whitney test statistic
and the Ansari-Bradley test statistic after scaling:
\begin{align*}
	\Lepage(\bx, \by) = \left(\frac{\WMW{\bx}{\by} - \mu_{\w}}{\sigma_{\w}}\right)^2 + 	\left(\frac{\ABT{\bx}{\by} - \mu_{\st}}{\sigma_{\st}}\right)^2,
\end{align*}
where $\WMW{\bx}{\by}$ denotes the Wilcoxon-Mann-Whitney test statistic,
$\ABT{\bx}{\by}$ denotes the Ansari-Bradley test statistic, 
and $\mu_{\w}, \mu_{\st}$ and $\sigma_{\w}^2, \sigma_{\st}^2$ are the mean and 
variance of the Wilcoxon-Mann-Whitney test statistic and the Ansari-Bradley test statistic under the null hypothesis.

\begin{table}
	\small
	\centering
	\caption{Estimated Type I error and power of the Lepage test  (Lep)
		and the proposed method (PM) based on Holm–Bonferroni method 
		without missing data. The sample $\bx$ is drawn from distribution $F(x)$, 
		while $\by$ is generated by first sampling $\bz \sim F(x)$ and then applying the transformation $\bz / a + b$. The sample sizes
		for $\bx$ and $\by$ are $\n$ and $\m$, respectively. 
		When $a = 1$ and $b = 0$, the table reports the estimated Type I error; 
		otherwise, it reports the estimated power. A significance level of 
		$\alpha = 0.05$ is used. 
		Each entry corresponds to the proportion of null hypothesis 
		rejections over 1000 Monte Carlo replications. \\
	}
	{	\begin{tabular}{lrrrrrrrrrr}
			%	\hline
			$F(x)$ & \multicolumn{2}{c}{Parameters} & \multicolumn{2}{c}{$n=m=50$} & \multicolumn{2}{c}{$n=m=75$} &   \multicolumn{2}{c}{$n=m=100$} &   \multicolumn{2}{c}{$n =m=125$}\\
			%\cmidrule{2-4} \cmidrule{6-8} 
			& a & b& Lep & PM  & Lep & PM & Lep & PM & Lep & PM\\ 
			%	\hline
			Normal & 1    & 0    & 0.039 & 0.040 & 0.049 & 0.050 & 0.055 & 0.053 & 0.039 & 0.035 \\
			&1    & 0.25 & 0.167 & 0.168 & 0.228 & 0.241 & 0.311 & 0.306 & 0.390 & 0.398 \\
			&1    & 0.5  & 0.556 & 0.568 & 0.753 & 0.755 & 0.873 & 0.879 & 0.931 & 0.941 \\
			&1    & 0.75 & 0.906 & 0.916 & 0.979 & 0.980 & 0.999 & 0.999 & 1.000 & 1.000 \\
			\cline{2-11}
			&1.25 & 0    & 0.173 & 0.169 & 0.218 & 0.221 & 0.326 & 0.331 & 0.389 & 0.403 \\
			&1.5  & 0    & 0.461 & 0.466 & 0.671 & 0.676 & 0.821 & 0.811 & 0.903 & 0.901 \\
			&1.75 & 0    & 0.750 & 0.771 & 0.917 & 0.920 & 0.972 & 0.979 & 0.992 & 0.995 \\
			\cline{2-11}
			&1.25 & 0.25 & 0.334 & 0.293 & 0.494 & 0.428 & 0.626 & 0.554 & 0.730 & 0.669 \\
			&1.5  & 0.5  & 0.887 & 0.835 & 0.984 & 0.961 & 0.996 & 0.991 & 1.000 & 0.999 \\
			&1.75 & 0.75 & 1.000 & 0.999 & 1.000 & 1.000 & 1.000 & 1.000 & 1.000 & 1.000 \\
			\\
			Skewed &1    & 0    & 0.041 & 0.037 & 0.045 & 0.046 & 0.052 & 0.046 & 0.052 & 0.048 \\
			&1    & 0.25 & 0.492 & 0.466 & 0.663 & 0.651 & 0.798 & 0.777 & 0.901 & 0.892 \\
			&1    & 0.5  & 0.974 & 0.972 & 0.999 & 0.997 & 1.000 & 1.000 & 1.000 & 1.000 \\
			&1    & 0.75 & 1.000 & 1.000 & 1.000 & 1.000 & 1.000 & 1.000 & 1.000 & 1.000 \\
						\cline{2-11}
			&1.25 & 0    & 0.227 & 0.216 & 0.301 & 0.270 & 0.414 & 0.380 & 0.488 & 0.440 \\
			&1.5  & 0    & 0.584 & 0.526 & 0.793 & 0.740 & 0.917 & 0.877 & 0.955 & 0.923 \\
			&1.75 & 0    & 0.867 & 0.822 & 0.979 & 0.947 & 0.997 & 0.991 & 1.000 & 0.998 \\
						\cline{2-11}
			&1.25 & 0.25 & 0.853 & 0.876 & 0.965 & 0.971 & 0.998 & 0.998 & 0.998 & 0.998 \\
			&1.5  & 0.5  & 1.000 & 1.000 & 1.000 & 1.000 & 1.000 & 1.000 & 1.000 & 1.000 \\
			&1.75 & 0.75 & 1.000 & 1.000 & 1.000 & 1.000 & 1.000 & 1.000 & 1.000 & 1.000 \\
			\\
			Student's $t$ &1    & 0    & 0.042 & 0.041 & 0.052 & 0.052 & 0.053 & 0.041 & 0.053 & 0.053 \\
			&1    & 0.25 & 0.132 & 0.138 & 0.173 & 0.178 & 0.218 & 0.212 & 0.252 & 0.260 \\
			&1    & 0.5  & 0.385 & 0.396 & 0.572 & 0.591 & 0.710 & 0.727 & 0.797 & 0.823 \\
			&1    & 0.75 & 0.736 & 0.742 & 0.902 & 0.913 & 0.972 & 0.974 & 0.991 & 0.991 \\
			\cline{2-11}
			&1.25 & 0    & 0.143 & 0.144 & 0.198 & 0.192 & 0.241 & 0.241 & 0.315 & 0.299 \\
			&1.5  & 0    & 0.356 & 0.366 & 0.521 & 0.510 & 0.620 & 0.627 & 0.745 & 0.763 \\
			&1.75 & 0    & 0.631 & 0.635 & 0.806 & 0.817 & 0.925 & 0.928 & 0.963 & 0.971 \\
			\cline{2-11}
			&1.25 & 0.25 & 0.237 & 0.214 & 0.361 & 0.327 & 0.471 & 0.408 & 0.539 & 0.465 \\
			&1.5  & 0.5  & 0.762 & 0.686 & 0.927 & 0.882 & 0.978 & 0.947 & 0.993 & 0.986 \\
			&1.75 & 0.75 & 0.983 & 0.971 & 0.998 & 0.996 & 1.000 & 1.000 & 1.000 & 1.000 \\
			\\
	\end{tabular}}
	\label{tab:AB:1}
	
	Lep, Lepage test. PM, Proposed method. Normal, standard normal distribution; Skewed Normal, skewed normal distribution with shape parameter $\lambda = 4$, 
	as illustrated in Equation~\eqref{eqn:skewedNormal}; Student's $t$, Student's $t$ distribution with degrees of freedom $\nu=3$.
\end{table}

An alternative approach for constructing the location-scale testing
with missing data method is to adapt the Lepage test following the 
similar approach considered in Section~\ref{ABtestwithmissingdata}, 
based on the tight bounds for the test statistics in the 
presence of missing data.
However, obtaining the bounds for the Lepage test statistic 
may be technically challenging. Moreover, it is not immediately 
clear that the Lepage test constructed such would have greater 
power than the proposed method based on the Holm-Bonferroni
method.

We cannot directly compare the two approaches directly 
in the presence of missing data, as
the bounds of the Lepage test statistic
in the presence of missing data are not available.
Consequently, we evaluate the two approaches 
empirically \emph{without} the missing data.

We consider experiments where the sample $\bx$ is drawn from a distribution $F(x)$, while $\by$ is generated by first sampling $\bz \sim F(x)$ and then applying the transformation $\by = \bz / a + b$. We evaluate the proposed method against the Lepage test across various choices of $a$, $b$, and distribution $F(x)$.

Table~\ref{tab:AB:1} shows that both the Lepage test
and the proposed method control the Type I error,
while have comparable power under various alternative hypotheses.
When the data are normally distributed, the Lepage test 
seems to be more powerful when $a \neq 1$ and $b \neq 0$. 
However, when $a = 1$, $b \neq 0$, 
or $a \neq 1$, $b = 0$, the proposed method
appears to be slightly more powerful, although
the results are all very similar.

Apart from the standard normal distribution, we also consider
the skewed normal distribution \citep{azzalini1985class} to incorporate
the case when the skewness of the distribution is non-zero. 
The skewed normal distribution is a family of distributions that extends the normal distribution by incorporating an additional shape parameter $\lambda$ to regulate the skewness. Specifically, the density function of a random variable $\bz$ following skewed normal distribution is defined as
\begin{align} \label{eqn:skewedNormal}
	\phi(\z;\lambda) = 2\phi(\z)\Phi(\lambda\z),~\z \in \mathbb{R}
\end{align}
where $\phi$ and $\Phi$ are the standard normal density and distribution, respectively.

When the data follow a skewed normal distribution, 
the proposed method appears to have similar or better 
performances than the Lepage test
when $a \neq 1$ and $b \neq 0$. Otherwise
the proposed method seems to have slightly less power.

Finally, we consider the cases when the data following 
student's $t$-distributions to incorporate the case 
when the distributions have heavy tails, The conclusion
for this case is similar to that when the data are normally distributed.

The experiment results in Table~\ref{tab:AB:1} 
empirically demonstrate that the Lepage test and the proposed method have similar performances under various alternatives, when data are completely observed.
These results might suggest that the proposed method in the 
presence of missing data has comparable power to the method 
based on the bounds of the Lepage test statistic, although
this implication is not guaranteed.

In next section, we further evaluate the Type I error and 
statistical power of the proposed location-scale
testing method in the presence of missing data.

\subsection{Numerical simulations}

We perform numerical simulations for evaluating the 
proposed location-scale testing method. The first experiment 
is similar to the one considered for Figure~\ref{fig:AB:mnar1}
when the data are missing not at random, except for evaluating the power of the proposed location-scale test with missing data method, the samples of $\bx$ are drawn independently from N(0,1) while the samples of $\by$ are drawn independently from N(1,3).

\begin{figure}
	\centering
	\includegraphics[scale=0.365]{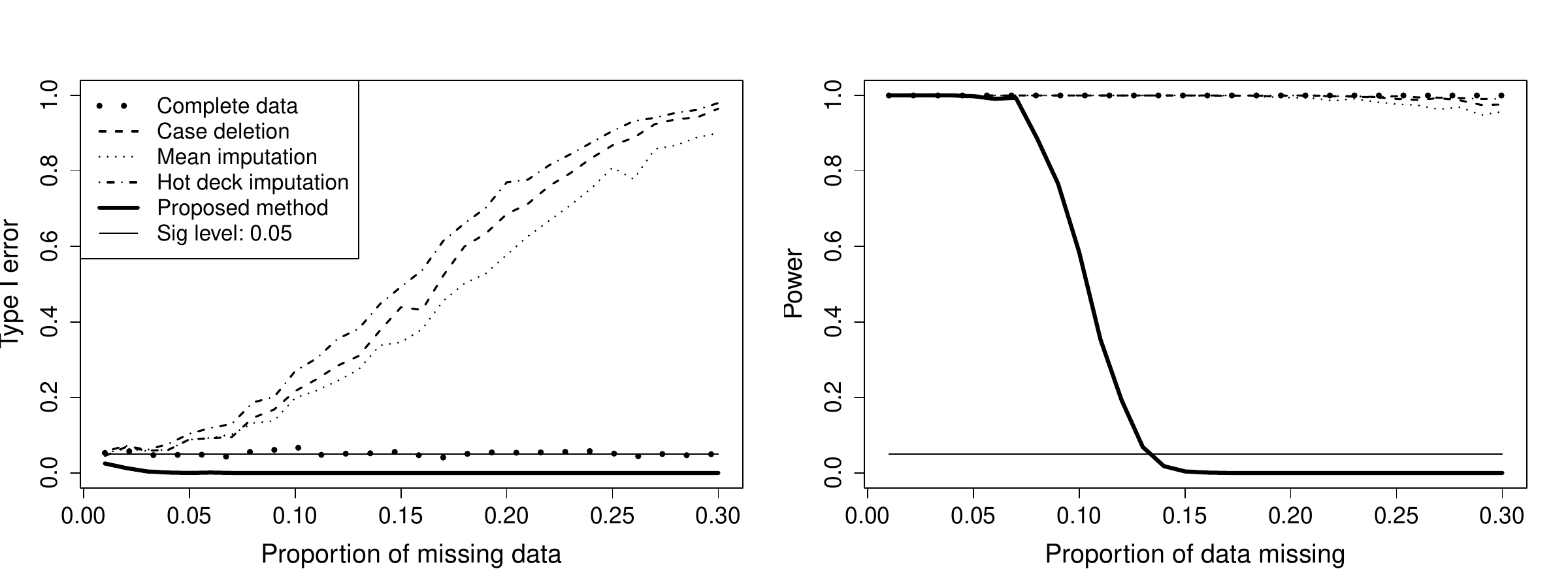}
	\caption{The Type I error and statistical power of the proposed location-scale test with the missing data method, compared to the combined Ansari-Bradley test and Wilcoxon-Mann-Whitney test using Holm-Bonferroni correction after mean imputation, hot deck imputation, or ignoring missing data. The data is missing not at random, according to the mechanism in Equation~\eqref{eqn:AB:missmechonex}, and \eqref{eqn:AB:missmechoney}.
	(Left) Type I error: $\mathrm{N}(0,1)$ vs $\mathrm{N}(0,1)$; (Right) Power: $\mathrm{N}(0,1)$ vs $\mathrm{N}(\mu,\sigma^2)$ with $\mu = 1$ and $\sigma = 3$. A significance level of $\alpha = 0.05$ was used for both figures. The total sample sizes are $n = 100$ and $m = 100$. 1000 trials are conducted.}
	\label{figure:4}
\end{figure}

\begin{figure}
	\includegraphics[width=\textwidth]{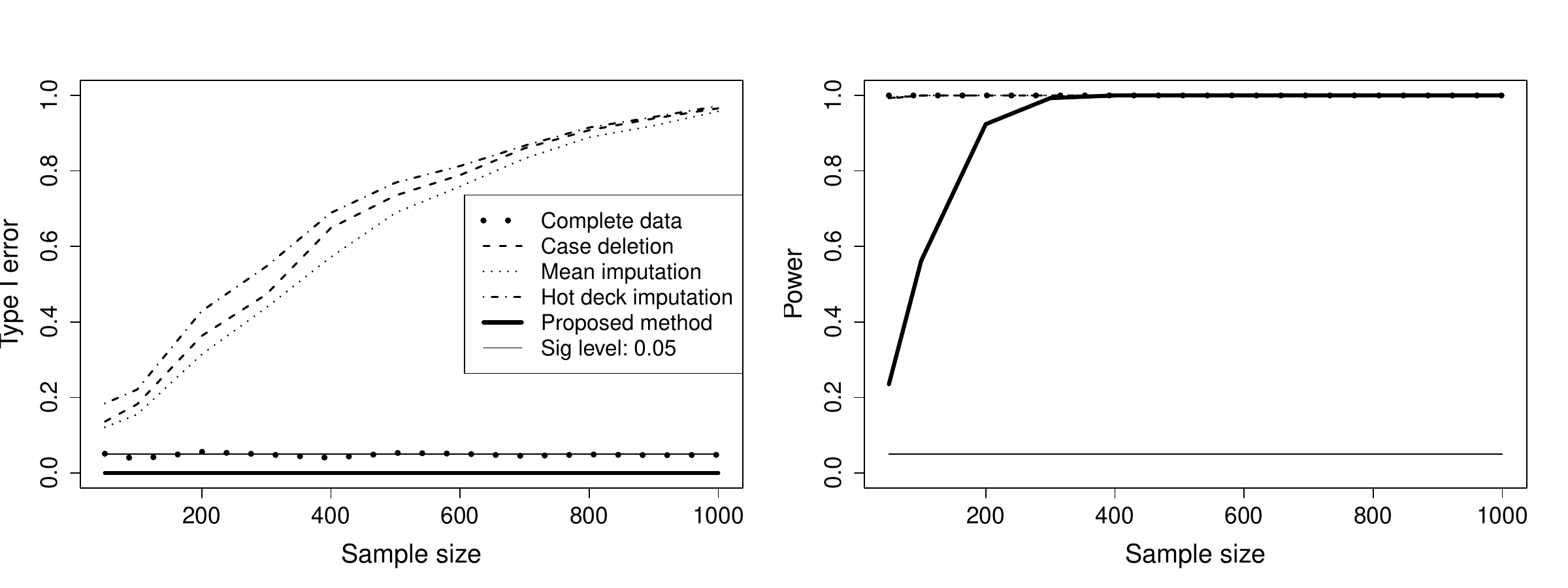}
	\caption{The Type I error and statistical power of the proposed 
		location-scale testing method, the proposed scale testing method,
		the proposed location testing method, and 
		the combination of the Ansari-Bradley test and the Wilcoxon-Mann-Whitney test using the Holm-Bonferroni method after the missing data is either known or has been imputed or ignored as the sample size increases.
		The data is missing not at random, according to the mechanism in Equation~\eqref{eqn:AB:missmechonex}, and \eqref{eqn:AB:missmechoney}.
		(Left) Type I error: $\mathrm{N}(0,1)$ vs $\mathrm{N}(0,1)$; 
		(Right) Power: $\mathrm{N}(0,1)$ vs $\mathrm{N}(\mu,\sigma^2)$ with the location parameter $\mu = 1$ and scale parameter $\sigma = 3$. For both figures, a significance threshold of $\alpha=0.05$ has been used and the proportion of missing data for both samples is $\misss = 0.1$, and $1000$ trials were used.}
	\label{fig:ABWMW:mnar:samplesize1}
\end{figure}

Figure~\ref{figure:4} shows similar results as Figure~\ref{fig:AB:mnar1}, where only the proposed method and the complete data method controls the Type I error, and the power of the proposed method drops to 0 when the proportion of missing data is larger than 15\%.

The second experiment considers a similar experiment
to the one for Figure~\ref{fig:AB:mnar:samplesize1}, where
the sample sizes are increasing. Figure~\ref{fig:ABWMW:mnar:samplesize1} 
shows the Type I error of case deletion, mean imputation and hot deck imputation methods asymptotically converge to 1 as the sample size increases with 10\% of missing data. In contrast, the proposed method controls the Type I error. The power of proposed method is 1 when the sample sizes are larger than 400.

\section{Applications to hepatitis C virus data}

To further illustrate our proposed location-scale testing method, we apply it to a hepatitis C virus (HCV) dataset that contains missing values. HCV infection is one of the most important causes of chronic liver disease worldwide \citep{yauganouglu2022hepatitis}. The infection is known to alter lipid metabolism \citep{gyamfi2019lipid}, and patients often exhibit reduced serum cholesterol (CHOL) levels compared to healthy controls, with levels declining further as liver injury progresses \citep{liou2022viral}. 

To investigate this effect, we analyze the publicly available HCV dataset from the UCI Machine Learning Repository \citep{hcv_data_571}, which contains data from 615 individuals with 14 features. These features include demographic variables such as age and gender, as well as a panel of laboratory measurements for blood donors (without HCV infection) and HCV patients at different disease stages (hepatitis, fibrosis, and cirrhosis), including CHOL, and other liver function biomarkers.

While most studies, such as  \citet{corey2009hepatitis} and \citet{sheridan2022infection}, emphasize the mean differences in CHOL levels across liver disease stages, our primary focus in this section is to evaluate whether the levels of CHOL differ significantly in either location or scale among HCV patients at three distinct disease stages: hepatitis, fibrosis, and cirrhosis. A difference in location means that one group tends to have overall higher or lower CHOL levels than another group. A difference in scale means that CHOL levels are more spread out in one group than another.

Briefly, hepatitis is the early stage of liver injury marked by inflammation; fibrosis is the progressive stage characterized by scar tissue accumulation; cirrhosis is the late stage involving advanced, irreversible scarring and substantial loss of liver function \citep{khullar2015hepatitis}.

The data for the three groups are shown in Table~\ref{tab:2}. Note that the original data set retains only two decimal place, resulting in several tied observations. For the purpose of this example, we do not adjust for ties; instead, we add small random perturbations at the third decimal place to break them.

\begin{table}[h]
	\centering
	\caption{Cholesterol (CHOL) levels of patients at different stages of HCV infection: hepatitis, fibrosis and cirrhosis. 
	Missing values are indicated by an asterisk *.}
	\setlength{\tabcolsep}{3pt}
	\begin{tabular}{lccccccccccccc}
		\hline
		Hepatitis & 6.008 & 3.903 & 3.206 & 4.280 & 3.201 & 3.606 & 6.110 & 6.281 & 5.490 & 5.372 & 4.140 & 5.734 & 4.894   \\ 
		(early)& 9.679 & 4.075 & 4.659 & 7.046 & 5.238 & 3.094 & 6.154 & 5.728 & 5.308 & 4.674 & 4.624 &  & \\
		\\
		Fibrosis & 5.421 & 4.209 & 4.018 & 4.220 & 3.503 & 5.202 & 4.705 & 4.009 & 5.859 & 4.645 & 4.453 & 4.523 & 4.709 \\ 
		(progress)& 4.818 & 4.316 & 6.195 & 4.898 & 3.107 & 4.208 & 5.006 & * \\
		\\
		Cirrhosis & 3.205 & 4.829 & 5.346 & 3.932 & 3.719 & * & 4.107 & 5.501 & 3.601 & 4.085 & 3.545 & 2.400 & 1.432 \\ 
		(late)& 3.922 & 3.687 & 3.615 & 3.516 & 4.885 & * & 3.296 & 5.170 & 3.820 & 3.783 & 3.987 & 4.518 & 6.308 \\
		& 3.026 & 3.633 & 4.206 & 5.305 \\
		\hline
	\end{tabular}
	\label{tab:2}
\end{table}

Notice that one value is missing in the fibrosis group and two values are missing in the cirrhosis group. Since these groups contain 21 and 30 observations, respectively, the corresponding proportions of missing data for the two groups are approximately 5\% and 7\%. The reason that the values are missing is not explained by its source \citep{hcv_data_571}.

\begin{table}[ht]
	\centering
	\caption{$p$-values of the proposed location-scale testing method, and the combined Ansari-Bradley test and Wilcoxon-Mann-Whitney test using the Holm-Bonferroni method after mean imputation, hot deck imputation, or ignoring missing data.}
	\begin{tabular}{lccc}
		\hline
		Methods &Hep. vs Fib. & Hep. vs Cir. & Fib. vs Cir.\\ 
		\hline
		Case Deletion & 0.0991 & 0.0070 & 0.0277 \\ 
		Mean Imputation & 0.0832 & 0.0060 & 0.0161 \\ 
		Hot Deck Imputation & 0.0803 & 0.0045 & 0.0135 \\ 
		\textbf{Prop. (min)} & 0.0573 & 0.0026 & 0.0063 \\ 
		\textbf{Prop. (max)} & 0.2203 & 0.0404 & 0.2389 \\ 
		\hline
	\end{tabular}
	\label{tab:4.3}
	
	Hep.: Hepatitis. Fib.: Fibrosis. Cir.: Cirrhosis.
	Prop.: proposed method. Proposed method
	provides both the minimum and maximum possible 
	$p$-values according to Proposition~\ref{prop:location-scale}.
\end{table}

To examine whether the CHOL levels differs either in location or scale across the three groups of HCV patients, we apply the combined Ansari-Bradley test and Wilcoxon-Mann-Whitney test with the Holm-Bonferroni method following mean imputation, hot deck imputation, or case deletion for missing data. The $p$-values of these methods are computed according to Definition~\ref{pvalue:holm}. The proposed method provides
both the minimum and maximum possible $p$-values in the presence of missing data, according to Proposition~\ref{prop:location-scale}. All the results are summarized in Table~\ref{tab:4.3}.

Using the proposed method, we find that CHOL levels in the hepatitis and fibrosis groups are not significantly different at the $\alpha=0.05$ significance level, regardless of the missing value present in the fibrosis group. In contrast, CHOL levels for the hepatitis and cirrhosis groups are significantly different in either location or scale, and this conclusion is unaffected by the missing data. Finally, for the fibrosis versus cirrhosis comparison our method does not yield significant result, since
the testing result depends on the values of missing data. By contrast, all other missing data methods yield significant results for this comparison.

In summary, our method identifies a significant difference only between the hepatitis and cirrhosis groups, corresponding to a comparison of patients with early-stage versus late-stage liver disease. Notably, this conclusion is robust to any possible values of the missing data. We emphasize that the purpose of this section is to illustrate the impact of missing data and to demonstrate the utility of the proposed method, rather than to provide clinical recommendations.

\section{Conclusion}

The main contribution of this paper is
the construction of a scale testing method
with controlled Type I error in the presence of 
missing data, without making any missing data
assumptions. This method is based on the
tight bounds of the Ansari-Bradley test
statistic in the presence of missing data. The bounds
are tight, meaning that they cannot be improved using the
observed data alone. Based on these bounds, we derive 
the conditions for rejecting the null hypothesis only when all possible
test statistics are significant, and show that this is equivalent
to rejecting the null hypothesis when the maximum possible
$p$-value is less than or equal to the significance level $\alpha$.
We then combine this proposed scale testing method 
with the location testing method from \citet{zeng2024two}
using the Holm-Bonferroni method for the location-scale testing.

The proposed methods are evaluated using numerical simulations.  
The results demonstrate that the proposed methods control the Type
I error, and have good statistical power when less than 10\% data
are missing, while other common
missing data methods such as case deletion, mean imputation
and hot deck imputation fail to control the Type I error. We also illustrate the proposed location-scale testing method on hepatitis C virus dataset where a subset of values is unobserved. 

While there could be multiple different directions for extending
the work in this paper, the following two topics represent
promising areas to be explored. The first is to extend the proposed
method to the case where data can be tied. While this paper assumes
the data are distinct, for discrete real values, it could happen
that two or more observations have the same value. One approach
for handling ties is to apply the mid-rank method \citep{lehmann1975} and derive 
the bounds of the Ansari-Bradley test statistic when this method is applied, as considered by \citet{zeng2024two}. The second extension
is to consider scale differences for multivariate data. Such an extension requires deriving bounds for the relevant test statistics in the presence of missing data. Multivariate rank-based statistics are natural candidates for this approach. However, rank-based statistics are not necessary: any 
statistic that admits valid bounds with missing data can be considered. 
For example, \citet{zeng2024mmd} extend the framework to the Maximum Mean Discrepancy (MMD) test \citep{gretton2012kernel} when 
the Laplacian kernel is applied.

\section*{Acknowledgement}
Yijin Zeng is funded by a Roth Studentship from the 
Department of Mathematics, Imperial College London and the EPSRC 
CDT in Statistics and Machine Learning.

\section*{Supplementary Material}

\begin{appendix}
	
\section{Lower bound with single value missing} \label{append:ab:proofofsinglemissngdata}
	
	We begin our proof by considering the case where only one 
	single value in $\bx$ is not
	observed. The main result for this section
	is Proposition~\ref{supp:prop:3}.
	%Since we consider only distinct values, it would be 
	%convenient to first define the following set formally:
	%
	%\begin{definition}
	%	% maybe define this at the begining of the thesis? it could 
	%	% be useful for WMW as well
	%	Let $\dnm{\n}{\m}$ denote the collection of all pairs $(\bx,\by)$, where
	%	$\bx = \{\xn{1}, \ldots, \xn{\n}\}$ and $\by = \{\yn{1}, \ldots, \yn{\m}\}$
	%	are sets of real values such that $|\bx| = \n$, $|\by| = \m$, and
	%	values in $\bx \cup \by$ are distinct. That is,	
	%	\begin{align*}
		%		\dnm{n}{m} = \left\{ (\bx, \by):~
		%		\bx = \{\xn{1}, \ldots, \xn{\n}\} \subset \mathbb{\br},~\by = \{\yn{1}, \ldots, \yn{\m}\} \subset \mathbb{\br}, \text{ and } \bx \cap \by = \emptyset
		%		\right\}.
		%	\end{align*}	
	%\end{definition}
	%Note that in the above definition of $\dnm{n}{m}$,  all values in $\bx \cup \by$ 
	%are distinct as long as $\bx \cap \by = \emptyset$, since $\bx$ and $\by$ are sets,
	%where duplicates within $\bx$ and $\by$ are not allowed.
	
	%Without loss of generality, we assume 
	%only $\bx$ has one missing value, while all values in $\by$ are %observed.
	We first prove the following lemma, which considers
	the implications if a single value $\x \in \bx$ is replaced with another
	value $\xs$ where the rank of $\xs$ is equal to the
	rank of $\x$ plus or minus one.
	
	\begin{lemma} \label{lemma:ab:1}
		Suppose $\bx = \{\xn{1}, \ldots, \xn{\n}\}$ 
		and $\by = \{\yn{1}, \ldots, \yn{\m}\}$ are samples of distinct real values. Let $\x \in \bx$ be any element in $\bx$. 
		Denote $\bxp = \bx \setminus \{\x\}$. Consider any real value $\xs$ that is distinct from all values  in $\bxp \cup \by$, and denote $\bxs = \{\xs\} \cup \bxp$. Then, if 
		$\rank{\x}{\bx \cup \by} = \rank{\xs}{\bxs \cup \by} + 1$
		we have $\x > \xs$, and $1 \ge \sum_{\z \in \bxp} \indicator{ \x > \z > \xs} \ge 0$.
		On the other hand, if $\rank{\x}{\bx \cup \by} = \rank{\xs}{\bxs \cup \by} - 1$,
		we have $\xs > \x$, and $1 \ge \sum_{\z \in \bxp} \indicator{ \xs > \z > \x} \ge 0$.
	\end{lemma}
	
	\begin{proof}
		To start, let us assume $\rank{\x}{\bx \cup \by} = \rank{\xs}{\bxs \cup \by} + 1$.
		According to the definition of rank, $\rank{\x}{\bx \cup \by} = \sum_{\z \in \bx \cup \by} \indicator{\x \ge \z} = 1 + \sum_{\z \in \bxp \cup \by}  \indicator{\x \ge \z}.$
		Since all values in $\bx \cup \by$ are distinct, we have 
		$\indicator{\x \ge \z} = \indicator{\x > \z}$ for any $\z \in \bxp \cup \by$.
		Hence, $\rank{\x}{\bx \cup \by} = 1 + \sum_{\z \in \bxp \cup \by}  \indicator{\x > \z}.$
		Similarly, we can show that $\rank{\xs}{\bxs \cup \by} = 1 +  \sum_{\z \in \bxp \cup \by}  \indicator{\xs > \z}.$
		Subsequently, since $\rank{\x}{\bx \cup \by} = \rank{\xs}{\bxs \cup \by} + 1$, 
		we have
		\begin{align*}
			&\sum_{\z \in \bxp \cup \by}  \indicator{\x > \z} = \sum_{\z \in \bxp \cup \by}  \indicator{\xs > \z} + 1\\
			\implies &\x > \xs.
		\end{align*}
		Further, we have
		\begin{align*}
			&\sum_{\z \in \bxp} \indicator{\x > \z} + \sum_{\z \in \by} \indicator{\x > \z} =
			\sum_{\z \in \bxp}  \indicator{\xs > \z} +  \sum_{\z \in \by}  \indicator{\xs > \z} + 1 \\
			\implies & \sum_{\z \in \bxp} \indicator{\x > \z} - \sum_{\z \in \bxp}  \indicator{\xs > \z}  = \sum_{\z \in \by}  \indicator{\xs > \z} -  \sum_{\z \in \by} \indicator{\x > \z} + 1\\
			\implies &\sum_{\z \in \bxp} \left( \indicator{\x > \z} -  \indicator{\xs > \z}  \right) = \sum_{\z \in \by} \left(\indicator{ \xs > \z} - \indicator{\x > \z} \right)+ 1.
		\end{align*}
		Since  $\x > \xs$, we have
		\begin{align*}
			\sum_{\z \in \bxp} \left( \indicator{\x > \z} -  \indicator{\xs > \z}  \right) \ge 0, \text{ and } \sum_{\z \in \by} \left(\indicator{ \xs > \z} - \indicator{\x > \z} \right)  \le 0.
		\end{align*}
		Therefore, we have $1 \ge \sum_{\z \in \bxp} \left( \indicator{\x > \z} -  \indicator{\xs > \z}  \right)   \ge 0$, which is equivalent to $1 \ge \sum_{\z \in \bxp} \indicator{ \x > \z > \xs} \ge 0$. Hence, we completes our proof when $\rank{\x}{\bx \cup \by} = \rank{\xs}{\bxs \cup \by} + 1$. 
		
		For the case when $\rank{\x}{\bx \cup \by} = \rank{\xs}{\bxs \cup \by} - 1$,
		note that we have $\rank{\xs}{\bxs \cup \by} = \rank{\x}{\bx \cup \by} + 1$,
		then using the result we have already proved above, 
		we have $\xs > \x$, and $1 \ge \sum_{\z \in \bxp} \indicator{ \xs > \z > \x} \ge 0$,
		which completes our proof.
	\end{proof}
	
	According to Lemma~\ref{lemma:ab:1},
	if $\x \in \bx$ is replaced with another
	value $\xs$ where the rank of $\xs$ is equal to the
	rank of $\x$ plus one, i.e. $\rank{\x}{\bx \cup \by} = \rank{\xs}{\bxs \cup \by} - 1$,
	then we must have the value of $\xs$ is strictly greater than $\x$, and
	perhaps less immediately obvious, there exists at most one value $\z \in \bxp$
	that is greater than $\x$ but smaller than $\xs$, i.e. 
	$1 \ge \sum_{\z \in \bxp} \indicator{ \xs > \z > \x} \ge 0$.
	However, when the rank of $\xs$ is equal to the
	rank of $\x$ minus one, the opposite conclusion 
	can be drawn from Lemma~\ref{lemma:ab:1}.
	
	Applying Lemma~\ref{lemma:ab:1}, we proceed by analyzing how the values of the
	Ansari-Bradley test statistic $\ABT{\bx}{\by}$ changes when $\x \in \bx$ 
	is replaced by $\xs$, still assuming that the rank of
	$\xs$ is equal to the rank of $\x$ plus or minus one. Additionally,
	we assume that the ranks of $\x$ and $\xs$ are either both greater than
	$(\bn+1)/2$, or both smaller than $(\bn+1)/2$.
	
	\begin{proposition} \label{prop:ab:1}
		Suppose $\bx = \{\xn{1}, \ldots, \xn{\n}\}$ 
		and $\by = \{\yn{1}, \ldots, \yn{\m}\}$ are samples of distinct real values. Let $\x \in \bx$ be any element in $\bx$. 
		Denote $\bxp = \bx \setminus \{\x\}$. Consider any real value $\xs$ that is distinct from all values  in $\bxp \cup \by$, and denote $\bxs = \{\xs\} \cup \bxp$.
		Assume that $(\rank{\x}{\bx \cup \by} - (\bn + 1)/2)(\rank{\xs}{\bxs \cup \by} - (\bn + 1)/2) \ge 0$. Then, if case 
		\begin{align*}
			\case{I}:~|\rank{\x}{\bx \cup \by} - (\bn + 1)/2| = |\rank{\xs}{\bxs \cup \by} - (\bn + 1)/2| + 1
		\end{align*}
		holds, we have $\ABT{\bx}{\by} \ge \ABT{\bxs}{\by}$. However,
		if case
		\begin{align*}
			\case{II}:~|\rank{\x}{\bx \cup \by} - (\bn + 1)/2| = |\rank{\xs}{\bxs \cup \by} - (\bn + 1)/2| - 1
		\end{align*}
		holds, we have $\ABT{\bx}{\by} \le \ABT{\bxs}{\by}$. 
	\end{proposition}	
	\begin{proof}
		We first prove the result for case \case{I}. Then, we show that the result
		for case \case{II} follows directly from the result for case \case{I}.
		
		Without loss of generality, assume $\x = \xln{1}$.
		According to the definition of the Ansari-Bradley test statistic, we have
		\begin{align*}
			\ABT{\bx}{\by} & = \sum_{i = 1}^{\n} \left|\rank{\xn{\iconstant}}{\bx \cup \by} - \frac{1}{2}(N+1)\right| \\
			& = \left|\rank{\xn{1}}{\bx \cup \by} - \frac{1}{2}(N+1)\right| + \sum_{i = 2}^{\n} \left|\rank{\xn{\iconstant}}{\bx \cup \by} - \frac{1}{2}(N+1)\right|.
		\end{align*}
		By the definition of rank, for any $\iconstant \in \{2,\ldots, \n\}$, $	\rank{\xn{\iconstant}}{\bx \cup \by} = \rank{\xn{\iconstant}}{\bxp \cup \by} + \indicator{\xn{\iconstant} \ge \xn{1}}.$
		Since $\xn{\iconstant} \neq \xn{1}$, we further have $\rank{\xn{\iconstant}}{\bx \cup \by} = \rank{\xn{\iconstant}}{\bxp \cup \by} + \indicator{\xn{\iconstant} > \xn{1}}$.
		Hence,
		\begin{align*}
			\ABT{\bx}{\by}	&= \left|\rank{\xn{1}}{\bx \cup \by} - \frac{1}{2}(N+1)\right|+ \sum_{i = 2}^{\n} \left|\rank{\xn{\iconstant}}{\bxp \cup \by} + \indicator{\xn{\iconstant} > \xn{1}} - \frac{1}{2}(N+1)\right|\\
			& = \left|\rank{\xn{1}}{\bx \cup \by} - \frac{1}{2}(N+1)\right| 
			+ \sum_{i = 2}^{\n} \indicator{\xn{\iconstant} > \xn{1}} \left|\rank{\xn{\iconstant}}{\bxp \cup \by} + 1 - \frac{1}{2}(N+1)\right|\\
			& + \sum_{i = 2}^{\n} \indicator{\xn{\iconstant} < \xn{1}} \left|\rank{\xn{\iconstant}}{\bxp \cup \by} - \frac{1}{2}(N+1)\right|.
		\end{align*}
		Similarly, we can show that
		\begin{align*}
			\ABT{\bxs}{\by} 
			& = \left|\rank{\xs}{\bxs \cup \by} - \frac{1}{2}(N+1)\right| 
			+ \sum_{i = 2}^{\n} \indicator{\xn{\iconstant} > \xs} \left|\rank{\xn{\iconstant}}{\bxp \cup \by} + 1 - \frac{1}{2}(N+1)\right|\\
			& + \sum_{i = 2}^{\n} \indicator{\xn{\iconstant} < \xs} \left|\rank{\xn{\iconstant}}{\bxp \cup \by} - \frac{1}{2}(N+1)\right|.
		\end{align*}
		For notation convenience, for any $\iconstant \in \{2,\ldots,\n\}$, let us denote
		\begin{align*}
			\alpha_i = \left|\rank{\xn{\iconstant}}{\bxp \cup \by} + 1 - \frac{1}{2}(N+1)\right|,
			~\beta_i = \left|\rank{\xn{\iconstant}}{\bxp \cup \by} - \frac{1}{2}(N+1)\right|.
		\end{align*}
		Then, we have
		\begin{align}
			\begin{split} \label{supp:prop:1:eqn:1}
				&\ABT{\bxs}{\by} - \ABT{\bx}{\by} \\
				&= \left|\rank{\xs}{\bxs \cup \by} - \frac{1}{2}(N+1)\right| - \left|\rank{\xn{1}}{\bx \cup \by} - \frac{1}{2}(N+1)\right| \\
				& + \sum_{i = 2}^{\n} \left( \indicator{\xn{\iconstant} > \xs} - \indicator{\xn{\iconstant} > \xn{1}} \right) \alpha_i + \sum_{i = 2}^{\n} (\indicator{\xn{\iconstant} < \xs}- \indicator{\xn{\iconstant} < \xn{1}} ) \beta_i.
			\end{split}
		\end{align}
		Notice that
		\begin{align*}
			|\rank{\xn{1}}{\bx \cup \by} - (\bn + 1)/2| = |\rank{\xs}{\bxs \cup \by} - (\bn + 1)/2| + 1.
		\end{align*}
		Hence,
		\begin{align*}
			&\ABT{\bxs}{\by} - \ABT{\bx}{\by}\\
			&= -1  + \sum_{i = 2}^{n} \left( \indicator{\xn{\iconstant} > \xs} - \indicator{\xn{\iconstant} > \xn{1}} \right) \alpha_i + \sum_{i = 2}^{n} (\indicator{\xn{\iconstant} < \xs}- \indicator{\xn{\iconstant} < \xn{1}} ) \beta_i.
		\end{align*}
		Since $(\rank{\xn{1}}{\bx \cup \by} - (\bn + 1)/2)(\rank{\xs}{\bxs \cup \by} - (\bn + 1)/2) \ge 0$, we have either the following two cases holds:
		\begin{align*}
			\case{i}: \rank{\xn{1}}{\bx \cup \by} > \rank{\xs}{\bxs \cup \by} \ge \frac{1}{2}(\bn + 1),~\text{and}~\rank{\xn{1}}{\bx \cup \by} = \rank{\xs}{\bxs \cup \by} + 1,\\
			\case{ii}: \frac{1}{2}(\bn + 1) \ge \rank{\xs}{\bxs \cup \by} > \rank{\xn{1}}{\bx \cup \by}~\text{and}~\rank{\xn{1}}{\bx \cup \by} = \rank{\xs}{\bxs \cup \by} - 1.
		\end{align*}
		Suppose case \case{i} holds, then according to Lemma \ref{lemma:ab:1}, we have $\xn{1} > \xs$.  Hence, 
		for any $\iconstant \in \{2, \ldots, \n\}$,
		\begin{alignat*}{2}
			&\indicator{\xn{\iconstant} > \xs} - \indicator{\xn{\iconstant} > \xn{1}} = \indicator{\xn{\iconstant} < \xs}- \indicator{\xn{\iconstant} < \xn{1}} = 0,~~&&\text{if}~\xn{\iconstant} >  \xn{1},\\
			&\indicator{\xn{\iconstant} > \xs} - \indicator{\xn{\iconstant} > \xn{1}} = 1,~ \indicator{\xn{\iconstant} < \xs}- \indicator{\xn{\iconstant} < \xn{1}} = -1,~~&&\text{if}~\xn{1} > \xn{\iconstant} >  \xs,\\
			&\indicator{\xn{\iconstant} > \xs} - \indicator{\xn{\iconstant} > \xn{1}} = \indicator{\xn{\iconstant} < \xs}- \indicator{\xn{\iconstant} < \xn{1}} = 0,~~&&\text{if}~\xn{\iconstant} <  \xs.
		\end{alignat*}
		Therefore, we have $\ABT{\bxs}{\by} - \ABT{\bx}{\by}= -1  + \sum_{i = 2}^{n} \indicator{\xn{1} > \xn{\iconstant} >  \xs} (\alpha_i -  \beta_i)$. 
		According to Lemma~\ref{lemma:ab:1}, we have $1 \ge \sum_{\iconstant = 2}^{\n} \indicator{ \xn{1} > \xn{\iconstant} > \xs}  \ge 0$. In other words, there is at most one $\xn{\iconstant}$ greater
		than $\xs$ but smaller than $\xn{1}$. By definition, we also have that 
		$1 \ge \alpha_i -  \beta_i \ge -1$ for any $\iconstant \in \{2,\ldots,\n\}$.
		Hence, we have proved $\ABT{\bxs}{\by} \le \ABT{\bx}{\by}$ under case \case{i}.
		
		Similarly,
		suppose case  \case{ii} holds, then according to Lemma \ref{lemma:ab:1}, we have $\xn{1} < \xs$.  Hence, 
		for any $\iconstant \in \{2, \ldots, \n\}$, we have
		\begin{alignat*}{2}
			&\indicator{\xn{\iconstant} > \xs} - \indicator{\xn{\iconstant} > \xn{1}} = \indicator{\xn{\iconstant} < \xs}- \indicator{\xn{\iconstant} < \xn{1}} = 0,~~&&\text{if}~\xn{\iconstant} >  \xs,\\
			&\indicator{\xn{\iconstant} > \xs} - \indicator{\xn{\iconstant} > \xn{1}} = -1,~ \indicator{\xn{\iconstant} < \xs}- \indicator{\xn{\iconstant} < \xn{1}} = 1,~~&&\text{if}~\xs > \xn{\iconstant} >  \xn{1},\\
			&\indicator{\xn{\iconstant} > \xs} - \indicator{\xn{\iconstant} > \xn{1}} = \indicator{\xn{\iconstant} < \xs}- \indicator{\xn{\iconstant} < \xn{1}} = 0,~~&&\text{if}~\xn{\iconstant} < \xn{1}.
		\end{alignat*}
		Therefore, we have $\ABT{\bxs}{\by} - \ABT{\bx}{\by}= -1  + \sum_{i = 2}^{n} \indicator{\xs > \xn{\iconstant} >  \xn{1}} (\beta_i -  \alpha_i)$.
		According to Lemma~\ref{lemma:ab:1}, we have $1 \ge \sum_{\iconstant = 2}^{\n} \indicator{ \xs > \xn{\iconstant} > \xn{1}}  \ge 0$.  In other words, there is at most one $\xn{\iconstant}$ greater
		than $\xn{1}$ but smaller than $\xs$. By definition, we also have that 
		$1 \ge \beta_i -  \alpha_i \ge -1$ for any $\iconstant \in \{2,\ldots,\n\}$.
		Hence, we have proved $\ABT{\bxs}{\by} \le \ABT{\bx}{\by}$ under case \case{ii},
		which completes our proof for the case \case{I} when  
		\begin{align*}
			|\rank{\xn{1}}{\bx \cup \by} - (\bn + 1)/2| = |\rank{\xs}{\bxs \cup \by} - (\bn + 1)/2| + 1.
		\end{align*}
		
		For the case \case{II} when
		\begin{align*}
			|\rank{\x}{\bx \cup \by} - (\bn + 1)/2| = |\rank{\xs}{\bxs \cup \by} - (\bn + 1)/2| - 1.
		\end{align*}
		Notice that 
		\begin{align*}
			|\rank{\xs}{\bxs \cup \by} - (\bn + 1)/2| = |\rank{\x}{\bx \cup \by} - (\bn + 1)/2| + 1.
		\end{align*}
		Then using the result for  the case \case{I}, we have
		$\ABT{\bxs}{\by} \ge \ABT{\bx}{\by}$, which completes our proof.
	\end{proof}
	
	Under the assumption that the ranks of $\x$ and $\xs$ are either both greater than
	$(\bn+1)/2$, or both smaller than $(\bn+1)/2$, 
	Proposition~\ref{prop:ab:1} asserts that 
	when the absolute distance between 
	the rank of the replaced value $\xs$ and $(\bn+1)/2$
	is one plus the absolute distance between 
	the rank of original value $\x$ and $(\bn+1)/2$,
	i.e. $|\rank{\xs}{\bxs \cup \by} - (\bn + 1)/2| = |\rank{\x}{\bx \cup \by} - (\bn + 1)/2| + 1$,
	the Ansari-Bradley test statistic is not decreasing.
	
	Proposition~\ref{prop:ab:1} appears to suggest that
	larger absolute distance between 
	the rank of the replaced value $\xs$ and $(\bn+1)/2$ 
	leads to larger value of the Ansari-Bradley test statistic.
	We shall prove this argument in Proposition \ref{supp:ab:prop:2}. 
	However, before doing so, we introduce the following lemma,
	which shows that when considering the values of the Ansari-Bradley
	test statistic after replacing a set of values, only the ranks of original
	values and replaced values matter.
	
	\begin{lemma} \label{supp:ab:lemma:2}
		Suppose $\bx = \{\xn{1}, \ldots, \xn{\n}\}$ 
		and $\by = \{\yn{1}, \ldots, \yn{\m}\}$ are samples of distinct real values. Let $\bxt = \{\xtn{1}, \ldots, \xtn{\n-\np}\}$ be
		a non-empty subset of $\bx$. Denote $\bxp = \bx \setminus \bxt$. 
		Consider a set $\bxh = \{\xsn{1},\ldots,\xsn{\n-\np}\}$ of real values,
		denote $\bxs = \bxh \cup \bxp$, and assume that $(\bxs,\by) \in \dnm{\n}{\m}$. 
		Then, if $$\{\rank{\xtn{1}}{\bx \cup \by}, \ldots, \rank{\xtn{\n-\np}}{\bx \cup \by}\}
		= \left\{\rank{\xsn{1}}{\bxs\cup\by}, \ldots, \rank{\xsn{\n-\np}}{\bxs\cup\by}\right\},$$
		we have $\ABT{\bx}{\by} = \ABT{\bxs}{\by}$.
	\end{lemma}

	\begin{proof}
		To start, let us denote $\bxp = \{\xpn{1}, \ldots, \xpn{\np}\}$, $\bz = \bx \cup \by$,
		and $\bzs = \bxs \cup \by$.
		Then, according to 
		the definition of the Ansari-Bradley test statistic, 
		\begin{align*}
			\ABT{\bx}{\by} &= \sum_{i = 1}^{\n} \left|\rank{\xn{\iconstant}}{\bz} - \frac{1}{2}(\bn +1)\right|\\
			& = \sum_{i = 1}^{\np} \left|\rank{\xpn{\iconstant}}{\bz} - \frac{1}{2}(\bn +1)\right| + \sum_{i = 1}^{\n - \np} \left|\rank{\xtn{\iconstant}}{\bz} - \frac{1}{2}(\bn +1)\right|.
		\end{align*}
		Similarly, we have
		\begin{align*}
			\ABT{\bxs}{\by} = \sum_{i = 1}^{\np} \left|\rank{\xpn{\iconstant}}{\bzs} - \frac{1}{2}(\bn +1)\right| 
			+ \sum_{i = 1}^{\n - \np} \left|\rank{\xsn{\iconstant}}{\bzs} - \frac{1}{2}(\bn +1)\right|.
		\end{align*}
		Since $\{\rank{\xtn{1}}{\bz}, \ldots, \rank{\xtn{\n-\np}}{\bz}\}
		= \left\{\rank{\xsn{1}}{\bzs}, \ldots, \rank{\xsn{\n-\np}}{\bzs}\right\}$,
		we have
		\begin{align*}
			\sum_{i = 1}^{\n - \np} \left|\rank{\xtn{\iconstant}}{\bz} - \frac{1}{2}(\bn +1)\right| = \sum_{i = 1}^{\n - \np} \left|\rank{\xsn{\iconstant}}{\bzs} - \frac{1}{2}(\bn +1)\right|.
		\end{align*}
		Thus, in order to prove $\ABT{\bx}{\by} = \ABT{\bxs}{\by}$, it is sufficient to show that
		\begin{align*}
			\rank{\xpn{\iconstant}}{\bz} = \rank{\xpn{\iconstant}}{\bzs},~\text{for any}~\iconstant \in \{1,\ldots,\np\}.
		\end{align*}
		
		Denote $\bzp = \bxp \cup \by$.
		Let $\xx \in \bxp$ be any value in $\bxp$, and denote 
		$\rank{\x}{\bzp} = \kkp$, then $\rank{\x}{\bz}$
		depends only on $\{\rank{\xtn{1}}{\bz}, \ldots, \rank{\xtn{\n-\np}}{\bz}\}$, such that 
		\begin{align*}
			\rank{\x}{\bz} = \kk \iff \sum_{\iconstant=1}^{\n - \np} \indicator{\rank{\xtn{\iconstant}}{\bz} < \kk} = \kk - \kkp.
		\end{align*}
		For the ``$\implies$", note that $\rank{\x}{\bzp} = \kkp$
		implies $\kkp$ elements in $\bzp$ smaller than $\x$, and 
		$\rank{\x}{\bz}$ implies $\kk$ elements in $\bz$ smaller than $\x$.
		Then there are $\kk - \kkp$ elements in $\bxp$ smaller than $\x$,
		hence the result.	
		For the ``$\impliedby$'', let us assume $\xtn{1} < \ldots < \xtn{\n-\np}$, then 
		\begin{align*}
			&\sum_{\iconstant=1}^{\n - \np} \indicator{\rank{\xtn{\iconstant}}{\bz} < \kk} = \kk - \kkp \\
			\iff & \{\rank{\xtn{1}}{\bz}, \ldots, \rank{\xtn{\kk-\kkp}}{\bz}\} < \kk,~\text{and}~\{\rank{\xtn{\kk-\kkp + 1}}{\bz}, \ldots, \rank{\xtn{\n -\np}}{\bz} \}\ge \kk \\
			\iff & \{\rank{\xtn{1}}{\bzp}, \ldots, \rank{\xtn{\kk-\kkp}}{\bzp}\} < \kkp,~\text{and}~\{\rank{\xtn{\kk-\kkp + 1}}{\bzp}, \ldots, \rank{\xtn{\n -\np}}{\bzp} \}\ge \kkp.
		\end{align*}
		Since $\rank{\x}{\bzp} = \kkp$, there are $\kk - \kkp$ elements in $\bxp$ smaller than $\x$.
		Hence, we have proven the ``$\impliedby$".
		
		Similarly, we have  
		\begin{align*}
			\rank{\x}{\bzs} = \kk \iff \sum_{\iconstant=1}^{\n - \np} \indicator{\rank{\xsn{\iconstant}}{\bzs} < \kk} = \kk - \kkp.
		\end{align*}
		Since $\{\rank{\xtn{1}}{\bz}, \ldots, \rank{\xtn{\n-\np}}{\bz}\}
		= \left\{\rank{\xsn{1}}{\bzs}, \ldots, \rank{\xsn{\n-\np}}{\bzs}\right\}$,
		we have
		\begin{align*}
			\rank{\xx}{\bz} = \rank{\xx}{\bzs},
		\end{align*}
		which completes the proof.
	\end{proof}
	
	Lemma~\ref{supp:ab:lemma:2} stresses that when considering
	the values of Ansari-Bradley test statistic, only the ranks
	of the values matter. With Lemma~\ref{supp:ab:lemma:2},
	we now proceed to prove the following proposition 
	which generalizes Proposition~\ref{prop:ab:1}.
	We continue to assume that the ranks of $\x$ and $\xs$ are either both greater than
	$(\bn+1)/2$, or both smaller than $(\bn+1)/2$. That is,
	$(\rank{\x}{\bx \cup \by} - (\bn + 1)/2)(\rank{\xs}{\bxs \cup \by} - (\bn + 1)/2) \ge 0$.
	However, unlike Proposition~\ref{prop:ab:1}, the following proposition  
	does not require the rank of $\xs$ to differ from that of $\x$ by exactly one.
	
	\begin{proposition} \label{supp:ab:prop:2}
		Suppose $\bx = \{\xn{1}, \ldots, \xn{\n}\}$ 
		and $\by = \{\yn{1}, \ldots, \yn{\m}\}$ are samples of distinct real values. Let $\x \in \bx$ be any element in $\bx$.  Denote $\bxp = \bx \setminus \{\x\}$.  Consider any real value $\xs$ that is distinct to 
		all values in $\bxp \cup \by$, and denote $\bxs = \{\xs\} \cup \bxp$. 
		Assume that $(\rank{\x}{\bx \cup \by} - (\bn + 1)/2)(\rank{\xs}{\bxs \cup \by} - (\bn + 1)/2) \ge 0$. Then, if case
		\begin{align*}
			\case{I}:~|\rank{\x}{\bx \cup \by} - (\bn + 1)/2| \ge |\rank{\xs}{\bxs \cup \by} - (\bn + 1)/2|
		\end{align*}
		holds, we have $\ABT{\bx}{\by} \ge \ABT{\bxs}{\by}$. However, if case
		\begin{align*}
			\case{II}: |\rank{\x}{\bx \cup \by} - (\bn + 1)/2| \le |\rank{\xs}{\bxs \cup \by} - (\bn + 1)/2|
		\end{align*}
		holds, we have $\ABT{\bx}{\by} \le \ABT{\bxs}{\by}$. 
	\end{proposition}	
	\begin{proof}
		We first prove the result for case \case{I}. Then, we show that the result
		for case \case{II} follows directly from the result for case \case{I}.
		
		Since $(\rank{\x}{\bx \cup \by} - (\bn + 1)/2)(\rank{\xs}{\bxs \cup \by} - (\bn + 1)/2) \ge 0$,
		then we have one of the following two cases holds:
		\begin{align*}
			&\case{i}~\rank{\x}{\bx \cup \by} \ge \rank{\xs}{\bxs \cup \by} \ge \frac{1}{2}(\bn + 1),\\
			\text{or}~&\case{ii}~\rank{\x}{\bx \cup \by} \le \rank{\xs}{\bxs \cup \by} \le \frac{1}{2}(\bn + 1).
		\end{align*}
		Note that if $\rank{\x}{\bx \cup \by} = \rank{\xs}{\bxs \cup \by}$, 
		then we have $\ABT{\bx}{\by} = \ABT{\bxs}{\by}$
		according to Lemma \ref{supp:ab:lemma:2}.
		Hence, we only need to consider the cases when
		$\rank{\xs}{\bxs \cup \by} \neq \rank{\x}{\bx \cup \by}.$
		
		%% Case (i)
		Suppose case $\case{i}$ holds. That is, assume that $\rank{\x}{\bx \cup \by} > \rank{\xs}{\bxs \cup \by} \ge \frac{1}{2}(\bn + 1)$. Then,
		without loss of generality, assume that
		$\rank{\x}{\bx \cup \by} = \rank{\xs}{\bxs \cup \by} + a$, 
		where $a \in \mathbb{N}^{+}$ is a positive integer. 
		Let $\xsn{1}, \ldots, \xsn{a}$ be real numbers distinct 
		to samples in $\bxp \cup \by$ such that
		\begin{align*}
			\rank{\xsn{\kk}}{\{\xsn{\kk}\} \cup \bxp \cup \by} = \rank{\x}{\bx \cup \by} - \kk, \text{for any}~\kk \in \{1,\ldots,a\}.
		\end{align*}  
		Denote $\x = \xsn{0}$. Then, for any $k \in \{0,\ldots,a\}$, we have
		\begin{align*}
			\rank{\xsn{\kk}}{\{\xsn{\kk}\} \cup \bxp \cup \by} \ge \frac{1}{2}(\bn + 1),
		\end{align*}
		and for any $k \in \{1,\ldots,a\}$,
		\begin{align*}
			\rank{\xsn{\kk}}{\{\xsn{\kk}\} \cup \bxp \cup \by} =\rank{\xsn{\kk-1}}{\{\xsn{\kk-1}\} \cup \bxp \cup \by} - 1.
		\end{align*}
		Subsequently, by applying Proposition \ref{prop:ab:1} between each $\{\xsn{\kk}\} \cup \bxp, \by$ and $\{\xsn{\kk-1}\} \cup \bxp, \by$ for $\kk \in \{1,\ldots,a\}$, it follows that
		\begin{align*}
			\ABT{\{\xsn{0}\} \cup \bxp}{\by} \ge \cdots \ge \ABT{\{\xsn{a}\} \cup \bxp}{\by}.
		\end{align*}
		Since $\x = \xsn{0}$, we have $	\ABT{\{\xsn{0}\} \cup \bxp}{\by}  = \ABT{\bx}{\by}.$
		Further, since
		\begin{align*}
			\rank{\xsn{a}}{\{\xsn{a}\} \cup \bxp \cup \by} = \rank{\x}{\bx \cup \by} - a = \rank{\xs}{\bxs \cup \by},
		\end{align*}
		then according to Lemma \ref{supp:ab:lemma:2}, we have $\ABT{\{\xsn{a}\} \cup \bxp}{\by} = \ABT{\bxs}{\by}$. Therefore,
		\begin{align*}
			\ABT{\bx}{\by} = 	\ABT{\{\xsn{0}\} \cup \bxp}{\by} \ge \cdots \ge \ABT{\{\xsn{a}\} \cup \bxp}{\by} = \ABT{\bxs}{\by}.
		\end{align*}
		which proves our result when case $\case{i}$ holds.
		Case $\case{ii}$ can be proved similarly, thus it is omitted here.
		
		Subsequently, for case \case{II}, note that
		\begin{align*}
			|\rank{\x}{\bx \cup \by} - (\bn + 1)/2| \le |\rank{\xs}{\bxs \cup \by} - (\bn + 1)/2|
		\end{align*}
		is equivalent to
		\begin{align*}
			|\rank{\xs}{\bxs \cup \by} - (\bn + 1)/2| \ge |\rank{\x}{\bx \cup \by} - (\bn + 1)/2|.
		\end{align*}
		Then, by applying the result for case \case{I}, we have $\ABT{\bxs}{\by} \ge \ABT{\bx}{\by}$,
		which completes our proof.
	\end{proof}
	
	Proposition~\ref{supp:ab:prop:2} asserts that 
	when the absolute distance between the rank 
	of the replaced value $\xs$ and $(\bn+1)/2$
	is larger than the absolute distance between the rank 
	of the original value $\x$ and $(\bn+1)/2$, i.e. 
	$|\rank{\xs}{\bxs \cup \by} - (\bn + 1)/2| \ge |\rank{\x}{\bx \cup \by} - (\bn + 1)/2|$,
	the values of Ansari-Bradley test statistic with $\xs$
	is greater or equal to that with $\x$. This
	result generalizes Proposition~\ref{prop:ab:1}, by dropping
	the assumption that the rank of $\xs$ differs from that of $\x$ 
	by exactly one. However, Proposition~\ref{supp:ab:prop:2}
	still assumes that the ranks of $\x$ and $\xs$ are either both greater than
	$(\bn+1)/2$, or both smaller than $(\bn+1)/2$. 
	
	In order to obtain more general results regarding the 
	values of the Ansari-Bradley test statistic with one single value missing.
	We first prove the following lemma.  
	Unlike Proposition~\ref{prop:ab:1} and
	Proposition~\ref{supp:ab:prop:2}, the
	following lemma does not assume 
	the ranks of $\x$ and $\xs$ are both larger 
	or both smaller than $(\bn+1)/2$.

	\begin{lemma} \label{lemma:ab:3}
		Suppose $\bx = \{\xn{1}, \ldots, \xn{\n}\}$ 
		and $\by = \{\yn{1}, \ldots, \yn{\m}\}$ are samples of distinct real values. 
		Denote $\bn = \n + \m$ and assume $\bn$ is even. 
		Suppose $\x \in \bx$ is any value in $\bx$ and denote $\bxp = \bx \setminus \{\x\}$. 
		Consider any real value $\xs$ that is distinct to all values in $\bxp \cup \by$,
		and denote $\bxs = \{\xs\} \cup \bxp$. Then, if 
		\begin{align*}
			\left|\rank{\xs}{\bxs \cup \by} - (\bn + 1)/2\right| = \left|\rank{\x}{\bx \cup \by} - (\bn + 1)/2\right| = 1/2
		\end{align*}
		we have $\ABT{\bx}{\by} = \ABT{\bxs}{\by}$.
	\end{lemma}
	
	\begin{proof}
		To start, if $\rank{\xs}{\bxs \cup \by} = \rank{\x}{\bx \cup \by}$.
		Then, we have our desired result $\ABT{\bx}{\by} = \ABT{\bxs}{\by}$
		according to Lemma \ref{supp:ab:lemma:2}.
		However, if $\rank{\xs}{\bxs \cup \by} \neq \rank{\x}{\bx \cup \by}$,
		since
		\begin{align*}
			\left|\rank{\xs}{\bxs \cup \by} - (\bn + 1)/2\right| = \left|\rank{\x}{\bx \cup \by} - (\bn + 1)/2\right| = 1/2,
		\end{align*}
		we have one of the following two cases is true
		\begin{align*}
			&\case{i}~\rank{\xs}{\bxs \cup \by} = (\bn + 1)/2 - 1/2,~\rank{\x}{\bx \cup \by}  = (\bn + 1)/2 + 1/2,\\
			\text{or}~&\case{ii}~\rank{\xs}{\bxs \cup \by} = (\bn + 1)/2 + 1/2,~\rank{\x}{\bx \cup \by}  = (\bn + 1)/2 -1/2.
		\end{align*}
		In the following, we consider only case $\case{i}$.
		Once case $\case{i}$ has been established, observe that
		case $\case{ii}$ is equivalent to
		\begin{align*}
			\rank{\x}{\bx \cup \by}  = (\bn + 1)/2 -1/2,~\rank{\xs}{\bxs \cup \by} = (\bn + 1)/2 + 1/2.
		\end{align*}
		Hence, we have $\ABT{\bxs}{\by} = \ABT{\bx}{\by}$ from 
		case $\case{i}$.
		
		Without loss of generality, let us assume that $\x = \xln{1}$. 
		Then, according to the definition of the Ansari-Bradley test statistic,
		\begin{align*}
			\ABT{\bx}{\by} & = \sum_{i = 1}^{\n} \left|\rank{\xn{\iconstant}}{\bx \cup \by} - \frac{1}{2}(N+1)\right| \\
			& = \left|\rank{\xn{1}}{\bx \cup \by} - \frac{1}{2}(N+1)\right| + \sum_{i = 2}^{\n} \left|\rank{\xn{\iconstant}}{\bx \cup \by} - \frac{1}{2}(N+1)\right|.
		\end{align*}
		By the definition of rank, we have $\rank{\xn{\iconstant}}{\bx \cup \by} = \rank{\xn{\iconstant}}{\bxp \cup \by} + \indicator{\xn{\iconstant} \ge \xn{1}}$ for any $\iconstant \in \{2,\ldots,\n\}$. 
		Since $\xn{\iconstant} \neq \xn{1}$, we further have
		$\rank{\xn{\iconstant}}{\bx \cup \by} = \rank{\xn{\iconstant}}{\bxp \cup \by} + \indicator{\xn{\iconstant} > \xn{1}}.$
		Hence,
		\begin{align*}
			\ABT{\bx}{\by}	
			& = \left|\rank{\xn{1}}{\bx \cup \by} - \frac{1}{2}(N+1)\right|
			+ \sum_{i = 2}^{\n} \left|\rank{\xn{\iconstant}}{\bxp \cup \by} + \indicator{\xn{\iconstant} > \xn{1}} - \frac{1}{2}(N+1)\right|\\
			& = \left|\rank{\xn{1}}{\bx \cup \by} - \frac{1}{2}(N+1)\right|
			+ \sum_{i = 2}^{\n} \indicator{\xn{\iconstant} > \xn{1}} \left|\rank{\xn{\iconstant}}{\bxp \cup \by} + 1 - \frac{1}{2}(N+1)\right|\\
			& + \sum_{i = 2}^{\n} \indicator{\xn{\iconstant} < \xn{1}} \left|\rank{\xn{\iconstant}}{\bxp \cup \by} - \frac{1}{2}(N+1)\right|.
		\end{align*}
		Similarly, we have
		\begin{align*}
			\ABT{\bxs}{\by} 
			& = \left|\rank{\xs}{\bxs \cup \by} - \frac{1}{2}(N+1)\right|
			+ \sum_{i = 2}^{\n} \indicator{\xn{\iconstant} > \xs} \left|\rank{\xn{\iconstant}}{\bxp \cup \by} + 1 - \frac{1}{2}(N+1)\right|\\
			& + \sum_{i = 2}^{\n} \indicator{\xn{\iconstant} < \xs} \left|\rank{\xn{\iconstant}}{\bxp \cup \by} - \frac{1}{2}(N+1)\right|.
		\end{align*}
		For notation convenience, for any $\iconstant \in \{2, \ldots, \n\}$, let us denote
		\begin{align} \label{A.2.6.eqn:1}
			\alpha_i = \left|\rank{\xn{\iconstant}}{\bxp \cup \by} + 1 - \frac{1}{2}(N+1)\right|,~\text{and}~\beta_i = \left|\rank{\xn{\iconstant}}{\bxp \cup \by} - \frac{1}{2}(N+1)\right|.
		\end{align}
		Subsequently,
		\begin{align}
			\begin{split} \label{supp:ab:lemma:3:eqn:1}
				&\ABT{\bxs}{\by} - \ABT{\bx}{\by} \\
				&= \left|\rank{\xs}{\bxs \cup \by} - \frac{1}{2}(N+1)\right| - \left|\rank{\xn{1}}{\bx \cup \by} - \frac{1}{2}(N+1)\right| \\
				& + \sum_{i = 2}^{\n} \left( \indicator{\xn{\iconstant} > \xs} - \indicator{\xn{\iconstant} > \xn{1}} \right) \alpha_i + \sum_{i = 2}^{\n} (\indicator{\xn{\iconstant} < \xs}- \indicator{\xn{\iconstant} < \xn{1}} ) \beta_i.
			\end{split}
		\end{align}
		
		Since case $\case{i}$ is true, we have
		\begin{align*}
			\rank{\xs}{\bxs \cup \by} = \frac{1}{2}(\bn + 1) - \frac{1}{2},~\rank{\xn{1}}{\bx \cup \by}  = \frac{1}{2}(\bn + 1) + \frac{1}{2}.
		\end{align*}
		Thus,
		\begin{align*}
			\left|\rank{\xs}{\bxs \cup \by} - \frac{1}{2}(N+1)\right| - \left|\rank{\xn{1}}{\bx \cup \by} - \frac{1}{2}(N+1)\right| = 0.
		\end{align*}
		Hence,
		\begin{align*}
			\ABT{\bxs}{\by} - \ABT{\bx}{\by}
			= \sum_{i = 2}^{\n} \left( \indicator{\xn{\iconstant} > \xs} - \indicator{\xn{\iconstant} > \xn{1}} \right) \alpha_i + \sum_{i = 2}^{\n} (\indicator{\xn{\iconstant} < \xs}- \indicator{\xn{\iconstant} < \xn{1}} ) \beta_i
		\end{align*}
		following Equation \eqref{supp:ab:lemma:3:eqn:1}.
		Since $\rank{\xn{1}}{\bx \cup \by} = \rank{\xs}{\bxs \cup \by} + 1$, we have $\xn{1} > \xs$ according to Lemma \ref{lemma:ab:1}.  Hence, for any $\iconstant \in \{2, \ldots,\n\}$,
		\begin{alignat*}{2}
			&\indicator{\xn{\iconstant} > \xs} - \indicator{\xn{\iconstant} > \xn{1}} = \indicator{\xn{\iconstant} < \xs}- \indicator{\xn{\iconstant} < \xn{1}} = 0,~~&&\text{if}~\xn{\iconstant} >  \xn{1}\\
			&\indicator{\xn{\iconstant} > \xs} - \indicator{\xn{\iconstant} > \xn{1}} = 1,~ \indicator{\xn{\iconstant} < \xs}- \indicator{\xn{\iconstant} < \xn{1}} = -1,~~&&\text{if}~\xn{1} > \xn{\iconstant} >  \xs\\
			&\indicator{\xn{\iconstant} > \xs} - \indicator{\xn{\iconstant} > \xn{1}} = \indicator{\xn{\iconstant} < \xs}- \indicator{\xn{\iconstant} < \xn{1}} = 0,~~&&\text{if}~\xn{\iconstant} <  \xs.
		\end{alignat*}
		This gives us
		\begin{align*}
			&\ABT{\bxs}{\by} - \ABT{\bx}{\by}= \sum_{i = 2}^{n} \indicator{\xn{1} > \xn{\iconstant} >  \xs} (\alpha_i -  \beta_i).
		\end{align*}
		According to Lemma \ref{lemma:ab:1}, we further have 
		\begin{align*}
			1 \ge \sum_{\iconstant = 2}^{\n} \indicator{ \xn{1} > \xn{\iconstant} > \xs}  \ge 0 .
		\end{align*}
		
		If $\sum_{\iconstant = 2}^{\n} \indicator{ \xn{1} > \xn{\iconstant} > \xs}  = 0$, then
		we have
		\begin{align*}
			\ABT{\bxs}{\by} - \ABT{\bx}{\by}= 0,
		\end{align*}
		which gives us the desired result.  
		
		However, if $\sum_{\iconstant = 2}^{\n} \indicator{ \xn{1} > \xn{\iconstant} > \xs}  = 1$, 
		then there exists one 
		and only one value $\xn{\jconstant} \in \bxp$ such that $\xn{1} > \xn{\jconstant} > \xs$. Hence, we have 
		\begin{align} \label{A.2.6.eqn:2}
			\ABT{\bxs}{\by} - \ABT{\bx}{\by}= \alpha_j -  \beta_j
		\end{align}
		Since $\xn{1} > \xn{\jconstant}$, we have
		$\rank{\xn{1}}{\bx \cup \by} > \rank{\xn{\jconstant}}{\bx \cup \by}$.
		Recall that $\rank{\xn{1}}{\bx \cup \by} = (\bn + 1)/2 + 1/2$.
		Hence, 
		\begin{align*}
			\frac{1}{2}(\bn + 1) + \frac{1}{2} > \rank{\xn{\jconstant}}{\bx \cup \by}.
		\end{align*}
		Additionally, since $\xs < \xn{\jconstant}$, it follows that
		\begin{align*}
			\frac{1}{2}(\bn + 1) - \frac{1}{2} = \rank{\xs}{\bxs \cup \by} \le \rank{\xn{\jconstant}}{\bx \cup \by}.
		\end{align*}
		Thus, we have
		\begin{align*}
			\rank{\xn{\jconstant}}{\bx \cup \by} = \frac{1}{2}(\bn + 1) - \frac{1}{2}.
		\end{align*}
		Further, notice that
		\begin{align*}
			&\rank{\xn{\jconstant}}{\bx \cup \by} = \rank{\xn{\jconstant}}{\bx' \cup \by}  + \indicator{\xn{\jconstant} > \xn{1}}.
		\end{align*}
		Since $\xn{1} > \xn{\jconstant} > \xs$, we have $\indicator{\xn{\jconstant} > \xn{1}} = 0$,
		which means
		\begin{align*}
			& \rank{\xn{\jconstant}}{\bx' \cup \by} = \rank{\xn{\jconstant}}{\bx \cup \by} = \frac{1}{2}(\bn + 1) - \frac{1}{2}.
		\end{align*}
		Then, by the definition of $\alpha_j$ and $\beta_j$
		in Equation~\eqref{A.2.6.eqn:1}, we have 
		$ \alpha_j = \beta_j = 1/2$, which proves the result $\ABT{\bxs}{\by} = \ABT{\bx}{\by}$ according to Equation~\eqref{A.2.6.eqn:2}.
		This completes our proof.
	\end{proof}
	
	We are now ready to prove the final result concerning the 
	lower bounds of the Ansari-Bradley test statistic when one
	single value in $\bx$ is missing. This result is proved by applying mainly Proposition~\ref{supp:ab:prop:2},
	and Lemma \ref{lemma:ab:3}.
	
	\begin{proposition} \label{supp:prop:3}
		Suppose $\bx = \{\xn{1}, \ldots, \xn{\n}\}$ 
		and $\by = \{\yn{1}, \ldots, \yn{\m}\}$ are samples of distinct real values.  Denote $\bn =\n + \m$. Suppose $\x \in \bx$ 
		is a value in $\bx$ and denote $\bxp = \bx \setminus \{\x\}$. Consider a real value
		$\xs$ that is distinct to all values in $\bxp \cup \by$, and denote 
		$\bxs = \{\xs\} \cup \bxp$. Then, if 
		\begin{align*}
			\left|\rank{\xs}{\bxs \cup \by} - \frac{1}{2}(\bn + 1)\right| =  \left\{ \begin{array}{lll}
				0, & \text{when} & \bn~\text{is odd,}  \\ 1/2, & \mbox{when} & \bn~\text{is even,}
			\end{array}\right.
		\end{align*}
		we have $\ABT{\bx}{\by} \ge \ABT{\bxs}{\by}$.
	\end{proposition}	
	
	\begin{proof}
		We first prove the case when $\bn$ is odd. In other
		words, we show that when
		$\rank{\xs}{\bxs \cup \by} = (\bn + 1)/2$,
		we have  $\ABT{\bx}{\by} \ge \ABT{\bxs}{\by}$.
		
		Suppose $\rank{\x}{\bx \cup \by} = (\bn + 1)/2$, then
		we have $\rank{\xs}{\bxs \cup \by} = \rank{\x}{\bx \cup \by}$.
		According to Lemma \ref{supp:ab:lemma:2}, we 
		have $\ABT{\bx}{\by} = \ABT{\bxs}{\by}$, which proves our result.
		However, if $\rank{\x}{\bx \cup \by} > (\bn + 1)/2$, or $\rank{\x}{\bx \cup \by} < (\bn + 1)/2$, then we have $\ABT{\bx}{\by} \ge \ABT{\bxs}{\by}$ according to Proposition \ref{supp:ab:prop:2}.
		Hence, we have shown that we always have  $\ABT{\bx}{\by} \ge \ABT{\bxs}{\by}$
		for the case when $\bn$ is odd.
		
		We now consider the case when $\bn$ is even. 	
		That is, if $\left|\rank{\xs}{\bxs \cup \by} - (\bn + 1)/2\right| = 1/2$,
		we have $\ABT{\bx}{\by} \ge \ABT{\bxs}{\by}$.
		
		Let $\xsn{1}$ and $\xsn{2}$ be values that are distinct to data in $\bxp \cup \by$,
		and suppose that
		\begin{align*}
			\rank{\xsn{1}}{\{\xsn{1}\} \cup \bxp \cup \by} = \frac{1}{2}(\bn + 1) + \frac{1}{2},\text{ and }
			\rank{\xsn{2}}{\{\xsn{2}\} \cup \bxp \cup \by} = \frac{1}{2}(\bn + 1) - \frac{1}{2}.
		\end{align*}
		Then,  according to Lemma \ref{lemma:ab:3}, we have
		\begin{align*}
			\ABT{\bxs}{\by} = \ABT{\{\xsn{1}\}  \cup \bxp}{\by} = \ABT{\{\xsn{2}\}  \cup \bxp}{\by}.
		\end{align*}
		
		Suppose $\rank{\x}{\bx \cup \by} = (\bn + 1)/2 - 1/2$, or $\rank{\x}{\bx \cup \by} = (\bn + 1)/2 + 1/2$. By applying Lemma~\ref{lemma:ab:3} directly, we have
		$\ABT{\bx}{\by} = \ABT{\bxs}{\by}$.
		
		However, if $\rank{\x}{\bx \cup \by} > (\bn + 1)/2 + 1/2$, then
		according to  Proposition \ref{supp:ab:prop:2},
		\begin{align*}
			\ABT{\bx}{\by} \ge \ABT{\{\xsn{1}\} \cup \bxp}{\by} = 	\ABT{\bxs}{\by}.
		\end{align*}
		Finally, if $\rank{\x}{\bx \cup \by} < (\bn + 1)/2 - 1/2$, then
		by applying Proposition \ref{supp:ab:prop:2} again,
		\begin{align*}
			\ABT{\bx}{\by} \ge \ABT{\{\xsn{2}\} \cup \bxp}{\by}  = \ABT{\bxs}{\by}.
		\end{align*}
		This completes our proof.
	\end{proof}

	\section{Lower bound with multiple missing values in $\bx$} \label{appendix:lowebound:ab:x}
	
	In this section, we generalize the results in
	Proposition~\ref{supp:prop:3} to the cases
	where multiple values in $\bx$ can be missing.
	The main results for this section is Theorem~\ref{supp:theorem:1}.
	
	We start by proving the following result:
	
	\begin{lemma} \label{supp:lemma:4.1}
		Suppose $\bx = \{\xn{1}, \ldots, \xn{\n}\}$ 
		and $\by = \{\yn{1}, \ldots, \yn{\m}\}$ are samples of distinct real values. Let $\bxt = \{\xtn{1}, \ldots, \xtn{\n - \np}\}$ be a 
		non-empty subset of $\bx$, and denote $\bxp = \bx \setminus \bxt$. 
		Suppose $\bxh = \{\xsn{1}, \ldots, \xsn{\n - \np}\}$, and denote
		$\bxs = \bxh \cup \bxsp$. Assume that all values in $\bxs \cup \bxp$ are distinct. Subsequently, for any $\xx  \in \bxp$, if
		\begin{align*}
			\rank{\xx}{\bx \cup \by} &> \max \{\rank{\xtn{1} }{\bx \cup \by}, \ldots, \rank{\xtn{\n - \np} }{\bx \cup \by} \},\\
			\text{and}~\rank{\xx}{\bx \cup \by} &> \max \{\rank{\xsn{1}}{\bxs \cup \by}, \ldots, \rank{\xsn{\n - \np}}{\bxs \cup \by}\},
		\end{align*}
		we have $\rank{\xx}{\bxs \cup \by} = \rank{\xx}{\bx \cup \by}$.
	\end{lemma}

	\begin{proof}
		First, we show that $\xx > \max \bxt$ and $\xx > \max \bxh$.	
		The inequality $\xx > \max \bxt$ comes directly from the assumption
		that
		\begin{align*}
			\rank{\xx}{\bx \cup \by} > \max \{\rank{\xtn{1} }{\bx \cup \by}, \ldots, \rank{\xtn{\n - \np} }{\bx \cup \by} \}.
		\end{align*} 
		The inequality $\xx > \max \bxh$ is not immediately obvious, so we prove it by contradiction. Suppose instead that $\max \bxh > \xx$. Without loss of generality, assume $\xsn{1} = \max \bxh > \xx$. Subsequently, according to the definition of rank,
		\begin{align*}
			\rank{\xsn{1}}{\bxs \cup \by} &= \sum_{\iconstant = 1}^{\n - \np}\indicator{\xsn{1} \ge \xsn{\iconstant}} + \sum_{\z \in \bxp \cup \by} \indicator{\xsn{1} \ge \z}\\
			& = \n - \np + \sum_{\z \in \bxp \cup \by} \indicator{\xsn{1} \ge \z}\\
			& \ge \n - \np + \sum_{\z \in \bxp \cup \by} \indicator{\xx \ge \z} \\
			& = \sum_{\iconstant = 1}^{\n - \np} \indicator{\xx \ge \xtn{\iconstant}} + \sum_{\z \in \bxp \cup \by} \indicator{\xx \ge \z} = \rank{\xx}{\bx \cup \by},
		\end{align*} 
		which contracts the assumption that 
		\begin{align*}
			\rank{\xx}{\bx \cup \by} > \max \{\rank{\xsn{1}}{\bxs  \cup \by}, \ldots, \rank{\xsn{\n - \np}}{\bxs \cup \by}\}.
		\end{align*}
		Therefore, it must hold that $\xx > \max \bxs$.
		
		Now we prove that $\rank{\xx}{\bxs \cup \by} = \rank{\xx}{\bx \cup \by}$
		using the above conclusions.
		Note that by the definition of rank,
		\begin{align*}
			&\rank{\xx}{\bx \cup \by} = \rank{\xx}{\bxp \cup \by} + \sum_{\iconstant = 1}^{\n - \np}\indicator{\xx \ge \xtn{\iconstant}}\\
			\implies  &\rank{\xx}{\bx \cup \by} = \rank{\xx}{\bxp \cup \by} + \n - \np,
		\end{align*}
		where the $``\implies"$ holds because $\xx > \max \bxt$.
		Similarly, 
		\begin{align*}
			&\rank{\xx}{\bxs \cup \by} = \rank{\xx}{\bxp \cup \by} + \sum_{\iconstant = 1}^{\n - \np}\indicator{\xx \ge \xsn{\iconstant}},\\
			\implies  &\rank{\xx}{\bxs \cup \by} = \rank{\xx}{\bxp \cup \by} + \n - \np,
		\end{align*}
		where the $``\implies"$ holds because $\xx > \max \bxs$. Hence,
		we have shown our desired result
		$\rank{\xx}{\bxs \cup \by} = \rank{\xx}{\bx \cup \by}$.
	\end{proof}
	
	The following lemma provides the same
	conclusion as Lemma~\ref{supp:lemma:4.1},
	but with different conditions.
	
	\begin{lemma} \label{supp:lemma:4.2}
		Suppose $\bx = \{\xn{1}, \ldots, \xn{\n}\}$ 
		and $\by = \{\yn{1}, \ldots, \yn{\m}\}$ are samples of distinct real values. Let $\bxt = \{\xtn{1}, \ldots, \xtn{\n - \np}\}$ be a 
		non-empty subset of $\bx$, and denote $\bxp = \bx \setminus \bxt$. 
		Suppose $\bxh = \{\xsn{1}, \ldots, \xsn{\n - \np}\}$, and denote
		$\bxs = \bxh \cup \bxsp$. Assume that all values in $\bxs \cup \bxp$ are distinct. Subsequently, for any $\xx  \in \bxp$, if
		\begin{align*}
			\rank{\xx}{\bx \cup \by} &< \min \{\rank{\xtn{1} }{\bx \cup \by}, \ldots, \rank{\xtn{\n - \np} }{\bx \cup \by} \},\\
			\text{and}~\rank{\xx}{\bx \cup \by} &< \min \{\rank{\xsn{1}}{\bxs \cup \by}, \ldots, \rank{\xsn{\n - \np}}{\bxs \cup \by}\},
		\end{align*}
		we have $\rank{\xx}{\bxs \cup \by} = \rank{\xx}{\bx \cup \by}$.
	\end{lemma}
	
	\begin{proof}
		The proof is similar to the proof for Lemma~\ref{supp:lemma:4.2}.	
		First, we show that $\xx < \min \bxt$ and $\xx  < \min \bxh$.
		The first inequality $\xx < \min \bxt$ comes directly from the assumption 
		that
		\begin{align*}
			\rank{\xx}{\bx \cup \by} &< \min \{\rank{\xtn{1} }{\bx \cup \by}, \ldots, \rank{\xtn{\n - \np} }{\bx \cup \by} \},
		\end{align*}
		The inequality $\xx < \min \bxh$ is not immediately obvious,
		so we prove it by contradiction. Suppose instead that $\xx > \min \bxh$.
		Without loss of generality, assume $\xsn{1} = \min \bxs < \xx$.
		Subsequently, according to the definition of rank,
		\begin{align*}
			\rank{\xsn{1}}{\bxs \cup \by} &= \sum_{\iconstant = 1}^{\n - \np}\indicator{\xsn{1} \ge \xsn{\iconstant}} + \sum_{\z \in \bxp \cup \by} \indicator{\xsn{1} \ge \z}\\
			& = 1 + \sum_{\z \in \bxp \cup \by} \indicator{\xsn{1} \ge \z}\\
			& \le 1 + \sum_{\z \in \bxp \cup \by} \indicator{\xx \ge \z} \\
			& < 0 + \sum_{\z \in \bxp \cup \by} \indicator{\xx \ge \z}\\
			& = \sum_{\iconstant = 1}^{\n - \np} \indicator{\xx \ge \xtn{\iconstant}} + \sum_{\z \in \bxp \cup \by} \indicator{\xx \ge \z} = \rank{\xx}{\bx \cup \by},
		\end{align*} 
		which contracts the assumption that 
		\begin{align*}
			\rank{\xx}{\bx \cup \by} < \min \{\rank{\xsn{1}}{\bxs \cup \by}, \ldots, \rank{\xsn{\n - \np}}{\bxs \cup \by}\}.
		\end{align*}
		Hence, we must have $\xx < \min \bxs$.
		
		Then, we prove
		$\rank{\xx}{\bxs \cup \by} = \rank{\xx}{\bx \cup \by}$
		using the above results.
		According to the definition of rank, we have
		\begin{align*}
			&\rank{\xx}{\bx \cup \by} = \rank{\xx}{\bxp \cup \by} + \sum_{\iconstant = 1}^{\n - \np}\indicator{\xx \ge \xtn{\iconstant}}\\
			\implies &\rank{\xx}{\bx \cup \by} = \rank{\xx}{\bxp \cup \by},
		\end{align*}
		where the $``\implies''$ holds because $\xx < \min \bxt$.
		Similarly, we have
		\begin{align*}
			&\rank{\xx}{\bxs \cup \by} = \rank{\xx}{\bxp \cup \by} + \sum_{\iconstant = 1}^{\n - \np}\indicator{\xx \ge \xsn{\iconstant}}\\
			\implies & \rank{\xx}{\bxs \cup \by} = \rank{\xx}{\bxp \cup \by}, 
		\end{align*}
		where the $``\implies''$ holds because $\xx < \min \bxh$.
		Hence,	we have shown our desired result
		$\rank{\xx}{\bxs \cup \by} = \rank{\xx}{\bx \cup \by}$.
	\end{proof}
	
	The following lemma provides similar
	results for Lemma~\ref{supp:lemma:4.1}
	and \ref{supp:lemma:4.2}, but is 
	concerning with the rank of $\xx$
	in observed values $\rank{\xx}{\bxp \cup \by}$,
	rather than the rank of $\xx$
	in all values $\rank{\xx}{\bx \cup \by}$.
	
	\begin{lemma} \label{supp:lemma:5}
		Suppose $\bx = \{\xn{1}, \ldots, \xn{\n}\}$ 
		and $\by = \{\yn{1}, \ldots, \yn{\m}\}$ are samples of distinct real values, $\bxp$ is a non-empty subset of $\bx$, and $|\bxp| = \np < \n$. Let $\bxh = \{\xsn{1}, \ldots, \xsn{\n -\np}\}$ be a set of distinct real values,
		and denote $\bxs = \bxh \cup \bxp$. Assume all values in 
		$\bxs \cup \by$ are distinct. Subsequently, for any $\xx \in \bxp$,
		\begin{align*}
			& \rank{\xx}{\bxp \cup \by} < \min \{\rank{\xsn{1}}{\bxs \cup \by}, \ldots, \rank{\xsn{\n - \np}}{\bxs \cup \by} \} \\
			\iff &\rank{\xx}{\bxs \cup \by} = \rank{\xx}{\bxp \cup \by}.
		\end{align*}
	\end{lemma}
	
	\begin{proof}
		We first prove the $``\implies"$. 
		Without loss of generality, let us assume that
		$\xsn{1} = \min \bxs.$ By definition, 
		$\rank{\xsn{1}}{\bxs \cup \by} = \sum_{\iconstant = 1}^{\n- \np} \indicator{\xsn{1} \ge \xsn{\iconstant}} + \sum_{\z \in \bxp \cup \by} \indicator{\xsn{1} \ge \z}$.
		Since $\xsn{1} \notin \bxp \cup \by$, we further have
		\begin{align*}
			\rank{\xsn{1}}{\bxs \cup \by} &= \sum_{\iconstant = 1}^{\n- \np} \indicator{\xsn{1} \ge \xsn{\iconstant}} + \sum_{\z \in \bxp \cup \by} \indicator{\xsn{1} > \z} \\
			& = 1 + \sum_{\z \in \bxp \cup \by} \indicator{\xsn{1} > \z}.
		\end{align*}
		Hence, $		\sum_{\z \in \bxp \cup \by} \indicator{\xsn{1} > \z} = \rank{\xsn{1}}{\bxs \cup \by} - 1.$ For any $\xx \in \bxp$, since
		\begin{align*}
			\rank{\xx}{\bxp \cup \by} < \min \{\rank{\xsn{1}}{\bxs  \cup \by}, \ldots, \rank{\xsn{\n - \np}}{\bxs \cup \by} \},
		\end{align*} 
		we have
		\begin{align*}
			\sum_{\z \in \bxp \cup \by} \indicator{\xsn{1} > \z} 		
			> \rank{\xx}{\bxp \cup \by} - 1 
			= \sum_{\z \in \bxp \cup \by} \indicator{\xx \ge \z} - 1 
			= \sum_{\z \in \bxp \cup \by} \indicator{\xx > \z}.
		\end{align*}
		Thus, $\xsn{1} = \min \bxs > \xx$,
		where $\xx \in \bxp$ is any value in $\bxp$.
		Using this result, we have
		\begin{align*}
			\rank{\xx}{\bxs \cup \by} = \sum_{\iconstant =1}^{\n - \np} \indicator{\xx \ge \xsn{\iconstant}} + \sum_{\z \in \bxp \cup \by} \indicator{\xx \ge \z} 
			=  \sum_{\z \in \bxp \cup \by} \indicator{\xx \ge \z}
			= \rank{\xx}{\bxp \cup \by},
		\end{align*}	
		which proves the $``\implies"$.
		
		We now prove the $``\impliedby"$. For any $\xx \in \bxp$, notice that
		\begin{align*}
			\rank{\xx}{\bxs \cup \by} &= \sum_{\iconstant =1}^{\n - \np} \indicator{\xx \ge \xsn{\iconstant}} + \sum_{\z \in \bxp \cup \by} \indicator{\xx \ge \z} \\
			& = \sum_{\iconstant =1}^{\n - \np} \indicator{\xx \ge \xsn{\iconstant}} + \rank{\xx}{\bxp \cup \by}.
		\end{align*}
		Hence,
		\begin{align*}
			& \rank{\xx}{\bxs \cup \by} = \rank{\xx}{\bxp \cup \by}\\
			\implies &\sum_{\iconstant =1}^{\n - \np} \indicator{\xx \ge \xsn{\iconstant}} = 0\\
			\implies &\xx < \min \{\xsn{1}, \ldots, \xsn{\n - \np}\} \\ 
			\implies &\rank{\xx}{\bxs \cup \by} < \min \{\rank{\xsn{1}}{\bxs \cup \by}, \ldots, \rank{\xsn{\n - \np}}{\bxs \cup \by} \}.
		\end{align*}
		By the definition of rank, we have $\rank{\xx}{\bxp \cup \by}  \le \rank{\xx}{\bxs \cup \by}$.
		Hence, we have
		\begin{align*}
			\rank{\xx}{\bxp \cup \by} < \min \{\rank{\xsn{1}}{\bxs \cup \by}, \ldots, \rank{\xsn{\n - \np}}{\bxs \cup \by} \}, 
		\end{align*}
		which proves the $``\impliedby"$.
	\end{proof}
	
	The following lemma considers the opposite
	condition as considered in Lemma~\ref{supp:lemma:5}.
	
	\begin{lemma} \label{supp:ab:lemma:6}
		Suppose $\bx = \{\xn{1}, \ldots, \xn{\n}\}$ 
		and $\by = \{\yn{1}, \ldots, \yn{\m}\}$ are samples of distinct real values and $\bxp \subset \bx$ is a subset of 
		$\bx$, with sample size $|\bxp| = \np < \n$. 
		Let $\bxh = \{\xsn{1}, \ldots, \xsn{\n -\np}\}$ be a set of real values,
		and denote $\bxs = \bxh \cup \bxp$. Assume that
		all values in $\bxs \cup \by$ are distinct. Then, for any $\xx \in \bxp$,
		\begin{align} 
			\begin{split} \label{supp:ab:lemma:6:eqn:1}
				&\rank{\xx}{\bxp \cup \by} \ge \min \{\rank{\xsn{1}}{\bxs \cup \by}, \ldots, \rank{\xsn{\n - \np}}{\bxs \cup \by} \} \\ 
				\iff & \rank{\xx}{\bxs \cup \by} > \rank{\xx}{\bxp \cup \by}.
			\end{split}
		\end{align}
		Additionally, suppose that the set ${\rank{\xsn{1}}{\bxs \cup \by}, \ldots, \rank{\xsn{\n - \np}}{\bxs \cup \by} }$ consists of consecutive integers. Then, the following equivalence holds:
		\begin{align}
			\begin{split} \label{supp:ab:lemma:6:eqn:2}
				&\rank{\xx}{\bxp \cup \by} \ge \min \{\rank{\xsn{1}}{\bxs \cup \by}, \ldots, \rank{\xsn{\n - \np}}{\bxs \cup \by} \}\\
				\iff & \rank{\xx}{\bxs \cup \by} = \rank{\xx}{\bxp \cup \by} + \n - \np.
			\end{split}
		\end{align}
	\end{lemma}
	
	\begin{proof}
		The $``\impliedby''$ of  \eqref{supp:ab:lemma:6:eqn:1} follows
		directly from Lemma~\ref{supp:lemma:5}. For the 
		$``\implies"$, we have $\rank{\xx}{\bxs \cup \by} \neq \rank{\xx}{\bxp \cup \by}$
		according to Lemma~\ref{supp:lemma:5}. By the definition
		of rank, we also have $ \rank{\xx}{\bxs \cup \by} \ge \rank{\xx}{\bxp \cup \by}$.
		Hence, we have $ \rank{\xx}{\bxs \cup \by} > \rank{\xx}{\bxp \cup \by}$,
		which proves the $``\implies"$.

		We now prove the $``\implies"$ of  \eqref{supp:ab:lemma:6:eqn:2}. 
		For any $\xx \in \bxp$ such that
		\begin{align*}
			\rank{\xx}{\bxp \cup \by} \ge \min \{\rank{\xsn{1}}{\bxs \cup \by}, \ldots, \rank{\xsn{\n - \np}}{\bxs \cup \by} \},
		\end{align*}
		we have $\rank{\xx}{\bxp \cup \by} \ge \rank{\xsn{1}}{\bxs \cup \by}$. 
		By the definition of rank, $\rank{\xx}{\bxs \cup \by} \ge  \rank{\xx}{\bxp \cup \by} $.
		Hence, $\rank{\xx}{\bxs \cup \by} \ge \rank{\xsn{1}}{\bxs \cup \bxp \cup \by}$.
		Further, since $\{\rank{\xsn{1}}{\bxs \cup \by}, \ldots, \rank{\xsn{\n - \np}}{\bxs \cup \by} \}$ consists of consecutive integers, and all values in $\bxs \cup \by$ are distinct, it follows that
		\begin{align*}
			\rank{\xx}{\bxs \cup \by} > \max \{\rank{\xsn{1}}{\bxs \cup \by}, \ldots, \rank{\xsn{\n - \np}}{\bxs \cup \by} \}.
		\end{align*}
		Hence, $\xx > \max \{\xsn{1}, \ldots, \xsn{\n - \np}\}$. Then, we have
		\begin{align*}
			\rank{\xx}{\bxs\cup \by} &= \sum_{\iconstant =1}^{\n - \np} \indicator{\xx \ge \xsn{\iconstant}} + \sum_{\z \in \bxp \cup \by} \indicator{\xx \ge \z} \\
			&= \n - \np + \sum_{\z \in \bxp \cup \by} \indicator{\xx \ge \z}\\
			&=  \n - \np + \rank{\xx}{\bxp \cup \by},
		\end{align*}	
		which proves the $``\implies"$ of \eqref{supp:ab:lemma:6:eqn:2}. 
		
		We now prove the $``\impliedby"$ of \eqref{supp:ab:lemma:6:eqn:2}.  
		Let $\xx \in \bxp$ be any value such that
		\begin{align*}
			\rank{\xx}{\bxs \cup \by} = \rank{\xx}{\bxp \cup \by} + \n - \np.
		\end{align*}
		By the definition of rank,
		\begin{align*}
			\rank{\xx}{\bxs \cup \by} &= \sum_{\iconstant =1}^{\n - \np} \indicator{\xx \ge \xsn{\iconstant}} + \sum_{\z \in \bxp \cup \by} \indicator{\xx \ge \z} \\
			& = \sum_{\iconstant =1}^{\n - \np} \indicator{\xx > \xsn{\iconstant}} + \rank{\xx}{\bxp \cup \by}.
		\end{align*}
		Hence,
		\begin{align*}
			&\rank{\xx}{\bxs \cup \by} =  \n - \np + \rank{\xx}{\bxp \cup \by} \\
			\implies & \sum_{\iconstant =1}^{\n - \np} \indicator{\xx > \xsn{\iconstant}}  = \n - \np \\
			\implies &\xx > \max \{\xsn{1}, \ldots, \xsn{\n - \np}\}.
		\end{align*}
		Without loss of generality, let us assume that  $\xsn{1} = \min \bxs$. Subsequently, since
		\begin{align*}
			\rank{\xsn{1}}{\bxs \cup \by} = \sum_{\iconstant = 1}^{\n- \np} \indicator{\xsn{1} \ge \xsn{\iconstant}} + \sum_{\z \in \bxp \cup \by} \indicator{\xsn{1} \ge \z} 
			= 1 + \sum_{\z \in \bxp \cup \by} \indicator{\xsn{1} > \z},
		\end{align*}
		and
		\begin{align*}
			\rank{\xx}{\bxp \cup \by} = \sum_{\z \in \bxp \cup \by} \indicator{\xx \ge \z}
			= 1 + \sum_{\z \in \bxp \cup \by} \indicator{\xx > \z},
		\end{align*}
		we have
		\begin{align*}
			\rank{\xx}{\bxp \cup \by} \ge \rank{\xsn{1}}{\bxs \cup \bxp \cup \by},
		\end{align*}
		which proves the $``\impliedby''$ of \eqref{supp:ab:lemma:6:eqn:2}.  
	\end{proof}
	
	The following lemma shows that if we
	replace values in $\bx$ with ranks
	in the right hand side of Equation~\eqref{supp:lemma:7:eqn:1}
	with values taking ranks
	in the right hand side of Equation~\eqref{supp:lemma:7:eqn:2},
	the value of the Ansari-Bradley statistic is not changed.
	Note that we now assume that the total sample
	size $\bn = \n + \m$ is odd.
	
	\begin{lemma} \label{supp:lemma:7}
		Suppose $\bx = \{\xn{1}, \ldots, \xn{\n}\}$ 
		and $\by = \{\yn{1}, \ldots, \yn{\m}\}$ are samples of distinct real values.  Let $\bn = \n + \m$ and assume $\bn$ is odd. 
		Suppose $\bxp \subset \bx$ is a non-empty subset of $\bx$ such that 
		$|\bxp| = \np$ is odd. Let  
		$\bxt = \bx \setminus \bxp$,
		and denote $\bxt = \{\xtn{1}, \ldots, \xtn{\n - \np}\}$.
		Consider a set 
		$\bxh = \{\xsn{1}, \ldots, \xsn{\n - \np}\}$ of 
		real values, and denote $\bxs = \bxh \cup \bxp$.
		Assume that all values in $\bxs \cup \by$ 
		are distinct. Subsequently, if
		\begin{align} \label{supp:lemma:7:eqn:1}
			\begin{split}
				&\left\{\rank{\xtn{1}}{\bx \cup \by} ,\ldots, \rank{\xtn{\n - \np}}{\bx \cup\by}\right\}\\
				&= \left\{-\frac{\n - \np}{2} + 1 + \frac{1}{2}(\bn + 1), \ldots, \frac{\n - \np}{2} + \frac{1}{2}(\bn+1) 	\right\},
			\end{split}
		\end{align}
		and
		\begin{align} \label{supp:lemma:7:eqn:2}
			\begin{split}
				&\left\{\rank{\xsn{1}}{ \bxs \cup \by } ,\ldots, \rank{\xsn{\n - \np}}{ \bxs \cup \by }\right\} \\
				&= \left\{-\frac{\n - \np}{2} + \frac{1}{2}(\bn + 1), \ldots, \frac{\n - \np}{2} - 1 + \frac{1}{2}(\bn+1) \right\},
			\end{split}
		\end{align}
		we have $\ABT{\bx}{\by} = \ABT{\bxs}{\by}$.
	\end{lemma}
	
	\begin{proof}
		To start, let us denote $\bnp = (\bn + 1)/2 - (\n - \np)/2$.
		By the definition of the Ansari-Bradley test statistic,
		\begin{align*}
			&\ABT{\bxs}{\by} 
			= \sum_{\xx \in \bxs}\left|\rank{\xx}{\bxs \cup \by} - \frac{1}{2}(N+1)\right| 
			+ \sum_{i \in \bxp} \left|\rank{\xx}{\bxs \cup \by} - \frac{1}{2}(N+1)\right|,\\
			\text{and }&\ABT{\bx}{\by} 
			=  \sum_{\xx \in \bxt}\left|\rank{\xx}{\bx \cup \by} - \frac{1}{2}(N+1)\right|
			+ \sum_{i \in \bxp} \left|\rank{\xx}{\bx \cup \by} - \frac{1}{2}(N+1)\right|.
		\end{align*}
		Note that
		\begin{align*}
			\sum_{\xx \in \bxs}\left|\rank{\xx}{\bxs \cup \by} - \frac{1}{2}(N+1)\right| &= \left|-\frac{\n - \np}{2} + 1\right| + \ldots + 0 + \ldots + \left|\frac{\n - \np}{2}\right|,\\
			\sum_{\xx \in \bxt}\left|\rank{\xx}{\bx \cup \by} - \frac{1}{2}(N+1)\right| &= \left|-\frac{\n - \np}{2} \right| + \ldots + 0 + \ldots + \left|\frac{\n - \np}{2} - 1\right| \\
			& = \left|-\frac{\n - \np}{2} + 1\right| + \ldots + 0 + \ldots + \left|\frac{\n - \np}{2}\right|.
		\end{align*}
		Hence, $\sum_{\xx \in \bxs}\left|\rank{\xx}{\bxs \cup \by} - \frac{1}{2}(N+1)\right| = \sum_{\xx \in \bxt}\left|\rank{\xx}{\bx \cup \by} - \frac{1}{2}(N+1)\right|$.
		Then, in order to prove our result, it is sufficient to show that
		\begin{align*}
			\sum_{\x \in \bxp} \left|\rank{\xx}{\bxs  \cup \by} - \frac{1}{2}(N+1)\right| = \sum_{\x \in \bxp} \left|\rank{\xx}{\bx \cup \by} - \frac{1}{2}(N+1)\right|.
		\end{align*}
		Notice that
		\begin{align*}
			\sum_{\x \in \bxp} \left|\rank{\xx}{\bxs \cup \by} - \frac{1}{2}(N+1)\right| 
			&= \sum_{\x \in \bxp} \indicator{\rank{\xx}{\bxp \cup \by} < \bnp } \left|\rank{\xx}{\bxs \cup \by} - \frac{1}{2}(N+1)\right| \\
			&+ \sum_{\x \in \bxp} \indicator{\rank{\xx}{\bxp \cup \by} \ge \bnp } \left|\rank{\xx}{\bxs \cup \by} - \frac{1}{2}(N+1)\right|.
		\end{align*}
		According to Lemma \ref{supp:lemma:5}, we have that 
		\begin{align*}
			&\sum_{\x \in \bxp} \indicator{\rank{\xx}{\bxp \cup \by} < \bnp } \left|\rank{\xx}{\bxs \cup \by} - \frac{1}{2}(N+1)\right|\\
			&= \sum_{\x \in \bxp} \indicator{\rank{\xx}{\bxp \cup \by} < \bnp } \left|\rank{\xx}{\bxp \cup \by} - \frac{1}{2}(N+1)\right|.
		\end{align*}
		Meanwhile, since $\left\{\rank{\xsn{1}}{ \bxs \cup \by } ,\ldots, \rank{\xsn{\n - \np}}{ \bxs \cup \by }\right\}
		= \left\{\bnp, \ldots, \bnp + \n -\np - 1\right\}$
		consists of consecutive integers, then according to Lemma \ref{supp:ab:lemma:6},
		\begin{align*}
			&\sum_{\x \in \bxp} \indicator{\rank{\xx}{\bxp \cup \by} \ge \bnp } \left|\rank{\xx}{\bxs \cup \by} - \frac{1}{2}(N+1)\right| \\
			&=\sum_{\x \in \bxp} \indicator{\rank{\xx}{\bxp \cup \by} \ge \bnp } \left|\rank{\xx}{\bxp \cup \by} + \n -\np - \frac{1}{2}(N+1)\right|.
		\end{align*}
		Hence, we have 
		\begin{align}
			\begin{split} \label{supp:lemma:7:eqn:3}
				&\sum_{\x \in \bxp} \left|\rank{\xx}{\bxs \cup \by} - \frac{1}{2}(N+1)\right| \\ 
				& = \sum_{\x \in \bxp} \indicator{\rank{\xx}{\bxp \cup \by} < \bnp } \left|\rank{\xx}{\bxp \cup \by} - \frac{1}{2}(N+1)\right| \\
				& + \sum_{\x \in \bxp} \indicator{\rank{\xx}{\bxp \cup \by} \ge \bnp } \left|\rank{\xx}{\bxp \cup \by} + \n - \np - \frac{1}{2}(N+1)\right|\\
				& = \sum_{\x \in \bxp} \indicator{\rank{\xx}{\bxp \cup \by} < \bnp } \left( \frac{1}{2}(N+1) - \rank{\xx}{\bxp \cup \by}\right) \\
				& + \sum_{\x \in \bxp} \indicator{\rank{\xx}{\bxp \cup \by} \ge \bnp } \left(\rank{\xx}{\bxp \cup \by} + \n - \np - \frac{1}{2}(N+1)\right).
			\end{split}
		\end{align}
		
		Similarly, by applying Lemma \ref{supp:lemma:5}
		and  Lemma \ref{supp:ab:lemma:6}, we can show that
		\begin{align*}
			&\sum_{\x \in \bxp} \left|\rank{\xx}{\bx \cup \by} - \frac{1}{2}(N+1)\right| \\ 
			& = \sum_{\x \in \bxp} \indicator{\rank{\xx}{\bxp \cup \by} < \bnp + 1 } \left|\rank{\xx}{\bxp \cup \by} - \frac{1}{2}(N+1)\right| \\
			& + \sum_{\x \in \bxp} \indicator{\rank{\xx}{\bxp \cup \by} \ge \bnp + 1 } \left|\rank{\xx}{\bxp \cup \by} + \n - \np - \frac{1}{2}(N+1)\right|\\
			& = \sum_{\x \in \bxp} \indicator{\rank{\xx}{\bxp \cup \by} < \bnp + 1 } \left( \frac{1}{2}(N+1) - \rank{\xx}{\bxp \cup \by}\right) \\
			& + \sum_{\x \in \bxp} \indicator{\rank{\xx}{\bxp \cup \by} \ge \bnp + 1 } \left(\rank{\xx}{\bxp \cup \by} + \n - \np - \frac{1}{2}(N+1)\right).
		\end{align*}
		Notice that
		\begin{align*}
			&\sum_{\x \in \bxp} \indicator{\rank{\xx}{\bxp \cup \by} < \bnp + 1 } \left( \frac{1}{2}(N+1) - \rank{\xx}{\bxp \cup \by}\right) \\
			&= \sum_{\x \in \bxp} \indicator{\rank{\xx}{\bxp \cup \by} < \bnp } \left( \frac{1}{2}(N+1) - \rank{\xx}{\bxp \cup \by}\right)\\
			&+ \sum_{\x \in \bxp} \indicator{\bnp \le \rank{\xx}{\bxp \cup \by} < \bnp + 1 } \left( \frac{1}{2}(N+1) - \rank{\xx}{\bxp \cup \by}\right).
		\end{align*}
		Meanwhile,
		\begin{align*}
			&\sum_{\x \in \bxp} \indicator{\rank{\xx}{\bxp \cup \by} \ge \bnp + 1 } \left(\rank{\xx}{\bxp \cup \by} + \n - \np - \frac{1}{2}(N+1)\right)\\
			& = \sum_{\x \in \bxp} \indicator{\rank{\xx}{\bxp \cup \by} \ge \bnp } \left(\rank{\xx}{\bxp \cup \by} + \n - \np - \frac{1}{2}(N+1)\right)\\
			& - \sum_{\x \in \bxp} \indicator{\bnp + 1 > \rank{\xx}{\bxp \cup \by} \ge \bnp } \left(\rank{\xx}{\bxp \cup \by} + \n - \np - \frac{1}{2}(N+1)\right).
		\end{align*}
		Hence, we can further decompose $\sum_{i \in \bxp} \left|\rank{\xx}{\bx \cup \by} - \frac{1}{2}(N+1)\right|$ as follows
		\begin{align*}
			&\sum_{i \in \bxp} \left|\rank{\xx}{\bx \cup \by} - \frac{1}{2}(N+1)\right| \\
			&= \sum_{i \in \bxp} \indicator{\rank{\xx}{\bxp \cup \by} < \bnp } \left( \frac{1}{2}(N+1) - \rank{\xx}{\bxp \cup \by}\right)\\
			&+ \sum_{i \in \bxp} \indicator{\bnp \le \rank{\xx}{\bxp \cup \by} < \bnp + 1 } \left( \frac{1}{2}(N+1) - \rank{\xx}{\bxp \cup \by}\right)\\
			&+ \sum_{i \in \bxp} \indicator{\rank{\xx}{\bxp \cup \by} \ge \bnp } \left(\rank{\xx}{\bxp \cup \by} + \n - \np - \frac{1}{2}(N+1)\right)\\
			& - \sum_{i \in \bxp} \indicator{\bnp + 1 > \rank{\xx}{\bxp \cup \by} \ge \bnp } \left(\rank{\xx}{\bxp \cup \by} + \n - \np - \frac{1}{2}(N+1)\right).
		\end{align*}
		Recall that in Equation \eqref{supp:lemma:7:eqn:3}, we showed that
		\begin{align*}
			&\sum_{\x \in \bxp} \left|\rank{\xx}{\bxs \cup \by} - \frac{1}{2}(N+1)\right|\\
			& = \sum_{\x \in \bxp} \indicator{\rank{\xx}{\bxp \cup \by} < \bnp } \left( \frac{1}{2}(N+1) - \rank{\xx}{\bxp \cup \by}\right) \\
			& + \sum_{\x \in \bxp} \indicator{\rank{\xx}{\bxp \cup \by} \ge \bnp } \left(\rank{\xx}{\bxp \cup \by} + \n - \np - \frac{1}{2}(N+1)\right).
		\end{align*}
		Hence, we further have
		\begin{align*}
			&\sum_{\x \in \bxp} \left|\rank{\xx}{\bx \cup \by} - \frac{1}{2}(N+1)\right| \\
			&= \sum_{\x \in \bxp} \left|\rank{\xx}{\bxs \cup \by} - \frac{1}{2}(N+1)\right|\\
			&+ \sum_{\x \in \bxp} \indicator{\bnp \le \rank{\xx}{\bxp \cup \by} < \bnp + 1 } \left( \frac{1}{2}(N+1) - \rank{\xx}{\bxp \cup \by}\right)\\
			& - \sum_{\x \in \bxp} \indicator{\bnp + 1 > \rank{\xx}{\bxp \cup \by} \ge \bnp } \left(\rank{\xx}{\bxp \cup \by} + \n - \np - \frac{1}{2}(N+1)\right).
		\end{align*}
		If $\sum_{i \in \bxp} \indicator{\bnp \le \rank{\xx}{\bxp \cup \by} < \bnp + 1 } = 0$
		then we have
		\begin{align*}
			\sum_{\x \in \bxp} \left|\rank{\xx}{\bx \cup \by} - \frac{1}{2}(N+1)\right| = \sum_{\x \in \bxp} \left|\rank{\xx}{\bxs \cup \by} - \frac{1}{2}(N+1)\right|,
		\end{align*}
		which proves our result. Otherwise,  we have
		\begin{align*}
			\sum_{\x \in \bxp} \indicator{\bnp \le \rank{\xx}{\bxp \cup \by} < \bnp + 1 } = 1,
		\end{align*}
		and $\rank{\xx}{\bxp \cup \by} = \bnp$. Subsequently,
		\begin{align*}
			&\sum_{\x \in \bxp} \indicator{\bnp \le \rank{\xx}{\bxp \cup \by} < \bnp + 1 } \left( \frac{1}{2}(N+1) - \rank{\xx}{\bxp \cup \by}\right)\\
			& - \sum_{\x \in \bxp} \indicator{\bnp + 1 > \rank{\xx}{\bxp \cup \by} \ge \bnp } \left(\rank{\xx}{\bxp \cup \by} + \n - \np - \frac{1}{2}(N+1)\right)\\
			& = \left( \frac{1}{2}(N+1) - \bnp \right) - \left(\bnp + \n - \np - \frac{1}{2}(N+1)\right) \\
			& = \left( \frac{1}{2}(N+1) + \frac{\n - \np}{2} - \frac{1}{2}(\bn + 1) \right) - \left(-\frac{\n - \np}{2} + \frac{1}{2}(\bn + 1) + \n - \np - \frac{1}{2}(N+1)\right)\\
			& = \frac{\n - \np}{2} -\frac{\n - \np}{2} = 0.
		\end{align*}
		Hence, we obtain our desired result
		\begin{align*}
			\sum_{\x \in \bxp} \left|\rank{\xx}{\bx \cup \by} - \frac{1}{2}(N+1)\right| 
			= \sum_{\x \in \bxp} \left|\rank{\xx}{\bxs \cup \by} - \frac{1}{2}(N+1)\right|.
		\end{align*}
	\end{proof}
	
	The following proposition considers
	the ranks of imputation which would
	allow us to take the lower bound
	of the Ansari-Bradley statistic, when
	the total sample size $\bn = \n + \m$
	is odd.
	
	\begin{proposition} \label{supp:prop:4}
		Suppose $\bx = \{\xn{1}, \ldots, \xn{\n}\}$ 
		and $\by = \{\yn{1}, \ldots, \yn{\m}\}$ are samples of distinct real values. 
		Denote $\bn = \n + \m$ and assume $\bn$ is odd. 
		Let $\bxp \subset \bx$ be a non-empty subset of $\bx$ with size $|\bxp| = \np$. Consider
		a set $\bxh = \{\xsn{1},\ldots,\xsn{\n -\np}\}$ of real values,
		and denote $\bxs =\bxh \cup\bxp$. Assume that all values in $\bxs\cup \by$ are distinct,
		and denote $\mysetr = \left\{\rank{\xsn{1}}{\bxs \cup \by} ,\ldots, \rank{\xsn{\n - \np}}{\bxs \cup \by}\right\}$. Subsequently, if $\n - \np$ is even, 
		then we have $\ABT{\bxs}{\by} \le \ABT{\bx}{\by}$, provided that either
		\begin{align*}
			&\mysetr = \left\{-\frac{\n - \np}{2} + 1 + \frac{1}{2}(\bn + 1), \ldots, \frac{\n - \np}{2} + \frac{1}{2}(\bn+1) \right\}, \\
			\text{ or }& \mysetr = \left\{-\frac{\n - \np}{2} + \frac{1}{2}(\bn + 1), \ldots, \frac{\n - \np}{2} - 1 + \frac{1}{2}(\bn+1) \right\}.
		\end{align*}
		However, if $\n - \np$ is odd then we have $\ABT{\bxs}{\by} \le \ABT{\bx}{\by}$,
		provided that
		\begin{align*}
			\mysetr = 	\left\{-\frac{\n - \np - 1}{2} + \frac{1}{2}(\bn + 1), \ldots, \frac{\n - \np - 1}{2} + \frac{1}{2}(\bn+1) \right\}.
		\end{align*}
	\end{proposition}
	
	\begin{proof}
		We prove the proposition by mathematical induction on $\n - \np$.
		Consider the base case when $\n - \np = 1$.
		We are given that $\rank{\xsn{1}}{\bxs \cup \by} = (\bn + 1)/2$, 
		which immediately follows that 
		$\left|\rank{\xsn{1}}{\bxs \cup \by} - (\bn + 1)/2 \right| = 0$. 
		Hence, we have
		$\ABT{\bxs}{\by} \le \ABT{\bx}{\by}$ 
		according to Proposition \ref{supp:prop:3}, which proves
		the base case.
		
		Let $\kk \in \{1, \ldots, \n - 1\}$, and suppose that the statement of Proposition~\ref{supp:prop:4} holds for $\n - \np = \kk$.
		We show that the proposition also holds for $\n - \np = \kk + 1$.
		We prove this result for the following two cases:
		\begin{align*}
			&\text{case} \case{I}: \kk \text{ is odd},\\
			\text{and } &\text{case} \case{II}: \kk \text{ is even},
		\end{align*}
		separately.
		
		Suppose case $\case{I}$ holds. We show that the proposition also holds for $\n - \np = \kk + 1$.
		Specifically, we show that $\ABT{\bxs}{\by} \le \ABT{\bx}{\by}$
		provided either
		\begin{align*}
			&\mysetr = \left\{-\frac{\n - \np}{2} + 1 + \frac{1}{2}(\bn + 1), \ldots, \frac{\n - \np}{2} + \frac{1}{2}(\bn+1) \right\}, \\
			\text{ or }& \mysetr = \left\{-\frac{\n - \np}{2} + \frac{1}{2}(\bn + 1), \ldots, \frac{\n - \np}{2} - 1 + \frac{1}{2}(\bn+1) \right\},
		\end{align*}
		where $\n - \np = \kk + 1$. By Lemma~\ref{supp:lemma:7}, these two cases
		yield the same value of the Ansari-Bradley test statistic
		$\ABT{\bxs}{\by}$. Therefore, denote
		\begin{align*}
			&\mysetrln{1} = \left\{-\frac{\n - \np}{2} + 1 + \frac{1}{2}(\bn + 1), \ldots, \frac{\n - \np}{2} + \frac{1}{2}(\bn+1) \right\}, \\
			\text{ or }& \mysetrln{2} = \left\{-\frac{\n - \np}{2} + \frac{1}{2}(\bn + 1), \ldots, \frac{\n - \np}{2} - 1 + \frac{1}{2}(\bn+1) \right\},
		\end{align*}
		it is sufficient to show $\ABT{\bxs}{\by} \le \ABT{\bx}{\by}$ when $\mysetr = \mysetrln{1}$ or $\mysetr = \mysetrln{2}$.
		
		Let us denote $\bxt = \bx \setminus \bxp$, and $\bxt = \{\xtn{1}, \ldots, \xtn{\kk+1}\}$.
		Consider a set $\bzs = \{\zsn{1},\ldots,\zsn{\kk}\}$ of distinct real values,
		and denote $\bz = \bzs \cup \{\xtn{1}\} \cup \bxp$.
		Assume that all values in $\bz \cup \by$ are distinct, and
		\begin{align*}
			\left\{\rank{\zsn{1}}{\bz \cup \by} ,\ldots, \rank{\zsn{\kk}}{\bz\cup \by}\right\} 
			=\left\{-\frac{\kk-1}{2} + \frac{1}{2}(\bn + 1), \ldots, \frac{\kk-1}{2} + \frac{1}{2}(\bn+1) \right\}.
		\end{align*}
		By the induction hypothesis, i.e. 
		the Proposition \ref{supp:prop:4} is correct when $\n - \np = \kk$,
		it follows that
		\begin{align} \label{supp:prop:4:eqn:2}
			\ABT{\bz}{\by} \le \ABT{\bx}{\by}.
		\end{align}	
		
		Since  $\zsn{1},\ldots,\zsn{\kk}$ takes the ranks 
		between $-{(\kk-1)}/{2} + (\bn + 1)/2$ to 
		${(\kk-1)}/2 + (\bn+1)/2$, the rank of $\xtn{1}$ in $\bz \cup \by$
		is either smaller than  $-{(\kk-1)}/{2} + (\bn + 1)/2$ 
		or greater than ${(\kk-1)}/2 + (\bn+1)/2$. 
		In other words, we have either	 
		\begin{align*}
			\rank{\xtn{1}}{\bz \cup \by} < -\frac{\kk-1}{2} + \frac{1}{2}(\bn + 1), \text{ or } \rank{\xtn{1}}{\bz \cup \by} > \frac{\kk-1}{2} + \frac{1}{2}(\bn + 1).
		\end{align*}
		Below we prove that  we have our desired result
		$\ABT{\bxs}{\by} \le \ABT{\bx}{\by}$ for both cases. 
		
		Suppose $\rank{\xtn{1}}{\bz \cup \by} < -{(\kk-1)}/{2} + (\bn + 1)/2$.
		Then, consider a real value $\xs$ that is distinct to values in 
		$\bzs \cup \bxp \cup \by$. Denote $\bzp = \{\xs\} \cup \bzs \cup \bxp$,
		and assume that
		\begin{align*}
			\rank{\xs}{\bzp \cup \by} = -\frac{\kk-1}{2} - 1 + \frac{1}{2}(\bn + 1).
		\end{align*}
		Notice that 
		$\rank{\xs}{\bzp \cup \by} < (\bn+1)/2$ and
		$\rank{\xtn{1}}{\bz \cup \by} <  (\bn + 1)/2$, and 
		\begin{align*}
			\left|\rank{\xs}{\bzp \cup \by} - \frac{1}{2}(\bn+1)\right| \le \left|\rank{\xtn{1}}{\bz \cup \by} - \frac{1}{2}(\bn+1)\right|.
		\end{align*}	
		According to Proposition \ref{supp:ab:prop:2}, we have 
		\begin{align*}
			\ABT{\bzp}{\by} \le \ABT{\bz}{\by} \le^{\text{ Inequality }  \eqref{supp:prop:4:eqn:2}} \ABT{\bx}{\by}.
		\end{align*}
		To complete our prove for this case, it then suffices to show that
		$\ABT{\bxs}{\by} = 	\ABT{\bzp}{\by}$.
		
		Recall that for any $\iconstant \in \{1,\ldots,\kk\}$, we have
		\begin{align*}
			\rank{\zsn{\iconstant}}{\bz \cup \by} > -\frac{\kk-1}{2} - 1 + \frac{1}{2}(\bn + 1).
		\end{align*}
		Then, for any $\iconstant \in \{1,\ldots,\kk\}$, 
		\begin{align*}
			&\rank{\zsn{\iconstant}}{\bz \cup \by} \ge -\frac{\kk-1}{2} + \frac{1}{2}(\bn + 1) > \rank{\xtn{1}}{\bz \cup \by} , \\
			\text{and }
			&\rank{\zsn{\iconstant}}{\bz \cup \by} > \rank{\xs}{\bzp \cup \by}.
		\end{align*}
		According to Lemma \ref{supp:lemma:4.1},
		for any $\iconstant \in \{1,\ldots,\kk\}$,
		\begin{align*}
			&\rank{\zsn{\iconstant}}{\bz \cup \by}  = \rank{\zsn{\iconstant}}{\bzp \cup \by}\\
			\implies &\{\rank{\zsn{1}}{\bzp \cup \by}, \ldots, \rank{\zsn{\kk}}{\bzp \cup \by}\}
			= \left\{-\frac{\kk-1}{2} + \frac{1}{2}(\bn + 1), \ldots, \frac{\kk-1}{2} + \frac{1}{2}(\bn+1) \right\}.
		\end{align*}
		Hence, we have
		\begin{align*}
			\mysetrln{2} = \rank{\xs}{\bzp \cup \by} \cup \{\rank{\zsn{1}}{\bzp \cup \by}, \ldots, \rank{\zsn{\kk}}{\bzp \cup \by}\}.
		\end{align*}
		According to Lemma~\ref{supp:ab:lemma:2}, we have
		\begin{align*}
			\ABT{\bxs}{\by} = 	\ABT{\bzp}{\by},
		\end{align*}
		which completes our proof for this case when $\rank{\xtn{1}}{\bz \cup \by} < -{(\kk-1)}/{2} + (\bn + 1)/2$.

		The case when $\rank{\xtn{1}}{\bz \cup \by} > {(\kk-1)}/2 + (\bn + 1)/2$
		can be proved similarly. Consider a real value $\xs$ that is distinct to 
		all values in $\bzs \cup \bxp \cup \by$. Denote $\bzp = \{\xs\} \cup \bzs \cup \bxp$,
		and assume that
		\begin{align*}
			\rank{\xs}{\bzp \cup \by} = \frac{\kk-1}{2} + 1 + \frac{1}{2}(\bn + 1).
		\end{align*}
		Note that $	\rank{\xs}{\bzp \cup \by} \ge (\bn+1)/2$ and
		$\rank{\xtn{1}}{\bz \cup \by} \ge (\bn + 1)/2$, and
		\begin{align*}
			\left|\rank{\xs}{\bzp \cup \by} - \frac{1}{2}(\bn + 1)\right| \le \left|\rank{\xtn{1}}{\bz \cup \by} - \frac{1}{2}(\bn + 1)\right|.
		\end{align*}
		Then, according to Proposition \ref{supp:ab:prop:2},
		\begin{align*}
			\ABT{\bzp}{\by} \le \ABT{\bz}{\by} \le^{\text{ Inequality }  \eqref{supp:prop:4:eqn:2}} \ABT{\bx}{\by}.
		\end{align*}
		To complete our prove for this case, it then suffices to show that
		$\ABT{\bxs}{\by} = 	\ABT{\bzp}{\by}$.
		
		Recall that for any $\iconstant \in \{1,\ldots,\kk\}$,
		\begin{align*}
			\rank{\zsn{\iconstant}}{\bz \cup \by} < \frac{\kk-1}{2} + 1 + \frac{1}{2}(\bn + 1).
		\end{align*}
		Then for any $\iconstant \in \{1,\ldots,\kk\}$, we have
		\begin{align*}
			&\rank{\xs}{\bzp \cup \by} = \frac{\kk-1}{2} +1 + \frac{1}{2}(\bn + 1) > \rank{\zsn{\iconstant}}{\bz \cup \by} , \\
			\text{and }&\rank{\xtn{1}}{\bz \cup \by}  > \rank{\zsn{\iconstant}}{\bz \cup \by} .
		\end{align*}
		According to Lemma \ref{supp:lemma:4.2}, for any $\iconstant \in \{1,\ldots,\kk\}$,
		\begin{align*}
			&\rank{\zsn{\iconstant}}{\bz \cup \by}  = \rank{\zsn{\iconstant}}{\bzp  \cup \by}\\
			\implies &\{\rank{\zsn{1}}{\bzp \cup \by}, \ldots, \rank{\zsn{\kk}}{\bzp \cup \by}\}
			= \left\{-\frac{\kk-1}{2} + \frac{1}{2}(\bn + 1), \ldots, \frac{\kk-1}{2} + \frac{1}{2}(\bn+1) \right\}.
		\end{align*}
		Hence, we have
		\begin{align*}
			\mysetrln{1} = \rank{\xs}{\bzp \cup \by} \cup \{\rank{\zsn{1}}{\bzp \cup \by}, \ldots, \rank{\zsn{\kk}}{\bzp \cup \by}\}.
		\end{align*}
		According to Lemma~\ref{supp:ab:lemma:2}, we have
		\begin{align*}
			\ABT{\bxs}{\by} = 	\ABT{\bzp}{\by},
		\end{align*}
		which completes our proof for this case 
		when $\rank{\xtn{1}}{\bz \cup \by} > {(\kk-1)}/2 + (\bn + 1)/2$.

		Suppose case $\case{II}$ holds. In other words,
		$\kk$ is even.
		We show that the proposition also holds for $\n - \np = \kk + 1$.
		Specifically, we show that $\ABT{\bxs}{\by} \le \ABT{\bx}{\by}$
		provided
		\begin{align*}
			&\mysetr = \left\{-\frac{\n - \np - 1}{2} + \frac{1}{2}(\bn + 1), \ldots, \frac{\n - \np - 1}{2} + \frac{1}{2}(\bn+1) \right\},
		\end{align*}
		where $\n - \np = \kk + 1$. As before, let us denote $\bxt = \bx \setminus \bxp$, and $\bxt = \{\xtn{1}, \ldots, \xtn{\kk+1}\}$.
		Then, we show that $\ABT{\bxs}{\by} \le \ABT{\bx}{\by}$
		for the following two cases separately:
		\begin{align*}
			\case{i}: \rank{\xtn{1}}{ \{\xtn{1}\} \cup \bxp \cup \by} < -\frac{\kk}{2} + 1 + \frac{1}{2}(\bn + 1),\\
			\case{ii}: \rank{\xtn{1}}{ \{\xtn{1}\} \cup \bxp \cup \by} \ge -\frac{\kk}{2} + 1 + \frac{1}{2}(\bn + 1).
		\end{align*}
		
		Suppose case $\case{i}$ holds, i.e., assume that
		$\rank{\xtn{1}}{ \{\xtn{1}\} \cup \bxp \cup \by} < -{\kk}/2 + 1 + (\bn + 1)/2$.
		Consider a set $\bzs = \{\zsn{1},\ldots,\zsn{\kk}\}$ of distinct real values,
		and denote $\bz = \bzs \cup \{\xtn{1}\} \cup \bxp$.
		Assume that all values in $\bz$ are distinct,
		and
		\begin{align*}
			\begin{split}
				\left\{\rank{\zsn{1}}{\bz \cup \by} ,\ldots, \rank{\zsn{\kk}}{\bz \cup \by}\right\}
				=	\left\{-\frac{\kk}{2} + 1 + \frac{1}{2}(\bn + 1), \ldots, \frac{\kk}{2} + \frac{1}{2}(\bn+1) \right\}.
			\end{split}
		\end{align*}
		By the induction hypothesis, i.e. the 
		Proposition \ref{supp:prop:4} is correct when $\n - \np = \kk$,
		we have
		\begin{align} \label{supp:prop:4:eqn:4}
			\ABT{\bz}{\by} \le \ABT{\bx}{\by}.
		\end{align}
		
		Notice that  
		\begin{align*}
			\rank{\xtn{1}}{ \{\xtn{1}\} \cup \bxp \cup \by} < -\frac{\kk}{2} + 1 + \frac{1}{2}(\bn + 1) = \min 	\left\{\rank{\zsn{1}}{\bz \cup \by} ,\ldots, \rank{\zsn{\kk}}{\bz \cup \by}\right\}.
		\end{align*}
		According to Lemma~\ref{supp:lemma:5}, we have
		\begin{align*}
			\rank{\xtn{1}}{\bz \cup \by} = \rank{\xtn{1}}{ \{\xtn{1}\} \cup \bxp \cup \by} < -\frac{\kk}{2} + 1 + \frac{1}{2}(\bn + 1).
		\end{align*}
		Now consider a real value $\xs$. Denote $\bzp = \bzs \cup \{\xs\} \cup \bxp$.
		Assume that values in $\bzp$ are distinct and 
		$$\rank{\xs}{\bzp \cup \by} = -{\kk}/{2} + (\bn + 1)/2.$$ Notice that
		$	\rank{\xtn{1}}{\bz \cup \by} \le (\bn+1)/2$
		and $\rank{\xs}{\bzp \cup \by}  \le (\bn+1)/2$,
		and $	\rank{\xtn{1}}{\bz \cup \by} \le \rank{\xs}{\bzp \cup \by}$.			
		Then according to Proposition \ref{supp:ab:prop:2}, we have
		\begin{align*}
			\ABT{\bzp}{\by} \le \ABT{\bz}{\by} \le^{\text{ Inequality }  \eqref{supp:prop:4:eqn:4} } \ABT{\bx}{\by}.
		\end{align*}
		Then, for proving our result under this case, it suffices to prove
		$\ABT{\bxs}{\by} = \ABT{\bzp}{\by}$.
		
		Notice that for any $\iconstant \in \{1,\ldots,\kk\}$,
		\begin{align*}
			\rank{\zsn{\iconstant}}{\bz \cup \by} \ge -\frac{\kk}{2} + 1+ \frac{1}{2}(\bn + 1).
		\end{align*}
		Then, for any $\iconstant \in \{1,\ldots,\kk\}$, we have
		\begin{align*}
			\rank{\xs}{\bzp \cup \by} &= -\frac{\kk}{2} + \frac{1}{2}(\bn + 1) < \rank{\zsn{\iconstant}}{\bz \cup \by}  \\
			\rank{\xtn{1}}{\bz \cup \by} &< -\frac{\kk}{2} + 1 + \frac{1}{2}(\bn + 1) \le 	\rank{\zsn{\iconstant}}{\bz \cup \by} .
		\end{align*}
		According to Lemma \ref{supp:lemma:4.1}, for any $\iconstant \in \{1,\ldots,\kk\}$,
		\begin{align*}
			&\rank{\zsn{\iconstant}}{\bzp \cup \by}  = \rank{\zsn{\iconstant}}{\bz \cup \by} \\
			\implies&\{\rank{\zsn{1}}{\bzp \cup \by}, \ldots, \rank{\zsn{\kk}}{\bzp \cup \by}\} =  \left\{-\frac{\kk}{2} + 1 + \frac{1}{2}(\bn + 1), \ldots, \frac{\kk}{2} + \frac{1}{2}(\bn+1) \right\}.
		\end{align*}
		Hence, we have
		\begin{align*}
			\mysetr = 	\{\rank{\xs}{\bzp \cup \by}\} \cup \{\rank{\zsn{1}}{\bzp \cup \by}, \ldots, \rank{\zsn{\kk}}{\bzp \cup \by}\} .
		\end{align*}
		According to Lemma~\ref{supp:ab:lemma:2}, we have
		\begin{align*}
			\ABT{\bxs}{\by} = 	\ABT{\bzp}{\by},
		\end{align*}
		which completes our proof for case $\case{i}$.
		
		Case $\case{ii}$ can be proved similarly. Suppose case 
		$\case{ii}$ holds, i.e., assume that 
		$\rank{\xtn{1}}{ \{\xtn{1}\} \cup \bxp \cup \by} \ge -{\kk}/2 + 1 + (\bn + 1)/2$.
		Consider a set $\bzs = \{\zsn{1},\ldots,\zsn{\kk}\}$ of distinct real values,
		and denote $\bz = \bzs \cup \{\xtn{1}\} \cup \bxp$.
		Assume that all values in $\bz$ are distinct, and
		\begin{align*}
			\begin{split}
				\left\{\rank{\zsn{1}}{\bz \cup \by} ,\ldots, \rank{\zsn{\kk}}{\bz \cup \by}\right\}
				=	\left\{-\frac{\kk}{2} + \frac{1}{2}(\bn + 1), \ldots, \frac{\kk}{2} -1 + \frac{1}{2}(\bn+1) \right\}.
			\end{split}
		\end{align*}
		By the induction hypothesis, i.e. the 
		Proposition \ref{supp:prop:4} is correct when $\n - \np = \kk$,
		we have
		\begin{align} \label{supp:prop:4:eqn:6}
			\ABT{\bz}{\by} \le \ABT{\bx}{\by}.
		\end{align}
		
		Notice that
		\begin{align*}
			\rank{\xtn{1}}{ \{\xtn{1}\} \cup \bxp \cup \by} \ge -\frac{\kk}{2} + 1 + \frac{(\bn + 1)}{2}  	\ge \min \left\{\rank{\zsn{1}}{\bz \cup \by} ,\ldots, \rank{\zsn{\kk}}{\bz \cup \by}\right\}.
		\end{align*}
		Then according to Lemma \ref{supp:ab:lemma:6}, we have
		\begin{align*}
			\rank{\xtn{1}}{\bz \cup \by} = \rank{\xtn{1}}{ \{\xtn{1}\} \cup \bxp \cup \by} + \kk \ge \frac{\kk}{2} + 1 + \frac{1}{2}(\bn + 1).
		\end{align*}
		Now consider a real value $\xs$. Denote $\bzp = \bzs \cup \{\xs\} \cup \bxp$.
		Assume that values in $\bzp$ are distinct and
		\begin{align*}
			\rank{\xs}{\bzp \cup \by} = \frac{\kk}{2} + \frac{1}{2}(\bn + 1).
		\end{align*}
		Notice that $\rank{\xs}{\bzp \cup \by} \ge (\bn + 1)/2$,
		and $	\rank{\xtn{1}}{\bz \cup \by}  \ge (\bn + 1)/2$
		and $\rank{\xs}{\bzp \cup \by}  \le \rank{\xtn{1}}{\bz \cup \by}$.
		Then, according to Proposition \ref{supp:ab:prop:2}, we have
		\begin{align*}
			\ABT{\bzp}{\by} \le \ABT{\bz}{\by} \le^{\text{ Inequality }  \eqref{supp:prop:4:eqn:6}} \ABT{\bx}{\by}.
		\end{align*}
		Then, for proving our result under this case, it suffices to prove
		$\ABT{\bxs}{\by} = \ABT{\bzp}{\by}$.
		
		Notice that for any $\iconstant \in \{1,\ldots,\kk\}$,
		\begin{align*}
			\rank{\zsn{\iconstant}}{\bz \cup \by} \le \frac{\kk}{2} - 1 + \frac{1}{2}(\bn + 1).
		\end{align*}
		Thus,  for any $\iconstant \in \{1,\ldots,\kk\}$,
		\begin{align*}
			&\rank{\xs}{\bzp \cup \by}  = \frac{\kk}{2} + \frac{1}{2}(\bn + 1) > \rank{\zsn{\iconstant}}{\bz \cup \by}, \\
			\text{and }	&\rank{\xtn{1}}{\bz \cup \by} \ge \frac{\kk}{2} + 1 + \frac{1}{2}(\bn + 1) > \rank{\zsn{\iconstant}}{\bz \cup \by} .
		\end{align*}
		According to Lemma \ref{supp:lemma:4.2}, for any $\iconstant \in \{1,\ldots,\kk\}$,
		\begin{align*}
			&\rank{\zsn{\iconstant}}{\bz \cup \by}  = \rank{\zsn{\iconstant}}{\bzp \cup \by}\\
			\implies& \left\{\rank{\zsn{1}}{\bzp \cup \by} ,\ldots, \rank{\zsn{\kk}}{\bzp \cup \by}\right\}
			=	\left\{-\frac{\kk}{2} + \frac{1}{2}(\bn + 1), \ldots, \frac{\kk}{2} -1 + \frac{1}{2}(\bn+1) \right\}.
		\end{align*}
		Hence, we have
		\begin{align*}
			\mysetr = 	\{\rank{\xs}{\bzp \cup \by}\} \cup \{\rank{\zsn{1}}{\bzp \cup \by}, \ldots, \rank{\zsn{\kk}}{\bzp \cup \by}\} .
		\end{align*}
		According to Lemma~\ref{supp:ab:lemma:2}, we have
		\begin{align*}
			\ABT{\bxs}{\by} = 	\ABT{\bzp}{\by},
		\end{align*}
		which proves our result under case $\case{i}$, and
		completes our proof.
	\end{proof}
	
	Recall that Lemma \ref{supp:lemma:7} and Proposition \ref{supp:prop:4} both assume that the total sample size
	$\bn = \n + \m$ is odd. The following Lemma \ref{supp:lemma:8} and Proposition \ref{supp:prop:5} provides similar results when the
	total sample size $\bn$ is even. 
	
	The proofs for Lemma \ref{supp:lemma:8} and Proposition \ref{supp:prop:5} are similar to that of Lemma \ref{supp:lemma:7} and Proposition \ref{supp:prop:4}, respectively, differing only 
	in notation. Thus, the proofs are omitted here.
	
	\begin{lemma} \label{supp:lemma:8}
		Suppose $\bx = \{\xn{1}, \ldots, \xn{\n}\}$ 
		and $\by = \{\yn{1}, \ldots, \yn{\m}\}$ are samples of distinct real values. 
		Let $\bn = \n + \m$ and assume $\bn$ is even. 
		Suppose $\bxp \subset \bx$ is a non-empty subset of $\bx$ such that $|\bx| = \np$ is odd.
		Let $\bxt = \bx \setminus \bxp$ and denote $\bxt = \{\xtn{1}, \ldots, \xtn{\n - \np}\}$.
		Consider a set $\bxt = \{\xsn{1}, \ldots, \xsn{\n - \np}\}$ of real values,
		and denote $\bxs = \bxt \cup \bxp$. Assume 
		that values in $\bxs \cup \bxp$ are distinct. Then, if
		\begin{align*} 
			\begin{split}
				&\left\{\rank{\xtn{1}}{\bx\cup \by} ,\ldots, \rank{\xtn{\n - \np}}{\bx \cup\by}\right\}\\
				&= \left\{-\frac{\n - \np}{2} + 1 + \frac{1}{2}(\bn + 1), \ldots, \frac{\n - \np}{2} + \frac{1}{2}(\bn+1) \right\},
			\end{split}
		\end{align*}
		and
		\begin{align*}
			\begin{split}
				&\left\{\rank{\xsn{1}}{ \bxs \cup \by } ,\ldots, \rank{\xsn{\n - \np}}{ \bxs \cup \by }\right\} \\
				&= \left\{-\frac{\n - \np}{2} + \frac{1}{2}(\bn + 1), \ldots, \frac{\n - \np}{2} - 1 + \frac{1}{2}(\bn+1) \right\},
			\end{split}
		\end{align*}
		we have $\ABT{\bx}{\by} = \ABT{\bxs}{\by}.$
	\end{lemma}
	
	\begin{proof}
		The proof follows the same approach as that of Lemma \ref{supp:lemma:7}, differing only in notation. Thus, it is omitted here.
	\end{proof}
	
	\begin{proposition} \label{supp:prop:5}
		Suppose $\bx = \{\xn{1}, \ldots, \xn{\n}\}$ and $\by = \{\yn{1}, \ldots, \yn{\m}\}$ are samples of distinct real values. Let $\bn = \n + \m$ and assume $\bn$ is even. 
		Let $\bxp$ be a non-empty subset of $\bx$ and denote $|\bxp| = \np$. 
		Consider a set $\bxh = \{\xsn{1},\ldots,\xsn{\n -\np}\}$ of distinct real values,
		and denote $\bxs = \bxh \cup \bxp$. Assume all values in $\bxs \cup \by$
		are distinct. Denote $\mysetr = \left\{\rank{\xsn{1}}{\bxs \cup \by} ,\ldots, \rank{\xsn{\n - \np}}{\bxs\cup \by}\right\}$.
		Subsequently, if $\n - \np$ is even, we have
		$\ABT{\bxs}{\by} \le \ABT{\bx}{\by}$, provided
		\begin{align*}
			\mysetr = 	\left\{-\frac{\n - \np - 1}{2} + \frac{1}{2}(\bn + 1), \ldots, \frac{\n - \np - 1}{2} + \frac{1}{2}(\bn+1) \right\}.
		\end{align*}
		Similarly, if $\n - \np$ is odd, we have
		$\ABT{\bxs}{\by} \le \ABT{\bx}{\by}$, provided that either
		\begin{align*}
			&\mysetr = \left\{-\frac{\n - \np}{2} + \frac{1}{2}(\bn + 1), \ldots, \frac{\n - \np}{2} - 1 + \frac{1}{2}(\bn+1) \right\},\\
			\text{or }&\mysetr = \left\{-\frac{\n - \np}{2} + 1 + \frac{1}{2}(\bn + 1), \ldots, \frac{\n - \np}{2} + \frac{1}{2}(\bn+1) \right\}.
		\end{align*}	 
	\end{proposition}
	\begin{proof}
		The proof follows the same approach as that of Proposition \ref{supp:prop:4}, differing only in notation. Thus, it is omitted here.
	\end{proof}

	Proposition~\ref{supp:prop:4} and 
	Proposition~\ref{supp:prop:5} provides
	the ranks of the imputations for
	taking the minimum value of the
	Ansari-Bradley statistic. Below we consider
	the minimum value of the Ansari-Bradley
	after the missing data are imputed 
	according to Proposition~\ref{supp:prop:4} and 
	Proposition~\ref{supp:prop:5}.
	
	We first make the following definition,
	which classifies all possible configurations
	of the total sample sizes $\bn = \n + \m$
	and the number of missing values 
	based on the parity of $\bn$ and $\n - \np$.
	
	\begin{definition} \label{supp:def:AB:case}
		Let $\bn \ge \n \ge \np$ be positive integers. We define the 
		following four cases 
		$\bmcn{1}, \bmcn{2}, \bmcn{3}$ and $\bmcn{4}$ 
		according to the parity of $\bn$ and $\n - \np$:
		\begin{align*}
			&\bmcn{1}:~\text{$\bn$ is odd but $\n - \np$ is even},\\
			&\bmcn{2}:~\text{$\bn$ is odd and $\n - \np$ is odd},\\
			&\bmcn{3}:~\text{$\bn$ is even but $\n - \np$ is odd},\\
			&\bmcn{4}:~\text{$\bn$ is even and $\n - \np$ is even}.
		\end{align*}
	\end{definition}
	
	Then, we define the following four sets, which are
	useful for Proposition~\ref{supp:prop:6}.
	\begin{definition} \label{supp:def:AB:sets}
		Let $\bn \ge \n \ge \np$ be positive integers,
		we denote $\bmsn{1}, \bmsn{2}, \bmsn{3}, \bmsn{4}$ as
		\begin{align*}
			&\bmsn{1} = \left\{-\frac{\n - \np}{2} + \frac{1}{2}(\bn + 1), \ldots, \frac{\n - \np}{2} -1  + \frac{1}{2}(\bn+1) \right\},\\
			&\bmsn{2} = \left\{-\frac{\n - \np - 1}{2} + \frac{1}{2}(\bn + 1), \ldots, \frac{\n - \np - 1}{2} + \frac{1}{2}(\bn+1) \right\},\\
			&\bmsn{3} = \left\{-\frac{\n - \np}{2} + \frac{1}{2}(\bn + 1), \ldots, \frac{\n - \np}{2} -1 + \frac{1}{2}(\bn+1) \right\},\\
			&\bmsn{4} = \left\{-\frac{\n - \np - 1}{2} + \frac{1}{2}(\bn + 1), \ldots, \frac{\n - \np - 1}{2} + \frac{1}{2}(\bn+1) \right\},
		\end{align*}
		respectively.
	\end{definition}
	
	Using Definition~\ref{supp:def:AB:case} and 
	Definition~\ref{supp:def:AB:sets},
	the following proposition gives the explicit form of the Ansari–Bradley statistic after imputing the missing values in $\bx$ according to Propositions~\ref{supp:prop:4} and \ref{supp:prop:5}, across all cases defined in Definition~\ref{supp:def:AB:case}.
	
	\begin{proposition} \label{supp:prop:6}
		Suppose $\bx = \{\xn{1}, \ldots, \xn{\n}\}$ and $\by = \{\yn{1}, \ldots, \yn{\m}\}$ 
		are samples of distinct real values and $\bxp \subset \bx$ is a non-empty subset of 
		$\bx$ with sample size $|\bxp| = \np$. Denote $\bn = \n + \m$.
		Consider a set $\bxh = \{\xsn{1}, \ldots, \xsn{\n - \np}\}$ of distinct real values,
		and denote $\bxs = \bxh \cup \bxp$. Assume that all values in $\bxs \cup \by$ are distinct and
		denote $\mysetr = \{\rank{\xsn{1}}{\bxs \cup \by}, \ldots, \rank{\xsn{\n - \np}}{\bxs \cup \by}\}$.
		Subsequently, we have
		\begin{align*}
			\ABT{\bxs}{\by} 
			= \left\{ \begin{array}{lcl}
				\ABT{\bxp}{\by} + {(\n^2 - \np^2)}/{4} &~\text{if}
				&\bmcn{1} \text{ holds, and } \mysetr = \bmsn{1}, \\ 
				\ABT{\bxp}{\by} + {(\n^2 - \np^2 - 1)}/{4} &~\text{if}
				&\bmcn{2} \text{ holds, and } \mysetr = \bmsn{2},\\ 
				\ABT{\bxp}{\by} + {(\n^2 - \np^2 + 1)}/{4} &~\text{if}
				&\bmcn{3} \text{ holds, and } \mysetr = \bmsn{3},\\ 
				\ABT{\bxp}{\by} + {(\n^2 - \np^2)}/{4} &~\text{if}
				&\bmcn{4} \text{ holds, and } \mysetr = \bmsn{4},
			\end{array}\right.
		\end{align*}
	\end{proposition}
	
	\begin{proof}
		To start, let us denote $\bxp = \{\xpn{1}, \ldots, \xpn{\np}\}$.
		Below, we only consider the first case when
		$\bmcn{1} \text{ holds, and } \mysetr = \bmsn{1}$,
		because the other three cases can be proved following the same approach.
		
		By the definition of the Ansari-Bradley test statistic, we have
		\begin{align*}
			\ABT{\bxs}{\by} = \sum_{\iconstant=1}^{\n -\np} \left|\rank{\xsn{\iconstant}}{\bxs \cup \by} - \frac{\bn + 1}{2}\right|
			+ \sum_{\iconstant=1}^{\np} \left|\rank{\xpn{\iconstant}}{\bxs \cup \by} - \frac{\bn + 1}{2}\right|.
		\end{align*}
		Since $\mysetr = \bmsn{1}$,  
		\begin{align*}
			\sum_{\iconstant=1}^{\n -\np} \left|\rank{\xsn{\iconstant}}{\bxs \cup \by} - \frac{\bn + 1}{2}\right|= \left|-\frac{\n -\np}{2}\right| + \ldots + \left|\frac{\n - \np}{2} - 1\right|
			= \frac{(\n - \np)^2}{4}.
		\end{align*}
		%	\begin{align}  \label{supp:prop:6:eqn:2}
			%		\begin{split}
				%			& \sum_{\iconstant=1}^{\n -\np} \left|\rank{\xsn{\iconstant}}{\bxs \cup \by} - \frac{\bn + 1}{2}\right|= \left|-\frac{\n -\np}{2}\right| + \ldots + \left|\frac{\n - \np}{2} - 1\right| \\
				%			& = \frac{\n -\np}{2} + \ldots + 0 + \ldots + \frac{\n - \np}{2} - 1\\
				%			& = \frac{\n -\np}{2} + \ldots + 0 + \ldots + \frac{\n - \np}{2} - \frac{\n - \np}{2}\\
				%			& = \left(1 + \frac{\n - \np}{2} \right)\left(\frac{\n -\np}{2}\right) - \frac{\n - \np}{2}\\
				%			& = \frac{(\n - \np)^2}{4}.
				%		\end{split}
			%	\end{align}
		Hence, in order to prove our result under this case, it is sufficient
		to show that 
		\begin{align*}
			\sum_{\iconstant=1}^{\np} \left|\rank{\xpn{\iconstant}}{\bxs \cup \by} - \frac{\bn + 1}{2}\right| &= \ABT{\bxp}{\by} +  \frac{\n^2 - \np^2}{4} -  \frac{(\n - \np)^2}{4}\\
			&=  \ABT{\bxp}{\by} + \frac{(\n - \np)\np}{2}.
		\end{align*}
		In order to prove our result under this case, we first show the two equivalence
		\eqref{supp:prop:6:eqn:3} and \eqref{supp:prop:6:eqn:5} below.
		
		Note that 
		\begin{align*}
			\min \bmsn{1} = -\frac{\n - \np}{2} + \frac{1}{2}(\bn + 1) = \frac{-\n + \np + \n + \m + 1}{2} = \frac{\np + m + 1}{2}.
		\end{align*}
		Then, according to Lemma~\ref{supp:lemma:5}, for any $\iconstant \in \{1,\ldots, \np\}$, we have
		\begin{align*}
			\rank{\xpn{\iconstant}}{\bxp \cup \by} < \frac{\np + \m + 1}{2} \iff \rank{\xpn{\iconstant}}{\bxs \cup \by} = \rank{\xpn{\iconstant}}{\bxp \cup \by}.
		\end{align*}	
		Further, for any $\iconstant \in \{1,\ldots, \np\}$, we have
		\begin{align}  \label{supp:prop:6:eqn:3}
			\rank{\xpn{\iconstant}}{\bxp \cup \by} < \frac{\np + \m + 1}{2} 
			\iff \rank{\xpn{\iconstant}}{\bxs \cup \by} < \frac{\bn + 1}{2}.
		\end{align}	
		The $``\implies"$ follows from the fact that 
		$\rank{\xpn{\iconstant}}{\bxs \cup \by} = \rank{\xpn{\iconstant}}{\bxp \cup \by}$
		and $\n' + \m < \bn$. 
		For the ``$\impliedby$'', since values of $\bxs$ take all the ranks in $\bmsn{1}$,
		\begin{align*}
			&\rank{\xpn{\iconstant}}{\bxs \cup \by} < (\bn + 1)/2  \\
			\implies & \rank{\xpn{\iconstant}}{\bxs  \cup \by} < \min \bmsn{1} = -\frac{\n - \np}{2} + \frac{1}{2}(\bn + 1) = \frac{\np + m + 1}{2}.
		\end{align*}
		
		Notice that $\mysetr = \bmsn{1}$
		consists of consecutive integers, and
		$ \min \bmsn{1} = {(\np + m + 1)}/2$ as we have shown before.
		Then according to Lemma~\ref{supp:ab:lemma:6}, for any $\iconstant \in \{1,\ldots, \np\}$,
		\begin{align*}
			\rank{\xpn{\iconstant}}{\bxp \cup \by} \ge \frac{\np + \m + 1}{2}
			\iff \rank{\xpn{\iconstant}}{\bxs \cup \by} = \rank{\xpn{\iconstant}}{\bxp \cup \by} + \n - \np.
		\end{align*}
		Further, for any $\iconstant \in \{1,\ldots, \np\}$, we have
		\begin{align} \label{supp:prop:6:eqn:5}
			\rank{\xpn{\iconstant}}{\bxp \cup \by} \ge \frac{\np + \m + 1}{2}
			\iff  \rank{\xpn{\iconstant}}{\bxs \cup \by} \ge \frac{\bn + 1}{2}.
		\end{align}
		The ``$\implies$" follows from the fact that 
		$ \rank{\xpn{\iconstant}}{\bxs \cup \by} = \rank{\xpn{\iconstant}}{\bxp \cup \by} + \n - \np$
		and $ {(\np + \m + 1)}/{2} + \n - \np \ge  (\bn+1)/2$.
		For the ``$\impliedby$'', since values of $\bxs$ take all the ranks in $\bmsn{1}$,
		we have
		\begin{align*}
			&\rank{\xpn{\iconstant}}{\bxs \cup \by} \ge \frac{\bn + 1}{2}\\
			\implies & \rank{\xpn{\iconstant}}{\bxs \cup \by} > \max \bmsn{1} = \frac{\n - \np}{2} -1 + \frac{1}{2}(\bn+1).
		\end{align*}
		Since $\rank{\xpn{\iconstant}}{\bxs \cup \by} = \rank{\xpn{\iconstant}}{\bxp \cup \by} + \sum_{\iconstant = 1}^{\n -\np} \indicator{\xpn{\iconstant} \ge \xsn{\iconstant}}$,
		\begin{align*}
			&\rank{\xpn{\iconstant}}{\bxp \cup \by} =\rank{\xpn{\iconstant}}{\bxs \cup \by} -   \sum_{\iconstant = 1}^{\n -\np} \indicator{\xpn{\iconstant} \ge \xsn{\iconstant}}\\
			\implies & \rank{\xpn{\iconstant}}{\bxp \cup \by} > \frac{\n - \np}{2} -1 + \frac{1}{2}(\bn+1) - (\n -\np)\\
			\implies & \rank{\xpn{\iconstant}}{\bxp \cup \by} \ge \frac{\np + \m  + 1}{2}.
		\end{align*}
		
		After proving
		\eqref{supp:prop:6:eqn:3} and $\eqref{supp:prop:6:eqn:5}$,
		we first decompose $\sum_{\iconstant=1}^{\np} \left|\rank{\xpn{\iconstant}}{\bxs \cup \by} - (\bn + 1)/2\right|$ as such  
		\begin{align*}
			\begin{split}
				\sum_{\iconstant=1}^{\np} \left|\rank{\xpn{\iconstant}}{\bxs \cup \by} - \frac{\bn + 1}{2}\right|
				& = \sum_{\iconstant = 1}^{\np} \indicator{\rank{\xpn{\iconstant}}{\bxs \cup \by} < \frac{\bn + 1}{2}} \left(\frac{\bn + 1}{2} - \rank{\xpn{\iconstant}}{\bxs \cup \by}  \right)\\
				& + \sum_{\iconstant = 1}^{\np} \indicator{\rank{\xpn{\iconstant}}{\bxs \cup \by} \ge \frac{\bn + 1}{2}} \left(\rank{\xpn{\iconstant}}{\bxs \cup \by}  - \frac{\bn + 1}{2} \right).
			\end{split}
		\end{align*}	
		For notation convenience, we denote $\bnp = (\np + \m + 1)/2$ and
		\begin{align*}
			\rrln{\iconstant} = \rank{\xpn{\iconstant}}{\bxp \cup \by} , \text{ for any } \iconstant \in \{1,\ldots,\np\}.
		\end{align*}
		Then, by applying \eqref{supp:prop:6:eqn:3} and $\eqref{supp:prop:6:eqn:5}$, we further have
		\begin{align*}
			\sum_{\iconstant=1}^{\np} \left|\rank{\xpn{\iconstant}}{\bxs \cup \by} - \frac{\bn + 1}{2}\right|
			&= \sum_{\iconstant = 1}^{\np} \indicator{\rrln{\iconstant} < \bnp } \left(\frac{\bn + 1}{2} - \rrln{\iconstant}  \right)\\
			& +  \sum_{\iconstant = 1}^{\np} \indicator{\rrln{\iconstant} \ge \bnp } \left(\rrln{\iconstant} + \n - \np  - \frac{\bn + 1}{2} \right).
		\end{align*}	
		Notice that 
		\begin{align*}
			&\frac{\bn + 1}{2} = \frac{\n + \m + 1}{2} = \frac{\np + \m + 1}{2} + \frac{\n - \np}{2} = \bnp + \frac{n-\np}{2},\\
			\text{and }& \n - \np  - \frac{\bn + 1}{2} = \frac{2\n -2\np -\n -\m - 1}{2} = -\frac{ \np +\m + 1}{2} + \frac{\n - \np}{2} = -\np + \frac{n-\np}{2}. 
		\end{align*}
		Hence, we have
		\begin{align*}
			&\sum_{\iconstant=1}^{\np} \left|\rank{\xpn{\iconstant}}{\bxs  \cup \by} - \frac{\bn + 1}{2}\right|\\
			&= \sum_{\iconstant = 1}^{\np} \indicator{\rrln{\iconstant} < \bnp} \left(\bnp - \rrln{\iconstant} + \frac{\n -\np}{2} \right) + \sum_{\iconstant = 1}^{\np} \indicator{\rrln{\iconstant} \ge \bnp} \left(\rrln{\iconstant}  - \bnp + \frac{\n - \np}{2}\right)\\
			& = \sum_{\iconstant = 1}^{\np} \indicator{\rrln{\iconstant} <\bnp} \left(\bnp - \rrln{\iconstant}  \right) + \sum_{\iconstant = 1}^{\np} \indicator{\rrln{\iconstant} \ge \bnp} \left(\rrln{\iconstant}  - \bnp \right) + \frac{(\n - \np)\np}{2}\\
			& =  \sum_{\iconstant = 1}^{\np} \left|\rrln{\iconstant} - \bnp \right| + \frac{(\n - \np)\np}{2}
			=  \ABT{\bxp}{\by} + \frac{(\n - \np)\np}{2},
		\end{align*}
		which completes the proof of our result under Case (1).
	\end{proof}
	
	Finally, we combine the results of 
	Propositions~\ref{supp:prop:4}, \ref{supp:prop:5}
	and \ref{supp:prop:6}, and concludes
	this section. The following theorem  provides
	the lower bound of the Ansari-Bradley statistic
	in the presence of missing data when only 
	values in $\bx$ can  be missing.
	
	\begin{theorem} \label{supp:theorem:1}
		Suppose $\bx = \{\xn{1}, \ldots, \xn{\n}\}$ and $\by = \{\yn{1}, \ldots, \yn{\m}\}$ are 
		samples of distinct real values, and $\bxp \subset \bx$ is a subset of $\bx$ 
		with sample size $|\bx| = \np$, which are observed. Then, the minimum possible Ansari-Bradley test statistic, across all possible values of
		missing data, is given as follows:
		\begin{align*}
			\min_{\bx \setminus \bxp \in \mathbb{R}^{\n - \np}}\ABT{\bx}{\by} = \ABT{\bxs}{\by} 
			= \left\{ \begin{array}{lcl}
				\ABT{\bxp}{\by} + {(\n^2 - \np^2)}/{4} &~\text{if}
				&\bmcn{1} \text{ or } \bmcn{4} \text{ holds}, \\ 
				\ABT{\bxp}{\by} + {(\n^2 - \np^2 - 1)}/{4} &~\text{if}
				&\bmcn{2} \text{ holds},\\ 
				\ABT{\bxp}{\by} + {(\n^2 - \np^2 + 1)}/{4} &~\text{if}
				&\bmcn{3} \text{ holds}.
			\end{array}\right.
		\end{align*}
	\end{theorem}
	
	\begin{proof}
		Denote $\bn = \n + \m$. Then, the parity of $\bn$ 
		and $\n - \np$ satisfy either $\bmcn{1}$, $\bmcn{2}$,
		$\bmcn{3}$ and $\bmcn{4}$.
		Let us assume that $\bmcn{\kk}$ holds, where $\kk \in \{1,2,3,4\}$.
		Let $\bxh = \{\xsn{1}, \ldots, \xsn{\n - \np}\}$ be a set of
		distinct real values, and denote that $\bxs = \bxh \cup \bxp$.
		Assume that all values in $\bxs \cup \by$ are distinct and
		\begin{align*}
			\{\rank{\xsn{1}}{\bxs \cup \by}, \ldots, \rank{\xsn{\n - \np}}{\bxs \cup \by}\} = \bmsn{\kk}.
		\end{align*}
		Then, according to Proposition \ref{supp:prop:4} and \ref{supp:prop:5}, we have
		\begin{align*}
			\ABT{\bxs}{\by} \le \ABT{\bx}{\by},
		\end{align*}
		for any possible values of $\bx \setminus \bxp$. Hence, it follows that
		\begin{align*}
			\ABT{\bxs}{\by} \le \min_{\bx \setminus \bxp \in \mathbb{R}^{\n - \np}}\ABT{\bx}{\by}.
		\end{align*}
		Be definition, we have
		\begin{align*}
			\min_{\bx \setminus \bxp \in \mathbb{R}^{\n - \np}}\ABT{\bx}{\by}  \le \ABT{\bxs}{\by}.
		\end{align*}
		Hence,
		\begin{align*}
			\min_{\bx \setminus \bxp \in \mathbb{R}^{\n - \np}}\ABT{\bx}{\by} = \ABT{\bxs}{\by},
		\end{align*}
		where
		\begin{align*}
			\ABT{\bxs}{\by} 
			= \left\{ \begin{array}{lcl}
				\ABT{\bxp}{\by} + {(\n^2 - \np^2)}/{4} &~\text{if}
				&\bmcn{1} \text{ or } \bmcn{4} \text{ holds}, \\ 
				\ABT{\bxp}{\by} + {(\n^2 - \np^2 - 1)}/{4} &~\text{if}
				&\bmcn{2} \text{ holds},\\ 
				\ABT{\bxp}{\by} + {(\n^2 - \np^2 + 1)}/{4} &~\text{if}
				&\bmcn{3} \text{ holds}.
			\end{array}\right.
		\end{align*}
		according to Proposition \ref{supp:prop:6}. Hence, we conclude our result.
	\end{proof}

	\section{Lower bound with multiple missing values in $\by$} \label{appendix:lowebound:ab:y}
	
	The previous section considers the tight lower bound
	of the Ansari-Bradley test statistic when multiple values
	can be missing in $\bx$. 
	This section considers similar problem, but when 
	multiple values in $\by$ can be missing, while all
	values in $\bx$ are observed.
	The main results in this section is Theorem~\ref{supp:theorem:2}.
	
	We first show that the sum of the Ansari-Bradley test statistics in both directions $\ABT{\bx}{\by} + \ABT{\by}{\bx}$ is a constant depending only on the total sample sizes $\bn = \n + \m$.
	
	\begin{lemma} \label{supp:ab:lemma:1}
		Suppose $\bx = \{\xn{1}, \ldots, \xn{\n}\}$ and $\by = \{\yn{1}, \ldots, \yn{\m}\}$ are samples of distinct real values. Denote $\bn = \n + \m$.
		Then, the sum of the Ansari-Bradley test statistics $\ABT{\bx}{\by}$ and $\ABT{\by}{\bx}$ 
		is a constant of sample size $\bn$. More specifically, 
		\begin{align*}
			\ABT{\bx}{\by} + \ABT{\by}{\bx} = \sum_{\iconstant=1}^{\bn} \left|\iconstant - \frac{\bn+1}{2}\right| =  \left\{ \begin{array}{ll}
				{\bn^2}/{4} 
				& \bn~\text{is even,} \\ 
				{(\bn^2 - 1)}/{4}
				& \bn~\text{is odd.} 
			\end{array}\right.
		\end{align*}
	\end{lemma}
	
	\begin{proof}
		By definition of the Ansari-Bradley test statistic, we have 
		\begin{align*}
			&\ABT{\bx}{\by} = \sum_{\iconstant=1}^{\n} \left|\rank{\xn{\iconstant}}{\bx\cup\by} - \frac{\bn+1}{2}\right|,\\
			\text{and } &\ABT{\by}{\bx} = \sum_{\iconstant=1}^{\m} \left|\rank{\yn{\iconstant}}{\bx\cup\by} - \frac{\bn+1}{2}\right|.
		\end{align*}
		Since $\bx$ and $\by$ are samples of distinct values, each value in
		$\bx$ and $\by$ take a distinct rank in $\{1, \ldots, \bn\}$. Hence,
		we have our desired result
		\begin{align*}
			\ABT{\bx}{\by} + \ABT{\by}{\bx} = \sum_{\iconstant=1}^{\bn} \left|\iconstant - \frac{\bn+1}{2}\right|.
		\end{align*}
		When $\bn$ is odd,
		\begin{align*}
			&\sum_{\iconstant = 1}^{\bn} \left|\iconstant - \frac{1}{2}(\bn+1)\right|\\
			&= \left(\frac{1}{2}(\bn + 1) - 1\right) + \ldots + \left(\frac{1}{2}(\bn + 1) - \frac{1}{2}(\bn + 1) \right)  + \ldots + \left(\bn - \frac{1}{2}(\bn + 1)\right) \\
			&=\left(\frac{1}{2}(\bn - 1)\right) + \ldots + 0 + \ldots + \left( \frac{1}{2}(\bn - 1)\right)\\
			& = \left(1 + \frac{1}{2}(\bn - 1) \right) \frac{1}{2}(\bn -1)\\
			& = \frac{\bn^2 - 1}{4}.
		\end{align*}
		However, when $\bn$ is even,
		\begin{align*}
			& \sum_{\iconstant = 1}^{\bn} \left|\iconstant - \frac{1}{2}(\bn+1)\right| \\
			&= \left(\frac{1}{2}(\bn + 1) - 1\right) + \ldots + \left(\frac{1}{2}(\bn+1) - \frac{1}{2}\bn \right)  + \frac{1}{2} + \ldots + \frac{1}{2}(\bn - 1) \\
			& = \left(\frac{1}{2} + \frac{1}{2}(\bn - 1) \right) \frac{1}{2}\bn\\
			& = \frac{\bn^2}{4}.
		\end{align*}
		Hence, we complete our proof.
	\end{proof}
	
	This lemma shows that in order to minimize
	the Ansari-Bradley statistic $\ABT{\bx}{\by}$
	when values in $\by$ can be missing, we only
	need to maximize $\ABT{\by}{\bx}$ since 
	$\ABT{\bx}{\by} + \ABT{\by}{\bx}$ is a constant.
	
	We approach this problem by first considering
	the case when only one value in $\by$ can be missing.
	
	\begin{lemma} \label{supp:lemma:9}
		Suppose $\bx = \{\xn{1}, \ldots, \xn{\n}\}$ and $\by = \{\yn{1}, \ldots, \yn{\m}\}$ are samples of distinct real values. Denote $\bn =\n + \m$.  Suppose $\y \in \by$ is a value in $\by$, and denote 
		$\byp = \by \setminus \{\y\}$. Consider two real values $\ys$, $\yls$, and denote $\bys = \{\ys\}\cup \byp$, and $\byls = \{\yls\} \cup \byp$. 
		Then, if 
		\begin{align*}
			\yls < \min \bx \cup \byp,~\text{and}~\ys > \max \bx \cup \byp,
		\end{align*}
		we have $\ABT{\by}{\bx} \le \max \{\ABT{\by}{\bxls}, \ABT{\by}{\bxs} \}$.
	\end{lemma}	
	
	\begin{proof}
		This result follows from Proposition~\ref{supp:ab:prop:2}.
		Suppose $\y \ge (N+1)/2$. Then, according to 
		Proposition~\ref{supp:ab:prop:2}, we have
		$\ABT{\by}{\bx} \le \ABT{\bys}{\bx}$.
		However, if $\y < (N+1)/2$, then according to 
		Proposition~\ref{supp:ab:prop:2}, we have
		$\ABT{\by}{\bx} \le \ABT{\byls}{\bx}$. Overall, we have
		\begin{align*}
			\ABT{\by}{\bx} \le \max \{\ABT{\by}{\bxls}, \ABT{\by}{\bxs} \}
		\end{align*}
		as required, which completes our proof.
	\end{proof}
	
	Then, we generalize the result in Lemma~\ref{supp:lemma:9}
	into the cases where multiple values in $\by$ can be missing.
	We first make the following definition:
	
	\begin{definition} \label{def:ab:1}
		Suppose $\bx$ is a sample of distinct real values. 
		We define $\mysetx(\bx, \tconstant)$ to be a collection of 
		sets of	$\tconstant$ distinct values such that if $\bxh \in \mysetx(\bx,\tconstant)$,
		then all values in $\bxh$ are either
		smaller than $\min \bx$ or larger than
		$\max \bx$. That is,
		\begin{align*}
			\mysetx(\bx,\tconstant) = \{\{\xsn{1}, \ldots, \xsn{\tconstant}\}:~\xsn{\iconstant} \notin [\min \bx, \max \bx],
			\text{ and } \xsn{\iconstant} \neq \xsn{\jconstant} \text{ for } \iconstant \neq \jconstant \}.
		\end{align*} 
	\end{definition}
	
	Applying Definition~\ref{def:ab:1}, we generalize 
	Lemma~\ref{supp:lemma:9} into the following result:
	
	\begin{lemma} \label{supp:lemma:10}
		Suppose $\bx = \{\xn{1}, \ldots, \xn{\n}\}$ and $\by = \{\yn{1}, \ldots, \yn{\m}\}$ are samples of distinct real values. Let $\byp \subset \by$ be a non-empty subset of $\by$ with sample size $|\byp| = \mmp$. Then there exist $\byh \in \mysetx(\bx\cup\byp, \m -\mmp)$ such that $\ABT{\byh \cup \byp}{\bx} \ge \ABT{\by}{\bx}$.	 
	\end{lemma}
	
	\begin{proof}		
		We prove this result by mathematical induction on $\m - \mmp$.
		The base case when $\m - \mmp = 1$ is proved in Lemma~\ref{supp:lemma:9}
		since $\{\ys\}, \{\yls\} \in \mysetx(\bx \cup \byp,1)$.
		For the induction step, let $\kk \in \{1, \ldots, \m - 1\}$, and assume the lemma 
		is correct when $\m - \mmp = \kk$. We show that it still holds when $\m - \mmp = \kk+1$.
		
		Denote $\by \setminus \byp = \byt = \{\ytn{1}, \ldots, \ytn{\kk+1}\}$. 
		By assumption that Lemma~\ref{supp:lemma:10} is correct when $\m - \mmp = \kk$,
		there exist $\bys \in \mysetx(\bx \cup \{\ytn{1}\} \cup \byp, \kk)$
		such that 
		\begin{align*}
			\ABT{\bys \cup \{\ytn{1}\} \cup \byp }{\bx} \ge \ABT{\by}{\bx}.
		\end{align*}

		Further, consider two values $\ys$, $\yls$
		such that 
		\begin{align*}
			\yls < \min \bx \cup \bys \cup \byp, \text{ and } 	\ys > \max \bx \cup \bys \cup \byp.
		\end{align*}
		Then according to Lemma~\ref{supp:lemma:9}, we have
		\begin{align*}
			&\max \{\ABT{\{\yls\} \cup \bys \cup \byp}{\bx}, \ABT{\{\ys\} \cup \bys \cup \byp}{\bx} \} \ge \ABT{\bys \cup \{\ytn{1}\} \cup \byp}{\bx}\\
			\implies &	\max  \{\ABT{\{\yls\} \cup \bys \cup \byp}{\bx}, \ABT{\{\ys\} \cup \bys \cup \byp}{\bx} \}  \ge \ABT{\by}{\bx}.
		\end{align*}
		Since $\bys \in \mysetx(\bx \cup \{\ytn{1}\} \cup \byp, \kk)$, and
		we have that $\yls < \min \bys \cup \byp \cup \bx$ 
		and $\ys > \max \bys \cup \byp \cup \bx$. Hence we have
		$\{\ys\} \cup \bys \in \mysetx(\byp \cup \bx, \kk+1)$,
		and $\{\yls\} \cup \bys \in \mysetx(\byp \cup \bx, \kk+1)$,
		which completes our proof.
	\end{proof}
	
	The following Definition~\ref{def:ab:2} 
	further defines subsets of 
	the set $\mysetx(\bx,\tconstant)$ in Definition~\ref{def:ab:1} 
	subjecting to a integer $\kk$.
	
	\begin{definition} \label{def:ab:2}
		Suppose $\bx$ is a sample of distinct real values,
		and $\bxh \in \mysetx(\bx,\tconstant)$, where
		$\tconstant$ is a positive integer. We define
		$\mysety(\bx, \tconstant, \kk)$ to be a 
		subset of $\mysetx(\bx, \tconstant)$ such that if $\bxh \in \mysety(\bx, \tconstant, \kk)$,
		we have $\bxh \in \mysetx(\bx, \tconstant)$, and $\sum_{\x \in \bxh} \indicator{ \x > \max \bx} = \kk$. In other words,
		\begin{align*}
			\mysety(\bx, \tconstant, \kk) = \mysetx(\bx,\tconstant) \cap \left\{\{\xsn{1}, \ldots, \xsn{\tconstant}\}:~\sum_{\iconstant = 1}^\tconstant \indicator{ \xsn{\iconstant} > \max \bx} = \kk \right\}.
		\end{align*}	
	\end{definition}
	
	The following Definition~\ref{supp:def:func} 
	defines a function which is important for stating
	the results in Proposition~\ref{supp:ab:prop:7}.
	
	\begin{definition} \label{supp:def:func}
		Suppose $\bx = \{\xn{1}, \ldots, \xn{\n}\}$ and $\byp = \{\yn{1}, \ldots, \yn{\mmp}\}$ are samples of distinct real values, and $\m \in \mathbb{\bn}$ is a positive integer such that $\m \ge \mmp$. Denote $\bzp = \bx \cup \byp$, $\bnp = (\n + \mmp + 1)/2$, and $\bmp = (\m-\mmp)/2$. For any 
		given $\kk  \in \{0, \ldots, \m - \mmp\}$, denote $\sa = \min\left\{0, \kk - \bmp \right\}$, $\sbb = \max \left\{0, \kk - \bmp \right\}$.
		For any $\iconstant \in \{1,\ldots, \n\}$, let $A_{\iconstant} = \indicator{\rank{\xn{\iconstant}}{ \bzp} < \bnp + \sa},
		B_{\iconstant} =  \indicator{\bnp + \sa \le \rank{\xn{\iconstant}}{ \bzp} \le \bnp + \sbb}$,	and $C_{\iconstant} = \indicator{\rank{\xn{\iconstant}}{\bzp} >  \bnp + \sbb }$.
		Then, we define
		\begin{align*}
			\func{\bx, \byp, \m}{\kk} = (\kk - \bmp) \sum_{\iconstant = 1}^{\n} (A_{\iconstant} - C_{\iconstant} ) + \sgn{\kk - \bmp} \sum_{\iconstant = 1}^{\n} B_{\iconstant} \left(2\bnp + \kk - \bmp - 2\rank{\xn{\iconstant}}{\bzp} \right),
		\end{align*}
		where $\sgn{\x} = 1$ if $\x \ge 0$, otherwise $\sgn{\x} = -1$.
	\end{definition}
	
	The following proposition provides explicit form
	of the Ansari-Bradley statistic when the missing
	values in $\by$ are imputed with $\byh \in \mysety(\bx\cup\byp, \m-\mmp, \kk)$.
	
	\begin{proposition} \label{supp:ab:prop:7}
		Suppose $\bx = \{\xn{1}, \ldots, \xn{\n}\}$ and $\by = \{\yn{1}, \ldots, \yn{\m}\}$ are samples of distinct real values and $\byp \subset \by$ is a subset of $\by$ with 
		sample size $\card{\byp} = \mmp$. 
		Then, if $\byh \in \mysety(\bx\cup\byp, \m-\mmp, \kk)$,
		we have $\ABT{\bx}{\bys} = \ABT{\bx}{\byp} + \func{\bx, \byp, \m}{\kk}$,
		where $\bys = \byh \cup \byp$, and $\func{\bx, \byp, \m}{\kk}$ is defined in Definition \ref{supp:def:func}. 
	\end{proposition}

	\begin{proof}
		Since $\byh \in \mysety(\bx\cup\bxp, \m-\mmp, \kk)$, according to the 
		definition of $\byh$ in Definition~\ref{def:ab:2}, we have
		$\ysn{\iconstant} < \min \bx \cup \byp~\text{or}~\ysn{\iconstant} > \max \bx \cup \byp, \text{ for any } \iconstant \in \{1,\ldots, \m - \mmp\}$. Notice that 
		we also have $\kk = \sum_{\iconstant=1} ^{\m - \mmp}\indicator{ \ysn{\iconstant} > \max \bx \cup \byp}$. In other words, the ranks of values of $\byh$
		take $\kk$ maximum, and $\m - \mmp - \kk$ minimum distinct values 
		in $\{1,\ldots,\bn\}$, i.e.,
		\begin{align*}
			\left\{\rank{\ysn{1}}{\bx \cup \bys}, \ldots, \rank{\ysn{\m -\mmp}}{\bx \cup \bys} \right\}
			= \left\{1,\ldots, \m - \mmp - \kk \right\} \cup \left\{\bn-\kk + 1,\ldots,\bn\right\}.
		\end{align*}
		Let us divide $\byh$ into $\byhln{1}$ and $\byhln{2}$ such that values
		in $\byhln{1}$ take the minimum ranks, while values in 
		$\byhln{2}$ take the maximum ranks. That is, 
		\begin{align*}
			&\byhln{1} =  \left\{ \y \in \byh|	
			1 \le \rank{\y}{\bx \cup \bys} \le \m - \mmp - \kk \right\} ,\\
			\text{and }&\byhln{2} = \left\{ \y \in \byh|	
			\bn -\kk + 1 \le \rank{\y}{\bx \cup \bys} \le \bn \right\} .
		\end{align*}
		
		According to the definition of the Ansari-Bradley test statistic,
		\begin{align*}
			\ABT{\bx}{\bys} &= \sum_{\iconstant = 1}^{\n} \left|\rank{\xn{\iconstant}}{\bx \cup \bys} - \frac{1}{2}(\bn + 1)\right|\\
			& = \sum_{\iconstant = 1}^{\n} \left|\rank{\xn{\iconstant}}{\bx \cup \byp} + \sum_{\y \in \byh} \indicator{\xn{\iconstant} > \y} - \frac{1}{2}(\bn + 1)\right|\\
			& = \sum_{\iconstant = 1}^{\n} \left|\rank{\xn{\iconstant}}{\bx \cup \byp} + \sum_{\y \in \byhln{1}} \indicator{\xn{\iconstant} > \y} + \sum_{\y \in \byhln{2}} \indicator{\xn{\iconstant} > \y}- \frac{1}{2}(\bn + 1)\right|.
		\end{align*}
		By definitions of $\byhln{1}$ and $\byhln{2}$, we have
		\begin{align*}
			\sum_{\y \in \byhln{1}} \indicator{\xn{\iconstant} > \y} = \m - \mmp - \kk, \text{ and }\sum_{\y \in \byhln{2}} \indicator{\xn{\iconstant} > \y} = 0.
		\end{align*}
		Hence,
		\begin{align*}
			\ABT{\bx}{\bys}  &= \sum_{\iconstant = 1}^{\n} \left|\rank{\xn{\iconstant}}{\bx \cup \byp} + \m - \mmp - \kk - \frac{1}{2}(\bn + 1)\right|.
		\end{align*}
		Notice that
		\begin{align*}
			\frac{1}{2}(\bn + 1) = \frac{1}{2} (\n + \m - \mmp + \mmp + 1) = \frac{1}{2}(\n + \mmp + 1) + \frac{1}{2}(\m - \mmp).
		\end{align*}
		Hence, we further have
		\begin{align*}
			\ABT{\bx}{\bys} &= \sum_{\iconstant = 1}^{\n} \left|\rank{\xn{\iconstant}}{\bx \cup \byp} + \m - \mmp - \kk -  \frac{1}{2}(\n + \mmp + 1) -  \frac{1}{2}(\m - \mmp)\right|\\
			& = \sum_{\iconstant = 1}^{\n} \left|\rank{\xn{\iconstant}}{\bx \cup \byp} -  \frac{1}{2}(\n + \mmp + 1) + \frac{1}{2}(\m - \mmp) - \kk \right|.
		\end{align*}
		For notation ease, let us denote $\bnp = (\n + \mmp + 1)/2$, $\bmp = (\m - \mmp)/2$ and
		\begin{align*}
			\chi_{\iconstant} = \left|\rank{\xn{\iconstant}}{\bx \cup \byp} -  \bnp + \bmp - \kk \right|, \text{ for any } \iconstant \in \{1,\ldots,\n\}.
		\end{align*}
		Denote $\sa = \min\left\{0, \kk - \bmp\right\}$, 
		and $\sbb = \max \left\{0, \kk - \bmp \right\}$.
		For any $\iconstant \in \{1,\ldots,\n\}$, also denote that
		\begin{align*}
			A_{\iconstant} &= \indicator{\rank{\xn{\iconstant}}{\bx \cup \byp} < \bnp + \sa}, \\
			B_{\iconstant} &=  \indicator{\bnp + \sa \le \rank{\xn{\iconstant}}{\bx \cup \byp} \le \bnp + \sbb},\\
			C_{\iconstant} &= \indicator{\rank{\xn{\iconstant}}{\bx \cup \byp} > \bnp + \sbb }.
		\end{align*}
		Note that $A_{\iconstant} + B_{\iconstant} + C_{\iconstant} = 1$. Thus, we have
		\begin{align} \label{supp:prop:7:eqn:1}
			\begin{split}
				\ABT{\bx}{\bys} = \sum_{\iconstant = 1}^{\n} A_{\iconstant} \chi_{\iconstant} + \sum_{\iconstant = 1}^{\n} B_{\iconstant} \chi_{\iconstant} + \sum_{\iconstant = 1}^{\n} C_{\iconstant} \chi_{\iconstant}.
			\end{split}
		\end{align}
		Below we prove our results for the following two cases separately:
		\begin{align*}
			\text{Case } \case{i}: \kk - \bmp \ge 0, ~~
			\text{Case } \case{ii}: \kk - \bmp < 0.
		\end{align*}
		
		Suppose case \case{i} holds. That is, assume that $\kk - \bmp \ge 0$.
		Then, we have $\sa = 0$ and $\sbb = \kk - \bmp$. For the first term 
		on the right hand side of the Equation \eqref{supp:prop:7:eqn:1}, notice that
		\begin{align*}
			A_{\iconstant} = 1 \implies \rank{\xn{\iconstant}}{\bx \cup \byp} < \bnp
			\implies \rank{\xn{\iconstant}}{\bx \cup \byp} < \bnp + \kk - \bmp.
		\end{align*}
		Hence, when $A_{\iconstant} = 1$, we have 
		$\chi_{\iconstant} = \left(\bnp - \bmp + \kk - \rank{\xn{\iconstant}}{\bx \cup \byp} \right).$
		Therefore,
		\begin{align*}
			\sum_{\iconstant = 1}^{\n} A_{\iconstant} \chi_{\iconstant} &= \sum_{\iconstant = 1}^{\n} A_i \left(\bnp - \bmp + \kk - \rank{\xn{\iconstant}}{\bx \cup \byp} \right)\\
			& = \sum_{\iconstant = 1}^{\n} A_{\iconstant} \left(\bnp - \rank{\xn{\iconstant}}{\bx \cup \byp} \right) +  \sum_{\iconstant = 1}^{\n} A_{\iconstant} \left(\kk - \bmp \right).
		\end{align*}
		Similarly, for the second term on the right hand side
		of the Equation \eqref{supp:prop:7:eqn:1}, when $B_{\iconstant} = 1$,
		we have 
		\begin{align*}
			\rank{\xn{\iconstant}}{\bx \cup \byp} \le \bnp + \kk - \bmp
			\implies\chi_{\iconstant} = \left(\bnp -\bmp + \kk - \rank{\xn{\iconstant}}{\bx \cup \byp} \right).
		\end{align*}
		Hence,
		\begin{align*}
			\sum_{\iconstant = 1}^{\n} B_i \chi_{\iconstant} &= \sum_{\iconstant = 1}^{\n} B_{\iconstant} \left(\bnp - \bmp + \kk - \rank{\xn{\iconstant}}{\bx \cup \byp} \right)\\
			& =  \sum_{\iconstant = 1}^{\n} B_{\iconstant} \left(\rank{\xn{\iconstant}}{\bx \cup \byp} - \bnp \right) 
			+  \sum_{\iconstant = 1}^{\n} B_{\iconstant} \left(2\bnp - \bmp + \kk - 2\rank{\xn{\iconstant}}{\bx \cup \byp} \right).
		\end{align*}
		Similarly, for the third term on the right hand side of the Equation \eqref{supp:prop:7:eqn:1},
		when $C_{\iconstant} = 1$, 	we have 
		\begin{align*}
			\rank{\xn{\iconstant}}{\bx \cup \byp} > \bnp + \kk -\bmp
			\implies\chi_{\iconstant} = \left(\rank{\xn{\iconstant}}{\bx \cup \byp} -  \bnp + \bmp - \kk \right).
		\end{align*}
		Hence, we have
		\begin{align*}
			\sum_{\iconstant = 1}^{\n} C_{\iconstant} \chi_{\iconstant} &= \sum_{\iconstant = 1}^{\n} C_{\iconstant} \left(\rank{\xn{\iconstant}}{\bx \cup \byp} -  \bnp + \bmp - \kk \right) \\
			& =  \sum_{\iconstant = 1}^{\n} C_{\iconstant} \left(\rank{\xn{\iconstant}}{\bx \cup \byp} -  \bnp \right) + \sum_{\iconstant = 1}^{\n} C_{\iconstant} \left(\bmp - \kk \right).
		\end{align*}
		Combining the above results for the first, second and third terms 
		on the right hand side of the Equation \eqref{supp:prop:7:eqn:1} together, 
		\begin{align*}
			\ABT{\bx}{\bys} &= \sum_{\iconstant = 1}^{\n} A_{\iconstant} \left(\bnp - \rank{\xn{\iconstant}}{\bx \cup \byp} \right)  +  \sum_{\iconstant = 1}^{\n} A_{\iconstant} \left(\kk - \bmp \right)\\
			& +  \sum_{\iconstant = 1}^{\n} B_{\iconstant} \left(\rank{\xn{\iconstant}}{\bx \cup \byp} - \bnp \right)
			+  \sum_{\iconstant = 1}^{\n} B_{\iconstant} \left(2\bnp- \bmp + \kk - 2\rank{\xn{\iconstant}}{\bx \cup \byp} \right)\\
			& + \sum_{\iconstant = 1}^{\n} C_{\iconstant} \left(\rank{\xn{\iconstant}}{\bx \cup \byp} -  \bnp \right) +  \sum_{\iconstant = 1}^{\n} C_{\iconstant} \left(\bmp - \kk \right).
		\end{align*}
		According to the definition of the Ansari-Bradley test statistic,
		\begin{align*}
			\ABT{\bx}{\byp} &= \sum_{\iconstant = 1}^{\n} \left|\rank{\xn{\iconstant}}{\bx \cup \byp} - \frac{1}{2}(\n + \mmp + 1)\right|\\
			& = \sum_{\iconstant = 1}^{\n} \indicator{\rank{\xn{\iconstant}}{\bx \cup \byp} <  \bnp}\left(\bnp - \rank{\xn{\iconstant}}{\bx \cup \byp} \right)\\
			& + \sum_{\iconstant = 1}^{\n} \indicator{\rank{\xn{\iconstant}}{\bx \cup \byp} \ge \bnp}\left(\rank{\xn{\iconstant}}{\bx \cup \byp} - \bnp \right)\\
			& = \sum_{\iconstant = 1}^{\n} A_{\iconstant}\left(\bnp - \rank{\xn{\iconstant}}{\bx \cup \byp} \right)
			+ \sum_{\iconstant = 1}^{\n} (B_{\iconstant} + C_{\iconstant}) \left(\rank{\xn{\iconstant}}{\bx \cup \byp} - \bnp \right).
		\end{align*}
		Therefore, we have
		\begin{align*}
			\ABT{\bx}{\bys} & = \ABT{\bx}{\byp} +  \sum_{\iconstant = 1}^{\n}  A_{\iconstant}\left(\kk - \bmp \right) \\
			&+ \sum_{\iconstant = 1}^{\n} B_{\iconstant} \left(2\bnp - 2\bmp + \kk - 2\rank{\xn{\iconstant}}{\bx \cup \byp} \right)
			+  \sum_{\iconstant = 1}^{\n} C_{\iconstant} \left(\bmp - \kk \right)\\
			& = \ABT{\bx}{\byp} + (\kk - \bmp)\sum_{\iconstant = 1}^{\n} (A_{\iconstant} - C_{\iconstant}) + \sum_{\iconstant = 1}^{\n} B_{\iconstant} \left(2\bnp + \kk - 2\bmp - 2\rank{\xn{\iconstant}}{\bx \cup \byp} \right) \\
			& =  \ABT{\bx}{\byp} + \func{\bx, \byp, \m}{\kk},
		\end{align*}
		where the last equation holds because $\kk - \bmp \ge 0$. 
		This completes our proof for case \case{i}.
		The result for case \case{ii} can be proved similarly.
		
		Suppose case \case{ii} holds. That is, assume that
		$\kk - \bmp \le 0$. Then we have  $\sa = \kk - (\m - \mmp)/2$, and $\sbb = 0$. 
		Hence, for the first term on the right hand side of the Equation \eqref{supp:prop:7:eqn:1},
		we have
		\begin{align*}
			\sum_{\iconstant = 1}^{\n} \baa_{\iconstant} \chi_{\iconstant} &= \sum_{\iconstant = 1}^{\n} \baa_i \left(\bnp - \bmp + \kk - \rank{\xn{\iconstant}}{\bx \cup \byp} \right)\\
			& = \sum_{\iconstant = 1}^{\n} \baa_{\iconstant} \left(\bnp - \rank{\xn{\iconstant}}{\bx \cup \byp} \right) +  \sum_{\iconstant = 1}^{\n} \baa_{\iconstant} \left(\kk - \bmp \right).
		\end{align*}
		Similarly, for the second term on the right hand side of the Equation \eqref{supp:prop:7:eqn:1},
		\begin{align*}
			\sum_{\iconstant = 1}^{\n} \bbb_{\iconstant} \chi_{\iconstant} &= \sum_{\iconstant = 1}^{\n} \bbb_{\iconstant} \left(\rank{\xn{\iconstant}}{\bx \cup \byp} -  \bnp + \bmp - \kk \right)\\
			& =  \sum_{\iconstant = 1}^{\n} \bbb_{\iconstant} \left(\bnp - \rank{\xn{\iconstant}}{\bx \cup \byp} \right)
			+  \sum_{\iconstant = 1}^{\n} \bbb_{\iconstant} \left(2\rank{\xn{\iconstant}}{\bx \cup \byp} - 2\bnp + \bmp - \kk \right),
		\end{align*}
		and for the second term on the right hand side of the Equation \eqref{supp:prop:7:eqn:1},
		\begin{align*}
			\sum_{\iconstant = 1}^{\n} \bcc_{\iconstant} \chi_{\iconstant} &= \sum_{\iconstant = 1}^{\n} \bcc_{\iconstant} \left(\rank{\xn{\iconstant}}{\bx \cup \byp} -  \bnp + \bmp - \kk \right) \\
			& =  \sum_{\iconstant = 1}^{\n} \bcc_{\iconstant} \left(\rank{\xn{\iconstant}}{\bx \cup \byp} - \bnp \right) + \sum_{\iconstant = 1}^{\n} \bcc_{\iconstant} \left(\bmp - \kk \right).
		\end{align*}
		Hence, we have
		\begin{align*}
			\ABT{\bx}{\bys \cup \byp} &= \sum_{\iconstant = 1}^{\n} \baa_{\iconstant} \left(\bnp - \rank{\xn{\iconstant}}{\bx \cup \byp} \right) +  \sum_{\iconstant = 1}^{\n} \baa_{\iconstant} \left(\kk - \bmp \right)\\
			& + \sum_{\iconstant = 1}^{\n} \bbb_{\iconstant} \left(\bnp - \rank{\xn{\iconstant}}{\bx \cup \byp} \right)
			+  \sum_{\iconstant = 1}^{\n} \bbb_{\iconstant} \left(2\rank{\xn{\iconstant}}{\bx \cup \byp} - 2\bnp + \bmp - \kk \right)\\
			& + \sum_{\iconstant = 1}^{\n} \bcc_{\iconstant} \left(\rank{\xn{\iconstant}}{\bx \cup \byp} - \bnp \right) + \sum_{\iconstant = 1}^{\n} \bcc_{\iconstant} \left(\bmp - \kk \right).
		\end{align*}
		Notice that
		\begin{align*}
			\ABT{\bx}{\byp} &= \sum_{\iconstant = 1}^{\n} \left|\rank{\xn{\iconstant}}{\bx \cup \byp} - \frac{1}{2}(\n + \mmp + 1)\right|\\
			& = \sum_{\iconstant = 1}^{\n} \indicator{\rank{\xn{\iconstant}}{\bx \cup \byp} \le  \bnp}\left(\bnp - \rank{\xn{\iconstant}}{\bx \cup \byp} \right)\\
			& + \sum_{\iconstant = 1}^{\n} \indicator{\rank{\xn{\iconstant}}{\bx \cup \byp} >  \bnp}\left(\rank{\xn{\iconstant}}{\bx \cup \byp} - \bnp \right)\\
			& = \sum_{\iconstant = 1}^{\n} (\baa_{\iconstant} + \bbb_{\iconstant})\left(\bnp - \rank{\xn{\iconstant}}{\bx \cup \byp} \right)
			+ \sum_{\iconstant = 1}^{\n} \bcc_{\iconstant} \left(\rank{\xn{\iconstant}}{\bx \cup \byp} - \bnp \right).
		\end{align*}
		Therefore, we have
		\begin{align*}
			\ABT{\bx}{\bys} & = \ABT{\bx}{\byp} + \sum_{\iconstant = 1}^{\n}  \baa_{\iconstant}\left(\kk - \bmp \right) \\
			&+ \sum_{\iconstant = 1}^{\n} \bbb_{\iconstant} \left(2\rank{\xn{\iconstant}}{\bx \cup \byp} - 2\bnp + \bmp - \kk \right)
			+  \sum_{\iconstant = 1}^{\n} \bcc_{\iconstant} \left(\bmp - \kk \right)\\
			& = \ABT{\bx}{\byp} + (A_i - C_i) \sum_{\iconstant = 1}^{\n}\left(\kk - \bmp \right) - \sum_{\iconstant = 1}^{\n} B_{\iconstant} \left(2\bnp + \kk - 2\bmp - 2\rank{\xn{\iconstant}}{\bx \cup \byp} \right) \\
			& =  \ABT{\bx}{\byp} + \func{\bx, \byp, \m}{\kk},
		\end{align*}
		where the last equation holds because $\kk - \bmp < 0$. 
		This completes our proof for case~\case{ii}.
	\end{proof}
	
	We complete our results for the lower bounds
	of the Ansari-Bradley statistic when multiple values
	in $\by$ can be missing using the following result:
	
	\begin{theorem} \label{supp:theorem:2}
		Suppose $\bx = \{\xn{1}, \ldots, \xn{\n}\}$ and $\by = \{\yn{1}, \ldots, \yn{\m}\}$ are samples of distinct real values and $\byp \subset \by$ is a subset of $\by$ with sample size $\card{\byp} = \mmp$, which are observed. Then, the minimum possible Ansari-Bradley test statistic, across all unobserved values, is given as follows:
		\begin{align*}
			\min_{\by \setminus \byp \in \mathbb{R}^{\m - \mmp}}\ABT{\bx}{\by} = \ABT{\bx}{\byp} + \min_{\kk \in \{0,\ldots,\m - \mmp \}} \func{\bx, \byp, \m}{\kk} ,
		\end{align*}
		where $\func{\bx, \byp, \m}{\kk}$ is defined in Definition~\ref{supp:def:func}.
	\end{theorem}
	
	\begin{proof}
		To start, let us assume that when the unobserved values $\by \setminus \byp = \byt$,
		the Ansari-Bradley test statistic takes the minimum value. In other words,
		\begin{align*}
			\ABT{\bx}{\byt \cup \byp} =	\min_{\by \setminus \byp \in \mathbb{R}^{\m - \mmp}}\ABT{\bx}{\by}. 
		\end{align*}
		Now, consider $\m - \mmp + 1$ different
		sets $\byhun{1}, \ldots, \byhun{\m-\mmp+1} \in  \mysetx(\bx\cup\byp, \m -\mmp)$,
		such that 
		\begin{align*}
			\kk = \sum_{\y \in \byhun{\kk}} \indicator{ \yy > \max \bx \cup \byp}, \text{ where } \kk \in \{1, \ldots, \m - \mmp + 1\}. 
		\end{align*}
		By definition, we have
		\begin{align*}
			\ABT{\bx}{\byt \cup \byp} = \min_{\by \setminus \byp \in \mathbb{R}^{\m - \mmp}}\ABT{\bx}{\by} \le \min \{\ABT{\bx}{\byp \cup \byhun{1}}, \ldots, \ABT{\bx}{\byp \cup \byhun{\m-\mmp+1}}\}.
		\end{align*}
		Below we show that
		\begin{align*}
			\ABT{\bx}{\byt \cup \byp} \ge  \min \{\ABT{\bx}{\byp \cup \byhun{1}}, \ldots, \ABT{\bx}{\byp \cup \byhun{\m-\mmp+1}}\}.
		\end{align*}

		Note that according to Lemma~\ref{supp:lemma:10},
		there exist a set $\byh \in  \mysetx(\bx\cup\byp, \m -\mmp)$
		such that $\ABT{\byh \cup \byp}{\bx} \ge \ABT{\byt \cup \byp}{\bx}$.
		By Lemma~\ref{supp:ab:lemma:1},
		\begin{align*}
			\ABT{\byh \cup \byp}{\bx}  + \ABT{\bx}{\byh \cup \byp} = \ABT{\bx}{\byt \cup \byp} + \ABT{\byt \cup \byp}{\bx} = \sum_{\iconstant=1}^{\bn} \left|\iconstant - \frac{\bn+1}{2}\right|.
		\end{align*} 
		Then we also have 
		\begin{align*}
			&\ABT{\bx}{\byh \cup \byp} \le \ABT{\bx}{\byt \cup \byp}\\
			\implies&\min \{\ABT{\bx}{\byp \cup \byhun{1}}, \ldots, \ABT{\bx}{\byp \cup \byhun{\m-\mmp+1}}\} \le \ABT{\bx}{\byt \cup \byp}.
		\end{align*}
		Hence, we have our desired result
		\begin{align*}
			\ABT{\bx}{\byt \cup \byp} &= \min \{\ABT{\bx}{\byp \cup \byhun{1}}, \ldots, \ABT{\bx}{\byp \cup \byhun{\m-\mmp+1}}\} \\
			& = \ABT{\bx}{\byp} + \min_{\kk \in \{0,\ldots,\m - \mmp \}} \func{\bx, \byp, \m}{\kk} ,
		\end{align*}
		where the last equation follows from Proposition~\ref{supp:ab:prop:7}.
	\end{proof}
	
	\section{Lower and upper bounds under general missingness case}
	\label{appendix:general}
	
	This section combines the results from Section~\ref{appendix:lowebound:ab:x}
	and \ref{appendix:lowebound:ab:y} when 
	only values in $\bx$ can be missing, and
	only values in $\by$ can be missing, respectively.
	We derive the tight lower and upper bounds
	of the Ansari-Bradley test statistic in the presence
	of missing data without assuming any missingness
	patterns, as shown in Theorem~\ref{supp:theorem:3}
	in the end of this section.
	
	We start by proving the following lemma, which
	shows that if we replace a subset $\byt$ 
	of $\by$ such that for any $\y \in \byt$, $\rank{\y}{\bx \cup \by} \le (\bn+1)/2$ with $\byh = \mysety(\bx \cup \byp, \m - \mmp, 0)$,
	the values of the Ansari-Bradley statistic can only decrease
	or remain unchanged.
	
	\begin{lemma} \label{supp:lemma:12}
		Suppose $\bx = \{\xn{1}, \ldots, \xn{\n}\}$ and $\by = \{\yn{1}, \ldots, \yn{\m}\}$ are 
		samples of distinct real values. Denote $\bn = \n +\m$.
		Suppose $\byt \subset \by$ is a non-empty subset 
		of $\by$ such that for any $\y \in \byt$, $\rank{\y}{\bx \cup \by} \le (\bn+1)/2$.
		Denote $\byp = \by \setminus \byt$, and $|\byp| = \mmp$. Then, if $\byh = \mysety(\bx \cup \byp, \m - \mmp, 0)$, we have $\ABT{\bx}{\bys} \le \ABT{\bx}{\by}$, where
		$\bys = \byh \cup \byp$.
	\end{lemma}
	
	\begin{proof}	
		We prove the result by mathematical induction on $\m - \mmp$.
		Let us denote $\byh = \{\ysn{1}, \ldots, \ysn{\m -\mmp}\}$.
		For the base case when $\m - \mmp = 1$, since 
		$\byh = \mysety(\bx \cup \byp, 1, 0)$, we have
		$\ysn{1} < \max \bx \cup \byp$. Hence,
		$\rank{\ysn{1}}{\bx \cup \bys} = 1$.	 
		Then, according to Proposition~\ref{supp:ab:prop:2},
		we have $\ABT{\by}{\bx} \le \ABT{\bys}{\bx}$.
		By applying Lemma~\ref{supp:ab:lemma:1},
		we have our desired result $\ABT{\bx}{\bys} \le \ABT{\bx}{\by}$.
		
		Let $\kk \in \{1, \ldots, \m - 1\}$, and suppose that Lemma~\ref{supp:lemma:12} 
		holds for $\m - \mmp = \kk$. We show that the lemma also holds when 
		$\m - \mmp = \kk + 1$.
		
		Denote $\byt = \{\ytn{1}, \ldots, \ytn{\kk + 1}\}$,
		and without loss of generality, let us assume $\ytn{1} = \max \byt$. 
		Consider a set $\bzh  = \{\zsn{1}, \ldots, \zsn{\kk}\}$ and assume that
		$\bzh \in \mysety(\bx \cup \byp \cup \{\ytn{1}\}, \kk, 0)$.
		For notation convenience, denote $\bz = \bzh \cup \byp \cup \{\ytn{1}\}$.	
		Then, by the assumption that Lemma~\ref{supp:lemma:12} is correct 
		when $\m -\mmp = \kk$, we have
		\begin{align} \label{supp:lemma:12:eqn:1}
			\ABT{\bx}{\bz} \le \ABT{\bx}{\by}.
		\end{align}
		Since $\ytn{1} = \max \byt$, we have
		\begin{align*}
			\rank{\ytn{1}}{\bx \cup \by} > \max \{\rank{\ytn{2}}{\bx \cup \by}, \ldots, \rank{\ytn{\kk+1}}{\bx \cup \by}\} \ge \kk,
		\end{align*}
		where the last $``\ge"$ holds because $\{\rank{\ytn{2}}{\bx \cup \by}, \ldots, \rank{\ytn{\kk+1}}{\bx \cup \by}\}$ are $\kk$ distinct values in $\{1,\ldots,\bn\}$.
		Then, since $\bzh \in \mysety(\bx \cup \byp \cup \{\ytn{1}\}, \kk, 0)$, we have
		\begin{align*}
			&\zsn{\iconstant} < \min \bx \cup \byp \cup \{\ytn{1}\}, \text{ for any } \iconstant \in \{1,\ldots, \kk\}\\
			\implies & \{\rank{\zsn{1}}{\bx \cup \bz}, \ldots, \rank{\zsn{\kk}}{\bx \cup \bz}\} = \{1,\ldots,\kk\}\\
			\implies & 	\rank{\ytn{1}}{\bx \cup \by} > \max \{\rank{\zsn{1}}{\bx \cup \bz}, \ldots, \rank{\zsn{\kk}}{\bx \cup \bz}\} .
		\end{align*}
		Then, according to Lemma \ref{supp:lemma:4.2}, 	
		\begin{align*}
			\rank{\ytn{1}}{\bx \cup \bz } = \rank{\ytn{1}}{\bx \cup \by} < \frac{1}{2}(\bn + 1).
		\end{align*}
		Next, consider any real value $\{\ys\}  \in \mysety(\byp \cup \bzh, 1, 0)$,
		and denote $\bzp = \byp \cup \bzh \cup \{\ys\}$.
		Then, by applying the lemma when $\m - \mmp = 1$, we have
		\begin{align*}
			\ABT{\bx}{\bzp} \le \ABT{\bx}{\bz} \le^{\text{ Inequality }  \eqref{supp:lemma:12:eqn:1}} \ABT{\bx}{\by}.
		\end{align*}
		Now, in order to finish our proof, we only need to show that
		$\bzh \cup \{\ys\} \in \mysety(\bx \cup \byp, \kk + 1, 0)$.
		
		Since $\bzh \in \mysety(\bx \cup \byp \cup \{\ytn{1}\}, \kk, 0)$, we have
		$\zsn{\iconstant} < \min \bx \cup \byp, \text{ for any } \iconstant \in \{1,\ldots, \kk\}$.
		Combining this result with $\{\ys\}  \in \mysety(\byp \cup \bzh, 1, 0)$, hence 
		we have $\bzh \cup \{\ys\} \in \mysety(\bx \cup \byp, \kk + 1, 0)$, which
		completes our proof.
	\end{proof}
	
	The following lemma provides the same conclusion
	as Lemma~\ref{supp:lemma:12},  but now
	we replace the subset $\byt$ of $\by$ such that for any $\y \in \byt$, $\rank{\y}{\bx \cup \by} \ge (\bn+1)/2$ with $\byh = \mysety(\bx \cup \byp, \m - \mmp, 0)$,
	
	\begin{lemma} \label{supp:lemma:13}
		Suppose $\bx = \{\xn{1}, \ldots, \xn{\n}\}$ and $\by = \{\yn{1}, \ldots, \yn{\m}\}$ are 
		samples of distinct real values. Denote $\bn = \n +\m$.
		Suppose $\byt \subset \by$ is a non-empty subset 
		of $\by$ such that for any $\y \in \byt$, $\rank{\y}{\bx \cup \by} \ge (\bn+1)/2$.
		Denote $\byp = \by \setminus \byt$, and $|\byp| = \mmp$. Then, if $\byh = \mysety(\bx \cup \byp, \m - \mmp, \m-\mmp)$, we have $\ABT{\bx}{\bys} \le \ABT{\bx}{\by}$, where
		$\bys = \byh \cup \byp$.
	\end{lemma}
	
	\begin{proof}
		This result can be prove following the same approach 
		for proving Lemma~\ref{supp:lemma:12}. Thus it is omitted here.
	\end{proof}
	
	The following definition is used in Proposition~\ref{supp:prop:8}.
	
	\begin{definition} \label{supp:def:ab:z}
		Suppose $\bx$ and $\by$ are samples of distinct values,
		$\tconstant_1$ and $\tconstant_2$ are positive integers,
		and $\kk \in \{1,2,3,4\}$. Then we define
		\begin{align*}
			\mysetz(\bx, \by, \tconstant_1, \tconstant_2, \kk) = \left\{(\bxh, \byh):  
			\mysetr = \bmsn{\kk}, \byh \in \mysetx(\bxs \cup \by, \tconstant_2) \ \right\},
		\end{align*}
		where 
		$\mysetr = \{\rank{\xsn{1}}{\bxs \cup \bys}, \ldots, \rank{\xsn{\tconstant_1}}{\bxs  \cup \bys}\}$,
		$\bxh = \{\xsn{1}, \ldots, \xsn{\tconstant_1}\}$,
		$\bxs = \bxh \cup \bx$, 
		and $\bys = \byh \cup \by$.
	\end{definition}
	
	%\begin{definition}
	%	Suppose $\bx$ and $\by$ are samples of distinct values,
	%	$\tconstant_1, \tconstant_2 \in \mathbb{Z}_{>0}$ are positive integers,
	%	and $\kk \in {1, 2, 3, 4}$. We define
	%	\begin{align*}
		%		\mysetz(\bx, \by, \tconstant_1, \tconstant_2, \kk)
		%		= \left\{(\bxh, \byh):
		%			\begin{array}{l}
			%				\bxh = \{\xsn{1}, \ldots, \xsn{\tconstant_1}\},~
			%				\byh \in \mysetx(\bx \cup \bxh \cup \by, \tconstant_2), \\
			%				\left\{\rank{\xsn{1}}{\bx \cup \bxh \cup \by \cup \byh}, \ldots, \rank{\xsn{\tconstant_1}}{\bxs \cup \bys} \right\} = \bmsn{\kk}
			%			\end{array}
		%			\right\},
		%	\end{align*}
	%	where $\bxs = \bxh \cup \bx$, and $\bys = \byh \cup \by$
	%\end{definition}
	
	Applying Definition~\ref{supp:def:ab:z}, we
	show that there exist imputations
	$\bxh, \byh \in \mysetz(\bxp, \byp, \n-\np, \m-\mmp,\iconstant)$
	that minimizes the Spearman's footrule, 
	when $\bn$ and $\n - \np$ satisfy 
	condition $\bmcn{\iconstant}$ for $\iconstant = \{1,2,3,4\}$.
	
	\begin{proposition} \label{supp:prop:8}
		Suppose $\bx = \{\xn{1}, \ldots, \xn{\n}\}$ and 
		$\by = \{\yn{1}, \ldots, \yn{\m}\}$ are samples of distinct real values. 
		Let $\bxp \subset \bx, \byp \subset \by$ be observed subsets 
		of $\bx$, and $\by$ with sample sizes $|\bxp| = \np$, and $|\byp| = \mmp$. 
		Denote $\bn = \n + \m$, and suppose $\bn$ and $\n - \np$ satisfy 
		condition $\bmcn{\iconstant}$, where $\iconstant \in \{1,2,3,4\}$.
		Then there exist 
		$\bxh, \byh \in \mysetz(\bxp, \byp, \n-\np, \m-\mmp,\iconstant)$ such that 
		$\ABT{\bxs}{\bys} \le \ABT{\bx}{\by}$,
		where $\bxs = \bxh \cup \bxp$, and $\bys = \byh \cup \byp$.
	\end{proposition}
	
	\begin{proof}
		To start, let $\bxh = \{\xsn{1},\ldots,\xsn{\n -\np}\}$ be a set of
		distinct real values. Denote $\bxs = \bxh \cup \bxp$. Assume
		that values in $\bxs \cup \by$ are distinct, and 
		\begin{align*}
			\{\rank{\xsn{1}}{\bxs \cup \by}, \ldots, \rank{\xsn{\n - \np}}{\bxs \cup \by}\} = \bmsn{\iconstant}.
		\end{align*}
		Then, according to Proposition \ref{supp:prop:4} and Proposition \ref{supp:prop:5}, we have
		\begin{align} \label{supp:prop:8:eqn:1}
			\ABT{\bxs}{\by} \le \ABT{\bx}{\by},
		\end{align}
		for any $\iconstant \in \{1,2,3,4\}$.
		
		Denote $\byt = \by \setminus \byp$. 
		Let $s_* = \min \bmsn{\iconstant}$ and $s^* = \max \bmsn{\iconstant}$. 
		Since values in $\bxs$ take the ranks between 
		$s_*$ and $s^*$, for any $\y \in \byt$, we have
		$\rank{\y}{\bxs \cup \by} \ge s^* + 1~\text{or}~\rank{\y}{\bxs \cup \by} \le s_* - 1.$
		By definition of $\bmsn{\iconstant}$, for any $\iconstant \in \{1,2,3,4\}$, we have
		\begin{align*}
			&s_* \le -\frac{\n - \np}{2} + 1 + \frac{1}{2}(\bn + 1) \le -\frac{1}{2} + 1 + \frac{1}{2}(\bn + 1) = \frac{1}{2} +  \frac{1}{2}(\bn + 1),\\
			\text{and }&s^* \ge \frac{\n - \np - 1}{2} + \frac{1}{2}(\bn+1) \ge \frac{1}{2}(\bn+1).
		\end{align*}
		Hence, for any $\y \in \byt$, we have either
		\begin{align*}
			&\rank{\y}{\bxs \cup \by} \ge s^* + 1 > \frac{1}{2}(\bn+1), \text{ or }\rank{\y}{\bxs \cup \by} \le s_* - 1 < \frac{1}{2}(\bn+1).
		\end{align*}
		Thus, we can partition $\byt = \bytln{1} \cup \bytln{2}$, where
		\begin{align*}
			\bytln{1} &= \left\{\y| \y \in \byt, \rank{\y}{\bxs \cup \by} < \frac{1}{2}(\bn+1)\right\},\\
			\bytln{2} &= \left\{\y| \y \in \byt, \rank{\y}{\bxs \cup \by} > \frac{1}{2}(\bn+1)\right\}.
		\end{align*}
		Denote $|\bytln{1}| = \lln{1}$, and $|\bytln{2}| = \lln{2}$. Without loss
		of generality, denote $\bytln{1} = \{\ytn{1}, \ldots \ytn{\lln{1}}\}$,
		$\bytln{2} = \{\ytn{\lln{1}+1}, \ldots \ytn{\lln{1} + \lln{2}}\}$,
		and $\byt = \{\ytn{1}, \ldots \ytn{\lln{1}}, \ytn{\lln{1} + 1},
		\ldots, \ytn{\lln{1} + \lln{2}}\}$.

		Suppose $\byhln{1} \in \mysety(\bxs \cup (\by \setminus \bytln{1}), \lln{1}, 0)$. Then, according to Lemma~\ref{supp:lemma:12},
		\begin{align} \label{supp:prop:8:eqn:2}
			\ABT{\bxs}{ (\by \setminus \bytln{1}) \cup \byhln{1}} \le \ABT{\bxs}{\by} \le^{\text{ Inequality \eqref{supp:prop:8:eqn:1}}} \ABT{\bx}{\by}.
		\end{align}
		Since for any $\y \in \bytln{1}$, $\rank{\y}{\bxs \cup \by} \le s_* - 1$, 
		and for any $\x \in \bxh$, $\rank{\x}{\bxs\cup \by} \ge s_*$.
		Hence, for any $\x \in \bxh$, $\rank{\x}{\bxs \cup \by} > \max \{\rank{\y}{\bxs \cup \by}: \y \in \bytln{1}\}$. Since $|\bytln{1}| = \lln{1}$, and values in
		$\bytln{1}$ takes distinct values in $\{1,\ldots,\bn\}$, we have 
		\begin{align*}
			\rank{\x}{\bxs \cup \by} > \max \{\rank{\y}{\bxs \cup \by}: \y \in \bytln{1}\} \ge \lln{1}.
		\end{align*}
		Note that $\byhln{1} \in \mysety(\bxs \cup (\by \setminus \bytln{1}), \lln{1}, 0)$.
		Hence, for any $\y \in \bytln{1}$, we have $\y < \min \bxs \cup (\by \setminus \bytln{1})$,
		which implies
		\begin{align*}
			\left\{\rank{\ytn{1}}{\bxs \cup (\by \setminus \bytln{1}) \cup \byhln{1}}, \ldots, \rank{\ytn{\lln{1}}}{\bxs \cup (\by \setminus \bytln{1}) \cup \byhln{1}} \right\} = \{1,\ldots, \lln{1}\}.
		\end{align*}
		Hence, for any $\x \in \bxh$, we also have
		\begin{align*}
			\rank{\x}{\bxs \cup \by}
			> \left\{\rank{\ytn{1}}{\bxs \cup (\by \setminus \bytln{1}) \cup \byhln{1}}, \ldots, \rank{\ytn{\lln{1}}}{\bxs \cup (\by \setminus \bytln{1}) \cup \byhln{1}} \right\}.
		\end{align*}
		Then, by applying Lemma \ref{supp:lemma:4.1}, 
		\begin{align*}
			\rank{\x}{\bxs \cup \by} = \rank{\x}{\bxs \cup (\by \setminus \bytln{1}) \cup \byhln{1}}, \text{ for any } \x \in \bxh.
		\end{align*}
		Therefore
		\begin{align*}
			&\left\{ \rank{\xsn{1}}{\bxs \cup (\by \setminus \bytln{1}) \cup \byhln{1}}, \ldots, \rank{\xsn{\n - \np}}{\bxs \cup (\by \setminus \bytln{1}) \cup \byhln{1}} \right\} \\
			&= 	\{\rank{\xsn{1}}{\bxs \cup \by}, \ldots, \rank{\xsn{\n - \np}}{\bxs \cup \by}\}\\
			& = \bmsn{\iconstant}.
		\end{align*}
		Similarly, by applying Lemma \ref{supp:lemma:4.1}, we have
		\begin{align*}
			\rank{\yt}{\bxs \cup (\by \setminus \bytln{1}) \cup \byhln{1}} = \rank{\yt}{\bxs \cup \by} > \frac{1}{2} (\bn + 1), \text{ for any } \y \in \bytln{2}.
		\end{align*}
		Suppose $\byhln{2} \in \mysety(\bxs \cup \byp \cup \byhln{1}, \lln{2}, \lln{2})$. Then, according to Lemma \ref{supp:lemma:13},
		\begin{align*}
			\ABT{\bxs}{\byp \cup \byhln{1} \cup \byhln{2}} \le \ABT{\bxs}{(\by \setminus \bytln{1}) \cup \byhln{1}} \le^{\text{ Inequality }\eqref{supp:prop:8:eqn:2}} \ABT{\bx}{\by}.
		\end{align*}
		Subsequently, in order to complete our proof, 
		we need to show that 
		$\byhln{1} \cup \byhln{2} \in \mysetx(\bxs \cup \byp, \lln{1} + \lln{2})$,
		and $\left\{ \rank{\xsn{1}}{\bxs \cup \byp \cup \byhln{1} \cup \byhln{2}}, \ldots, \rank{\xsn{\n - \np}}{\bxs \cup \byp \cup \byhln{1} \cup \byhln{2}} \right\} = \bmsn{\iconstant}$.
		
		For the first part, notice that $\byhln{1} \in \mysety(\bxs \cup (\by \setminus \bytln{1}), \lln{1}, 0)$. Hence for any $\y \in \byhln{1}$, we have $\y < \min \bxs \cup \byp$. Note also 
		$\byhln{2} \in \mysety(\bxs \cup \byp \cup \byhln{1}, \lln{2}, \lln{2})$.
		Hence for any $\y \in \byhln{2}$, we have $\y > \max \bxs \cup \byp$. Therefore, we have
		$\byhln{1} \cup \byhln{2} \in \mysetx(\bxs \cup \byp, \lln{1} + \lln{2})$.
		
		For the second part, note that $\rank{\yt}{\bxs \cup (\by \setminus \bytln{1}) \cup \byhln{1}} > (\bn+1)/2$ for any $\y \in \bytln{2}$, and  
		$\left\{ \rank{\xsn{1}}{\bxs \cup (\by \setminus \bytln{1}) \cup \byhln{1}}, \ldots, \rank{\xsn{\n - \np}}{\bxs \cup (\by \setminus \bytln{1}) \cup \byhln{1}} \right\} = \bmsn{\iconstant}$ takes consecutive ranks between $s_*$ and $s^*$, where $s_* \le (\bn+1)/2 + 1/2$. Since ranks are distinct values in $\{1,\ldots, \bn\}$, for any $\x \in \bxh$, we have $$\rank{\x}{\bxs \cup (\by \setminus \bytln{1}) \cup \byhln{1}} < \min \{ \rank{\y}{\bxs \cup (\by \setminus \bytln{1}) \cup \byhln{1}}: \y \in \bytln{2} \}.$$ Note also that
		$\byhln{2} \in \mysety(\bxs \cup \byp \cup \byhln{1}, \lln{2}, \lln{2})$.
		By definition of $\mysety$, we have 
		\begin{align*}
			&\y > \max \bxs \cup \byp \cup \byhln{1}, \text{for any } \y \in \byhln{2}\\
			\implies& \min \{\rank{\y}{\bxs \cup \byp \cup\byhln{1} \cup \byhln{2}}: \y \in \bytln{2} \} > \rank{\x}{\bxs \cup (\by \setminus \bytln{1}) \cup \byhln{1}}, \text{ for any } \x \in \bxh.
		\end{align*}	 
		Then, by applying Lemma \ref{supp:lemma:4.2}, we obtain desired result $$\left\{ \rank{\xsn{1}}{\bxs \cup \byp \cup \byhln{1} \cup \byhln{2}}, \ldots, \rank{\xsn{\n - \np}}{\bxs \cup \byp \cup \byhln{1} \cup \byhln{2}} \right\} = \bmsn{\iconstant}.$$
		This completes our proof
	\end{proof}
	
	Before we present the final results, we prove the 
	following two lemmas. The first lemma
	considers imputations $\bxh, \byh \in \mysetz(\bxp, \byp, \n - \np, \m -\mmp, \iconstant)$, and
	provides the minimum and maximum possible
	number of imputed values in $\by$
	that are larger than all values in $ \bxh \cup \bxp \cup \byp$.
	In other words, the lemma provides lower
	and upper bounds
	for  $\kk = \sum_{\iconstant = 1}^{\m - \mmp} \indicator{\ysn{\iconstant} > \max \bxh \cup \bxp \cup \byp }$.
	
	\begin{lemma} \label{supp:ab:lemma:4}
		Suppose $\bx = \{\xn{1}, \ldots, \xn{\n}\}$ and $\by = \{\yn{1}, \ldots, \yn{\m}\}$ are samples of distinct real values. Let $\bxp \subset \bx$ and $\byp \subset \by$ be observed subsets with sizes $|\bxp| = \np$ and $|\byp| = \mmp$. 
		Denote $\bn = \n + \m$, and suppose
		$\bn, \n - \np$ satisfy condition $\bmcn{\iconstant}$,
		where $\iconstant \in \{1,2,3,4\}$.
		Suppose $\bxh, \byh \in \mysetz(\bxp, \byp, \n - \np, \m -\mmp, \iconstant)$.
		Denote $\byh = \{\ysn{1}, \ldots, \ysn{\m-\mmp}\}$, and 
		$\kk = \sum_{\iconstant = 1}^{\m - \mmp} \indicator{\ysn{\iconstant} > \max \bxh \cup \bxp \cup \byp }$, then we have
		\begin{align*}
			\left\{ \begin{array}{lcl}
				\sa \le \kk \le \sbb &~\text{if}~\iconstant \in \{1,3\}, \\ 
				\scc \le \kk \le \sdd &~\text{if}~\iconstant \in \{2,4\},
			\end{array}\right.
		\end{align*}
		where $\sa, \sbb, \scc, \sdd$ are constants of 
		sample sizes defined as
		\begin{align*}
			\sa &= \max \left\{{(\m -\np -2\mmp + 1)}/{2}, 0\right\},\\
			\sbb &=  \min \left\{{(\m + \np + 1)}/{2},  \m -\mmp \right\},\\
			\scc &=  \max \left\{{(\m -\np -2\mmp)}/{2}, 0 \right\},\\
			\text{and }\sdd &=  \min \left\{{(\m + \np)}/{2},  \m -\mmp \right\}.
		\end{align*} 
	\end{lemma}
	
	\begin{proof}
		To start, consider the cases when $\iconstant \in \{1,3\}$.
		That is, suppose $\bn$ and $\n - \np$ satisfy $\bmcn{1}$ or $\bmcn{3}$,
		and $\bxh, \byh \in \mysetz(\bxp, \byp, \n - \np, \m -\mmp, 1)$,
		or $\bxh, \byh \in \mysetz(\bxp, \byp, \n - \np, \m -\mmp, 3)$.
		Denote $\bxs = \bxh \cup \bxp$, $\bys = \byh \cup \byp$,
		and $\mysetr = \{\rank{\xsn{1}}{\bxs \cup \bys}, \ldots, \rank{\xsn{\n - \np}}{\bxs \cup  \bys}\}$. Then, by definition of $\mysetz$, we have 
		\begin{align*}
			\mysetr  = \bmsn{\iconstant} =  \left\{-\frac{\n - \np}{2} + \frac{1}{2}(\bn + 1), \ldots, \frac{\n - \np}{2} -1  + \frac{1}{2}(\bn+1) \right\},
		\end{align*}
		where $\iconstant = \{1,3\}$.
		Denote $s_* = -{(\n - \np)}/{2} + (\bn + 1)/2$,
		and $s^* = \{{(\n - \np)}/{2}  + (\bn+1)/2, \ldots, \bn\}$,
		as $\min \mysetr$ and $\max \mysetr$, respectively.
		Let $$\byhln{1} = \left\{\y \in \byh: \y > \max \bxs \cup \byp \right\}.$$
		Then $|\byhln{1}| = \kk$. By definition of $\byhln{1} $, for any $\y \in \byhln{1}$,
		\begin{align*}
			s^* < \rank{\y}{\bxs \cup\bys} \le \bn.
		\end{align*}
		Note that
		\begin{align*}
			|\{s_* + 1, \ldots\bn\}| = \bn - s^* = \frac{2\bn - \n + \np + 2 - \bn - 1}{2}=\frac{\m + \np + 1}{2},
		\end{align*}
		and since each value in $\byhln{1}$ take different values in 
		$\{s_* + 1, \ldots, \bn\}$, we have 
		$\kk = |\byhln{1}| \le (\m + \np + 1)/2$.
		By definition of $\bytln{1}$, we also have
		$\kk \le \m -\mmp$.
		Hence, $\kk \le \sbb$.
		
		Using similar approach, we can show that $a \le \kk$. 
		Let $$\byhln{2} = \left\{\y \in \byh: \y < \min \bxs \cup \byp \right\}.$$
		Since $\bxh, \byh \in \mysetz(\bxp, \byp, \n - \np, \m -\mmp, 1)$,
		or $\bxh, \byh \in \mysetz(\bxp, \byp, \n - \np, \m -\mmp, 3)$,
		we have $\byh \in \mysetx(\bxs \cup \byp, \m-\mmp)$. 
		By definition, for any $\y \in \byh$, we have 
		$\y < \min \bxs \cup \byp$ or $\y > \max \bxs \cup \byp$. 
		Hence, $|\byhln{2}| = \m - \mmp - \kk$.
		Also, by definition of $\byhln{2}$, for any $\y \in \byhln{2}$, 
		\begin{align*}
			1 \le \rank{\y}{\bxs \cup\bys} < s_*.
		\end{align*}
		Note that 
		\begin{align*}
			|\{1, \ldots, s_*-1\}| = s_* = -\frac{\n - \np}{2} + \frac{1}{2}(\bn + 1).
		\end{align*}
		and since each value in $\byhln{2}$ take different values in 
		$\{1, \ldots, s_* - 1\}$, we have 
		\begin{align*}
			&|\byhln{2}|  \le |\{1, \ldots, s_*-1\}| =	-\frac{\n - \np}{2} + \frac{1}{2}(\bn + 1) - 1\\
			\implies \kk &\ge \m -\mmp + \frac{\n - \np}{2} - \frac{1}{2}(\bn + 1) + 1
			= \frac{\m -\np -2\mmp + 1}{2}.
		\end{align*}
		By definition, we also have $\kk \ge 0$.
		Hence, $\kk \ge \sa$, which proves our cases
		when $\iconstant \in \{1,3\}$.
		
		The cases when $\iconstant \in \{2,4\}$ can be proved similarly. Thus, it is omitted here.
	\end{proof}

	The following lemma provides
	explicit form for the Ansari-Bradley statistic
	after being imputed with
	$(\bxh, \byh) \in \mysetz(\bxp, \byp, \n-\np, \m-\mmp, \iconstant)$.
	
	\begin{lemma} \label{supp:lemma:14}
		Suppose $\bx = \{\xn{1}, \ldots, \xn{\n}\}$ and 
		$\by = \{\yn{1}, \ldots, \yn{\m}\}$ are samples of distinct real values. 
		Let $\bxp \subset \bx, \byp \subset \by$ be subsets of $\bx$, $\by$ 
		with sample sizes $|\bxp| = \np$, and $|\byp| = \mmp$. 
		Denote $\bn = \n + \m$, and suppose $\bn$ and $\n - \np$ 
		satisfy $\bmcn{\iconstant}$, where $\iconstant \in \{1,2,3,4\}$.
		Suppose $(\bxh, \byh) \in \mysetz(\bxp, \byp, \n-\np, \m-\mmp, \iconstant)$,
		and denote $\sum_{\iconstant = 1}^{\m - \mmp} \indicator{\ysn{\iconstant} > \max \bxh \cup \bxp \cup \byp } = \kk$.
		Subsequently, we have
		\begin{align*}
			\ABT{\bxs }{\bys} =  \left\{ \begin{array}{lcl}
				\ABT{\bxp}{\byp} + \func{\bxp, \byp, \m}{\kk} + {(\n^2 - \np^2)}/{4} &~\text{if}
				&\iconstant \in \{1,4\}, \\ 
				\ABT{\bxp}{\byp} + \func{\bxp, \byp, \m}{\kk} + {(\n^2 - \np^2 - 1)}/{4} &~\text{if}
				&\iconstant = 2, \\ 
				\ABT{\bxp}{\byp} + \func{\bxp, \byp, \m}{\kk} + {(\n^2 - \np^2 + 1)}/{4} &~\text{if}
				&\iconstant = 3,
			\end{array}\right.
		\end{align*}
		where $\bxs = \bxp \cup \bxh$ and $\bys = \byh \cup \byp$.
	\end{lemma}
	
	\begin{proof}
		To start, by applying Proposition~\ref{supp:prop:6}, we have
		\begin{align*}
			\ABT{\bxs}{\bys \cup \byp} =  \left\{ \begin{array}{lcl}
				\ABT{\bxp}{\bys} + {(\n^2 - \np^2)}/{4}, &~\text{if}
				&\iconstant \in \{1,4\}, \\ 
				\ABT{\bxp}{\bys} + {(\n^2 - \np^2 - 1)}/{4} &~\text{if}
				&\iconstant = 2,\\ 
				\ABT{\bxp}{\bys} + {(\n^2 - \np^2 + 1)}/{4} &~\text{if}
				&\iconstant = 3.
			\end{array}\right.
		\end{align*}
		Then, by applying Proposition~\ref{supp:ab:prop:7} we have
		\begin{align*}
			\ABT{\bxp}{\bys} =  \ABT{\bxp}{\byp} + \func{\bxp, \byp, \m}{\kk}.
		\end{align*}
		Combining the above to results, we have our desired results.
	\end{proof}
	
	Finally, we prove the main results for the tight lower
	and upper bounds of the Ansari-Bradley statistics 
	in the presence of missing data.
	
	\begin{theorem} \label{supp:theorem:3}
		Suppose $\bx = \{\xn{1}, \ldots, \xn{\n}\}$ and $\by = \{\yn{1}, \ldots, \yn{\m}\}$ are samples of distinct real values. Let $\bxp \subset \bx$ and $\byp \subset \by$ be observed subsets with sizes $|\bxp| = \np$ and $|\byp| = \mmp$. Define $\sa = \max \left\{{(\m -\np -2\mmp + 1)}/{2}, 0\right\}$, $\sbb =  \min \left\{{(\m + \np + 1)}/{2},  \m -\mmp \right\}$,  $\scc =  \max \left\{{(\m -\np -2\mmp)}/{2}, 0\right\}$ and $\sdd =  \min \left\{{(\m + \np)}/{2},  \m -\mmp \right\}$. Then, 
		the minimum possible Ansari-Bradley test statistic over all
		possible missing values is:
		\begin{align*}
			\min_{\substack{\bx \setminus \bxp \in \mathbb{R}^{\n - \np}, \\ \by \setminus \byp \in \mathbb{R}^{\m - \mmp} }} \ABT{\bx}{\by} =  \ABT{\bxp}{\byp} + \left\{ \begin{array}{lcl}
				\min\limits_{\kk \in \{\sa, \ldots, \sbb\}} \func{\bxp, \byp, \m}{\kk} + {(\n^2 - \np^2)}/{4} &\text{~if}~\bmcn{1}, \\ 
				\min\limits_{\kk \in \{\scc, \ldots, \sdd\}} \func{\bxp, \byp, \m}{\kk} + {(\n^2 - \np^2 - 1)}/{4} &\text{~if}~\bmcn{2},\\ 
				\min\limits_{\kk \in \{\sa, \ldots, \sbb\}} \func{\bxp, \byp, \m}{\kk} + {(\n^2 - \np^2 + 1)}/{4} &\text{~if}~\bmcn{3},\\
				\min\limits_{\kk \in \{\scc, \ldots, \sdd\}} \func{\bxp, \byp, \m}{\kk} + {(\n^2 - \np^2)}/{4} &\text{~if}~\bmcn{4}, \\ 
			\end{array}\right.
		\end{align*}
		and the maximum possible Ansari-Bradley test statistic is:
		\begin{align*}
			\max_{\substack{\bx \setminus \bxp \in \mathbb{R}^{\n - \np}, \\ \by \setminus \byp \in \mathbb{R}^{\m - \mmp} }} \ABT{\bx}{\by}=  - \min_{\substack{\bx \setminus \bxp \in \mathbb{R}^{\n - \np}, \\ \by \setminus \byp \in \mathbb{R}^{\m - \mmp} }}  \ABT{\by}{\bx}   + \left\{ \begin{array}{ll}
				{\bn^2}/{4} 
				& \bn~\text{is even,} \\ 
				{(\bn^2 - 1)}/{4}
				& \bn~\text{is odd.} 
			\end{array}\right.
		\end{align*}
	\end{theorem}
	
	\begin{proof}	
		For notation convenience, let us denote
		\begin{align*}
			T_{\min}(\bx,\by)  = \min_{\substack{\bx \setminus \bxp \in \mathbb{R}^{\n - \np}, \\ \by \setminus \byp \in \mathbb{R}^{\m - \mmp} }} \ABT{\bx}{\by}, \text{ and }
			T_{\max}(\bx,\by)  = \max_{\substack{\bx \setminus \bxp \in \mathbb{R}^{\n - \np}, \\ \by \setminus \byp \in \mathbb{R}^{\m - \mmp} }} \ABT{\bx}{\by}.
		\end{align*}
		
		To start, denote $\bn = \n + \m$. Suppose 
		$\bn, \n - \np$ satisfy $\bmcn{\iconstant}$,
		where $\iconstant \in \{1,2,3,4\}$.
		Let us assume that when
		$\bx \setminus \bxp = \bxt$,
		and $\by \setminus \byp = \byt$,
		the Ansari-Bradley test statistic 
		takes the minimum values, i.e.,
		\begin{align*}
			\ABT{\bxp \cup \bxt}{\byp \cup \byt} = T_{\min}(\bx,\by) . 
		\end{align*}
		Consider $(\bxh, \byh) \in \mysetz(\bxp, \byp, \n-\np, \m-\mmp,\iconstant)$,
		and denote $\bxh = \{\xsn{1}, \ldots, \xsn{\n - \np}\}$,
		$\byh = \{\ysn{1}, \ldots, \ysn{\m -\mmp}\}$,
		and $\kk = \sum_{\iconstant = 1}^{\m - \mmp} \indicator{\ysn{\iconstant} > \max \bxh \cup \bxp \cup \byp }$. Then, 
		by Lemma~\ref{supp:lemma:14},
		\begin{align*}
			\ABT{\bxs}{\bys} =  \left\{ \begin{array}{lcl}
				\ABT{\bxp}{\byp} + \func{\bxp, \byp, \m}{\kk} + {(\n^2 - \np^2)}/{4}, &~\text{if}
				&\bmcn{1},~\text{or}~\bmcn{4} \\ 
				\ABT{\bxp}{\byp} + \func{\bxp, \byp, \m}{\kk} + {(\n^2 - \np^2 - 1)}/{4}, &~\text{if}
				&\bmcn{2}\\ 
				\ABT{\bxp}{\byp} + \func{\bxp, \byp, \m}{\kk} + {(\n^2 - \np^2 + 1)}/{4}, &~\text{if}
				&\bmcn{3}
			\end{array}\right.
		\end{align*}	
		By Lemma~\ref{supp:ab:lemma:4}, when $\bn, \n - \np$
		satisfy $\bmcn{1}$ or $\bmcn{3}$, $\kk \in \{a,\ldots,b \}$;
		while $\bn, \n - \np$ satisfy $\bmcn{2}$ or $\bmcn{4}$, $\kk \in \{c,\ldots,d\}$.
		Without loss of generality, let us assume that $\byh$ takes the 
		$\kk$ that minimizes $\func{\bxp, \byp, \m}{\kk}$. In order words,
		\begin{align*}
			\ABT{\bxs}{\bys}  =  \left\{ \begin{array}{lcl}
				\ABT{\bxp}{\byp} + \min\limits_{\kk \in \{\sa, \ldots, \sbb\}} \func{\bxp, \byp, \m}{\kk} + {(\n^2 - \np^2)}/{4}, &\text{~if}~\bmcn{1} \\ 
				\ABT{\bxp}{\byp} + \min\limits_{\kk \in \{\scc, \ldots, \sdd\}} \func{\bxp, \byp, \m}{\kk} + {(\n^2 - \np^2 - 1)}/{4}, &\text{~if}~\bmcn{2}\\ 
				\ABT{\bxp}{\byp} + \min\limits_{\kk \in \{\sa, \ldots, \sbb\}} \func{\bxp, \byp, \m}{\kk} + {(\n^2 - \np^2 + 1)}/{4}, &\text{~if}~\bmcn{3}\\
				\ABT{\bxp}{\byp} + \min\limits_{\kk \in \{\scc, \ldots, \sdd\}} \func{\bxp, \byp, \m}{\kk} + {(\n^2 - \np^2)}/{4}, &\text{~if}~\bmcn{4} \\ 
			\end{array}\right.
		\end{align*}
		By definition, we have 
		\begin{align*}
			T_{\min}(\bx,\by)  \le \ABT{\bxs}{\bys}  
		\end{align*}
		However, according to  Proposition~\ref{supp:prop:8}, there exist 
		$(\bxhln{1}, \byhln{1}) \in \mysetz(\bxp, \byp, \n-\np, \m-\mmp,\iconstant)$,
		such that 
		\begin{align*}
			\ABT{\bxp \cup \bxhln{1}}{\byp \cup \byhln{1}}  \le	\ABT{\bxp \cup \bxt}{\byp \cup \byt}. 
		\end{align*}
		Applying Lemma~\ref{supp:lemma:14},	 we have 
		\begin{align*}
			\ABT{\bxs}{\bys} \le \ABT{\bxp \cup \bxhln{1}}{\byp \cup \byhln{1}}.
		\end{align*}
		Combining the above results,
		\begin{align*}
			&\ABT{\bxs}{\bys} \le \ABT{\bxp \cup \bxhln{1}}{\byp \cup \byhln{1}} \le \ABT{\bxp \cup \bxt}{\byp \cup \byt} = T_{\min}(\bx,\by)  \le \ABT{\bxs}{\bys}  \\
			\implies& \ABT{\bxs}{\bys} = T_{\min}(\bx,\by).
		\end{align*}
		Then, according to  Lemma~\ref{supp:ab:lemma:1}, we have
		\begin{align*}
			\ABT{\bx}{\by} = - \ABT{\by}{\bx}  +  \left\{ \begin{array}{ll}
				{\bn^2}/{4} 
				& \bn~\text{is even,} \\ 
				{(\bn^2 - 1)}/{4}
				& \bn~\text{is odd.} 
			\end{array}\right.
		\end{align*}
		Hence, 
		\begin{align*}
			T_{\max}(X,Y) = - T_{\min}(Y,X) + \left\{ \begin{array}{ll}
				{\bn^2}/{4} 
				& \bn~\text{is even,} \\ 
				{(\bn^2 - 1)}/{4}
				& \bn~\text{is odd.} 
			\end{array}\right.
		\end{align*}
		This completes our proof.
	\end{proof}
	
	\section{Bounds of the $p$-values}
	\label{supp:bounds:p-value}
	This section provides results for the bounds of $p$-values.
	The following proposition provides bounds
	of $p$-values of the Ansari-Bradley test in the 
	presence of missing data. 

	\begin{proposition} \label{sup:prop:wmw:boundspvalue}
		Suppose that $\bx = \{\xln{1}, \cdots, \xln{\n}\}$ and 
		$\by = \{\yln{1}, \cdots, \yln{\m}\}$ are samples of distinct, 
		real-valued observations. Suppose that $\bxp \subset \bx$ is 
		a subset of $\np$ values in $\bx$, and suppose that $\byp \subset \by$ 
		is a subset of $\mmp$ values in $\by$. Defining $\bz = \bx \cup \by$ 
		and $\bzp = \bxp \cup \byp$ and supposing only $\bzp$ is known, then
		the $p$-value of the Ansari-Bradley test $\p(\WMW{\bx}{\by})$
		is bounded such that
		\begin{align*}
			\p(\WMW{\bx}{\by})  &\ge \pmin{\bx}{\by} = \min\{	\p(\WMWmin{\bx}{\by}), \p(\WMWmax{\bx}{\by})  \}, \\
			\p(\WMW{\bx}{\by})  &\le \pmax{\bx}{\by} = \left\{ \begin{array}{cl}
				1,~~~\text{ if } (\WMWmin{\bx}{\by} - \mu)(\WMWmax{\bx}{\by} - \mu) \le 0,\\
				\max\{	\p(\WMWmin{\bx}{\by}), \p(\WMWmax{\bx}{\by})  \},~~~\text{otherwise}.
			\end{array}\right.
		\end{align*}
	\end{proposition}
	
	\begin{proof}
		We first prove that 
		$\p(\WMW{\bx}{\by}) \ge \min\{	\p(\WMWmin{\bx}{\by}), \p(\WMWmax{\bx}{\by}) \}$.
		Note that if $\WMW{\bx}{\by} < \mu$, we have
		\begin{align*}
			\Phi\left( \frac{\WMW{\bx}{\by} - \mu}{\sigma} \right) < \Phi(0) = \frac{1}{2}.
		\end{align*}
		Hence,
		\begin{align*}
			\p(\WMW{\bx}{\by}) &= 2\min \left\{ \Phi\left( \frac{\WMW{\bx}{\by} - \mu}{\sigma} \right), 1 -  \Phi\left( \frac{\WMW{\bx}{\by} - \mu}{\sigma} \right)\right\} \\
			&=2  \Phi\left( \frac{\WMW{\bx}{\by} - \mu}{\sigma} \right).
		\end{align*}
		Since $\WMWmin{\bx}{\by} \le \WMW{\bx}{\by}$, and $\Phi(\x)$ is a monotonic
		increasing function, we have
		\begin{align*}
			\p(\WMW{\bx}{\by}) \ge 2  \Phi\left( \frac{\WMWmin{\bx}{\by} - \mu}{\sigma} \right) \ge 	\p(\WMWmin{\bx}{\by}).
		\end{align*}
		Following the same approach, we can show that
		if $\WMW{\bx}{\by} \ge \mu$, we have
		\begin{align*}
			\p(\WMW{\bx}{\by}) \ge 2  \left( 1 - \Phi\left( \frac{\WMWmax{\bx}{\by} - \mu}{\sigma} \right) \right)\ge 	\p(\WMWmax{\bx}{\by}).
		\end{align*}
		Hence, we have proven that 
		\begin{align*}
			\p(\WMW{\bx}{\by}) \ge \min\{	\p(\WMWmin{\bx}{\by}), \p(\WMWmax{\bx}{\by}) \}.
		\end{align*}
		
		Next, we prove that
		\begin{align*}
			\p(\WMW{\bx}{\by})  &\le \left\{ \begin{array}{cl}
				1,~~~\text{ if } (\WMWmin{\bx}{\by} - \mu)(\WMWmax{\bx}{\by} - \mu) \le 0,\\
				\min\{	\p(\WMWmin{\bx}{\by}), \p(\WMWmax{\bx}{\by})  \},~~~\text{otherwise}.
			\end{array}\right.
		\end{align*}
		By definition, we have $\p(\WMW{\bx}{\by}) \le 1$. Hence the result must be true
		if $(\WMWmin{\bx}{\by} - \mu)(\WMWmax{\bx}{\by} - \mu) \le 0$.
		
		If, however, $(\WMWmin{\bx}{\by} - \mu)(\WMWmax{\bx}{\by} - \mu) > 0$,
		then we must have either
		\begin{align*}
			&\WMWmin{\bx}{\by}, \WMWmax{\bx}{\by} > \mu,\\
			\text{or }& \WMWmin{\bx}{\by}, \WMWmax{\bx}{\by} < \mu.
		\end{align*}
		Suppose $\WMWmin{\bx}{\by}, \WMWmax{\bx}{\by} > \mu$. Since
		$\WMW{\bx}{\by} > \WMWmin{\bx}{\by}$, we have
		$\WMW{\bx}{\by} > \mu$.
		Subsequently,
		\begin{align*}
			\p(\WMW{\bx}{\by}) &=2  \left(1 - \Phi\left( \frac{\WMW{\bx}{\by} - \mu}{\sigma} \right) \right) \\
			& \le 2  \left(1 - \Phi\left( \frac{\WMWmin{\bx}{\by} - \mu}{\sigma} \right) \right)\\
			& = \p(\WMWmin{\bx}{\by}) \\
			&\le \max\{	\p(\WMWmin{\bx}{\by}), \p(\WMWmax{\bx}{\by})  \}. 
		\end{align*}
		Similarly, when $\WMWmin{\bx}{\by}, \WMWmax{\bx}{\by} < \mu$,
		we can use the same approach and prove that
		\begin{align*}
			\p(\WMW{\bx}{\by}) \le \p(\WMWmax{\bx}{\by})
			\le \max\{	\p(\WMWmin{\bx}{\by}), \p(\WMWmax{\bx}{\by})  \}.
		\end{align*}
		Hence, we conclude our result.	
	\end{proof}
	
	Now, we prove that all possible Ansari-Bradley
	test statistics are significant is equivalent 
	to the maximum possible $p$-value is smaller than
	or equal to the significant level $\alpha$.
	\begin{proposition} \label{supp:prop:wmw:pvalueequivalent}
		Suppose that $\bx = \{\xln{1}, \cdots, \xln{\n}\}$ and 
		$\by = \{\yln{1}, \cdots, \yln{\m}\}$ are 
		partially observed samples of distinct, 
		real-valued observations. 
		Then for any given significance level
		$\alpha < 1$, 
		\begin{align*}
			\text{ Condition~\ref{condition:wmw:1} or \ref{condition:wmw:2} hold} \iff \pmax{\bx}{\by} \le \alpha,
		\end{align*}
		where $\pmax{\bx}{\by}$ is defined in Proposition~\ref{prop:wmw:boundspvalue}.
		Hence, the $p$-value of the proposed location testing 
		method is the maximum possible $p$-value $\pmax{\bx}{\by}$ over all possible values of missing data.
	\end{proposition}
	\begin{proof}
		We first prove that ``$\implies$'' holds.
		Suppose Condition~\ref{condition:wmw:1} or \ref{condition:wmw:2} holds.
		Then, if Condition~\ref{condition:wmw:1} is true,
		\begin{align*}
			&\WMWmin{\bx}{\by}, \WMWmax{\bx}{\by} > \mu\\
			\implies &\pmax{\bx}{\by} =  \max\{	\p(\WMWmin{\bx}{\by}), \p(\WMWmax{\bx}{\by})  \}.
		\end{align*}
		According to Condition~\ref{condition:wmw:1}, we also have
		\begin{align*}
			\Phi\left( \frac{\WMWmin{\bx}{\by} - \mu}{\sigma} \right) \ge \Phi\left( \Phi^{-1}\left(1 - \frac{\alpha}{2}\right)\right) = 1 - \frac{\alpha}{2},\\
			\Phi\left( \frac{\WMWmax{\bx}{\by} - \mu}{\sigma} \right) \ge \Phi\left( \Phi^{-1}\left(1 - \frac{\alpha}{2}\right)\right) = 1 - \frac{\alpha}{2}.
		\end{align*}
		Hence, 
		\begin{align*}
			\p(\WMWmin{\bx}{\by}) &= 2\min \left\{ \Phi\left( \frac{\WMWmin{\bx}{\by} - \mu}{\sigma} \right), 1 -  \Phi\left( \frac{\WMWmin{\bx}{\by} - \mu}{\sigma} \right)\right\}\\
			& \le \alpha,
		\end{align*}
		and
		\begin{align*}
			\p(\WMWmax{\bx}{\by}) &= 2\min \left\{ \Phi\left( \frac{\WMWmax{\bx}{\by} - \mu}{\sigma} \right), 1 -  \Phi\left( \frac{\WMWmax{\bx}{\by} - \mu}{\sigma} \right)\right\}\\
			& \le \alpha.
		\end{align*}
		Therefore, when Condition~\ref{condition:wmw:1} is true, we have
		\begin{align*}
			\pmax{\bx}{\by} =  \max\{	\p(\WMWmin{\bx}{\by}), \p(\WMWmax{\bx}{\by})  \} \le \alpha.
		\end{align*}
		
		The result when Condition~\ref{condition:wmw:2} is true
		can be proved using the same approach. Thus, it is omitted here.
		Hence, we have proved that ``$\implies$'' is true.
		
		We now prove that ``$\impliedby$'' holds.
		If $\pmax{\bx}{\by} \le \alpha$, then we
		have
		\begin{align*}
			(\WMWmin{\bx}{\by} - \mu)(\WMWmax{\bx}{\by} - \mu) > 0,
		\end{align*}
		and
		\begin{align*}
			\max\{	\p(\WMWmin{\bx}{\by}), \p(\WMWmax{\bx}{\by})  \} \le \alpha. 
		\end{align*}
		
		Note that
		\begin{align*}
			(\WMWmin{\bx}{\by} - \mu)(\WMWmax{\bx}{\by} - \mu) > 0
		\end{align*}
		implies either
		\begin{align*}
			\WMWmin{\bx}{\by}, \WMWmax{\bx}{\by} > \mu, \text{ or } \WMWmin{\bx}{\by}, \WMWmax{\bx}{\by} < \mu.
		\end{align*}
		
		Suppose $\WMWmin{\bx}{\by}, \WMWmax{\bx}{\by} < \mu$.
		We then have
		\begin{align*}
			\p(\WMWmin{\bx}{\by}) &= 2\Phi\left( \frac{\WMWmin{\bx}{\by} - \mu}{\sigma} \right),
		\end{align*}
		and
		\begin{align*}
			\p(\WMWmax{\bx}{\by}) &= 2\Phi\left( \frac{\WMWmax{\bx}{\by} - \mu}{\sigma} \right).
		\end{align*}
		Since
		\begin{align*}
			\max\{	\p(\WMWmin{\bx}{\by}), \p(\WMWmax{\bx}{\by})  \} \le \alpha, 
		\end{align*}
		we have
		\begin{align*}
			\Phi\left( \frac{\WMWmin{\bx}{\by} - \mu}{\sigma} \right), \Phi\left( \frac{\WMWmax{\bx}{\by} - \mu}{\sigma} \right) \le \frac{\alpha}{2}.
		\end{align*}
		Hence, we have shown Condition~\ref{condition:wmw:2}
		\begin{align*}
			\frac{\WMWmin{\bx}{\by} - \mu}{\sigma}  \le \Phi^{-1}\left( \frac{\alpha}{2}\right)\text{ and } \frac{\WMWmax{\bx}{\by}  - \mu}{\sigma}  \le \Phi^{-1}\left( \frac{\alpha}{2}\right)
		\end{align*}
		is true.
		
		Following the same approach, 	
		when $\WMWmin{\bx}{\by}, \WMWmax{\bx}{\by} < \mu$, 
		we can show Condition~\ref{condition:wmw:1} is true. Hence, 
		we have proven ``$\impliedby$''. This completes our proof.
	\end{proof}
	
	\section{Additional simulation results}
	
	In this section, we provide additional simulation 
	results for evaluating the Type I error and statistical
	power of the proposed method.

	\subsection{Scale testing for data missing completely at random}
	\label{addtionalexp:ab:1}
	
\begin{figure}
		\includegraphics[width=\textwidth]{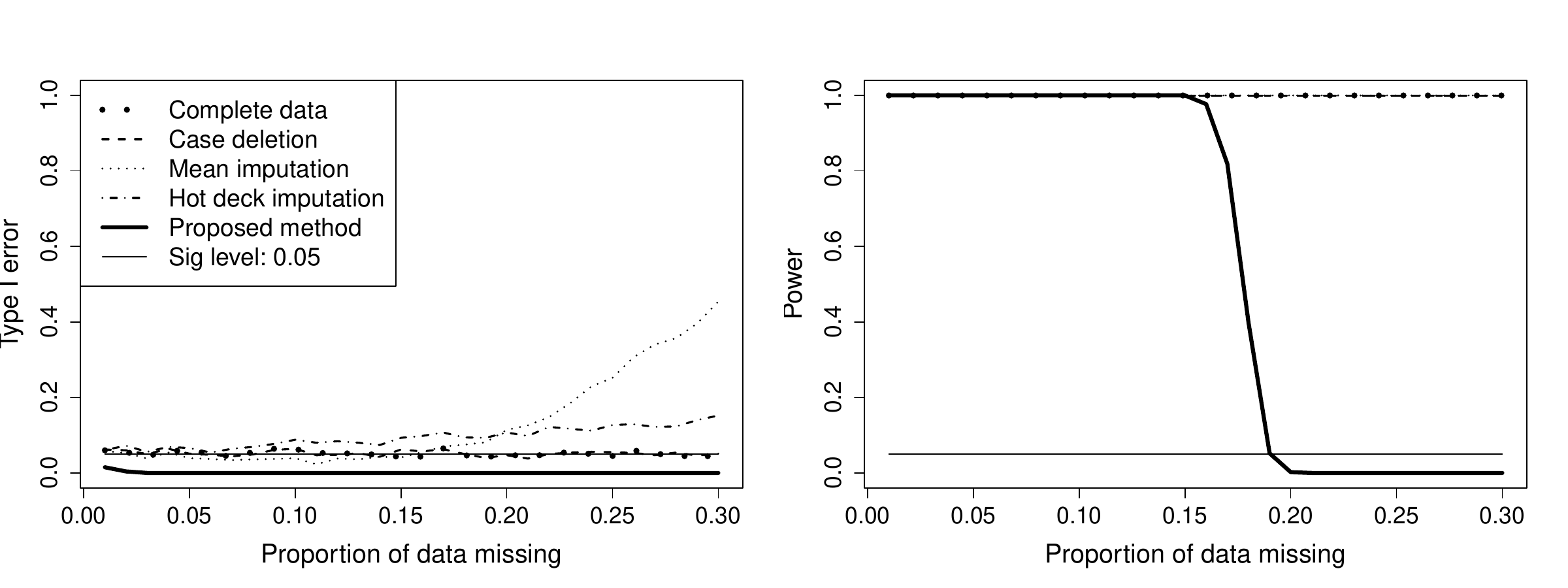}
		\caption{The Type I error and statistical power of the proposed method and 
			the standard Ansari-Bradley test after the missing data is either known or
			has been imputed or ignored as the proportion of missing data increases.
			The data is missing completely at random (MCAR).
			(Left) Type I error: $\mathrm{N}(0,1)$ vs $\mathrm{N}(0,1)$; 
			(Right) Power: $\mathrm{N}(0,1)$ vs $\mathrm{N}(0,\sigma^2)$,
			with the scale parameter $\sigma = 3$. For both figures, 
			a significance threshold of $\alpha=0.05$ has been used and the total
			sample sizes are $\n=500$, $\m=500$, and $1000$ trials were used.}
		\label{fig:ab:mcar2}
	\end{figure}

\begin{figure}
		\includegraphics[width=\textwidth]{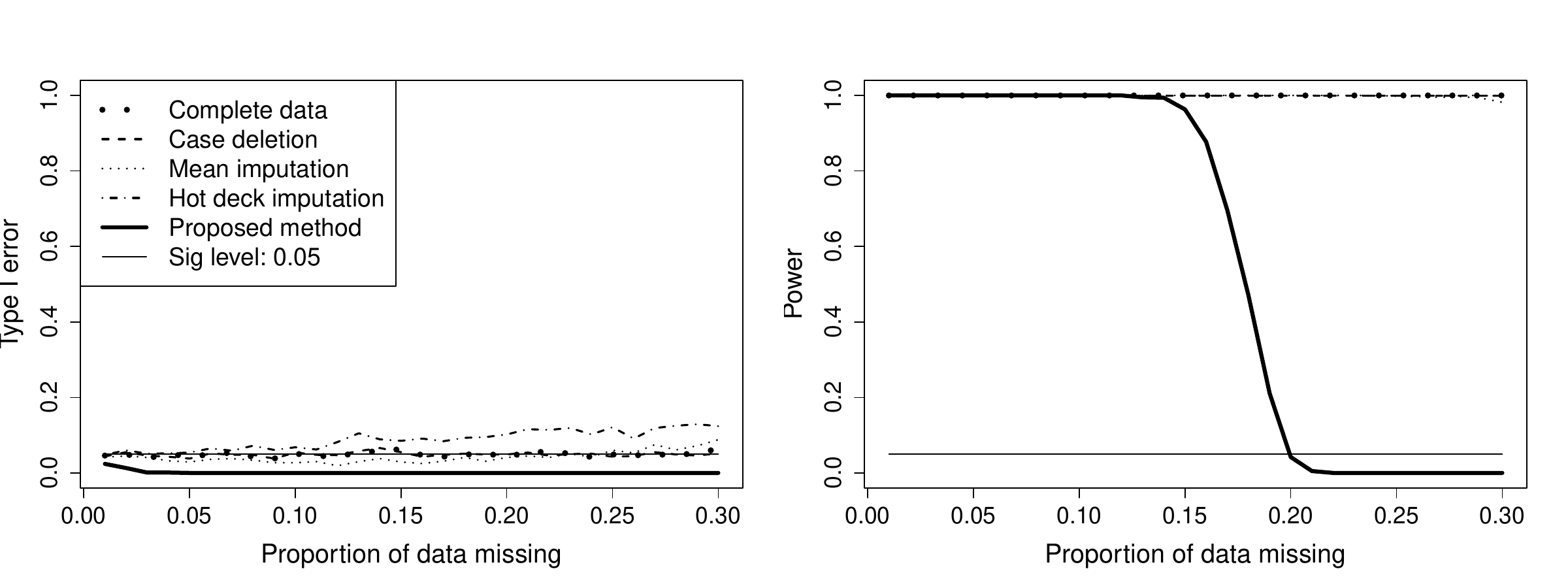}
		\caption{The Type I error and statistical power of the proposed method and 
			the standard Ansari-Bradley test after the missing data is either known or
			has been imputed or ignored as the proportion of missing data increases.
			The data is missing completely at random (MCAR).
			(Left) Type I error: $\mathrm{N}(0,1)$ vs $\mathrm{N}(0,1)$; 
			(Right) Power: $\mathrm{N}(0,1)$ vs $\mathrm{N}(0,\sigma^2)$,
			with the scale parameter $\sigma = 5$. For both figures, 
			a significance threshold of $\alpha=0.05$ has been used and the total
			sample sizes are $\n=100$, $\m=100$, and $1000$ trials were used.}
		\label{fig:ab:mcar3}
	\end{figure}

	This section considers similar experiments 
	as that for Figure~\ref{fig:AB:mcar1} when data
	are missing completely at random (MCAR),
	but with different
	sample sizes and scale parameters.
	The results are shown in
	Figure~\ref{fig:ab:mcar2} and Figure~\ref{fig:ab:mcar3}.
	Figure~\ref{fig:ab:mcar2} considers larger
	sample sizes $\n = \m =500$, and
	shows similar results
	to Figure~\ref{fig:AB:mcar1} with
	larger power.
	
	Figure~\ref{fig:ab:mcar3} shows the increased power of 
	the proposed method by considering the first sample consists of observations
	from $\mathrm{N}(0,1)$, but the second sample consists of observations 
	from $\mathrm{N}(0,\sigma^2)$ with $\sigma = 5$, rather than $3$
	used in Figure~\ref{fig:AB:mcar1}.
	
	\subsection{Scale testing for data missing not at random}
	\label{addtionalexp:ab:2}
	
\begin{figure}
		\includegraphics[width=\textwidth]{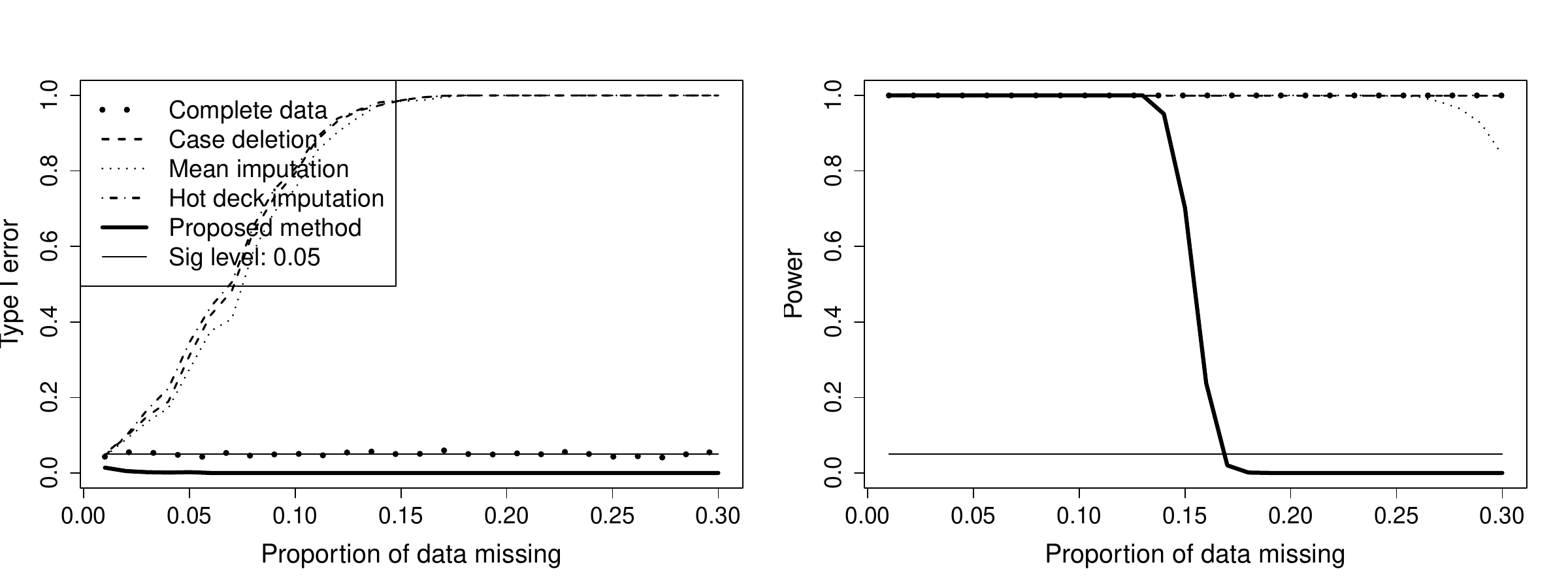}
		\caption{The Type I error and statistical power of the proposed method and 
			the standard Ansari-Bradley test after the missing data is either known or
			has been imputed or ignored as the proportion of missing data increases.
			The data is missing not at random (MNAR), according to the mechanism in 
			Equation~\eqref{eqn:AB:missmechonex}, and \eqref{eqn:AB:missmechoney}.
			(Left) Type I error: $\mathrm{N}(0,1)$ vs $\mathrm{N}(0,1)$; 
			(Right) Power: $\mathrm{N}(0,1)$ vs $\mathrm{N}(0,\sigma^2)$,
			with scale parameter $\sigma = 3$. For both figures, 
			a significance threshold of $\alpha=0.05$ has been used and the total
			sample sizes are $\n=500$, $\m=500$, and $1000$ trials were used.}
		\label{fig:ab:mnar2}
	\end{figure}

\begin{figure}
		\includegraphics[width=\textwidth]{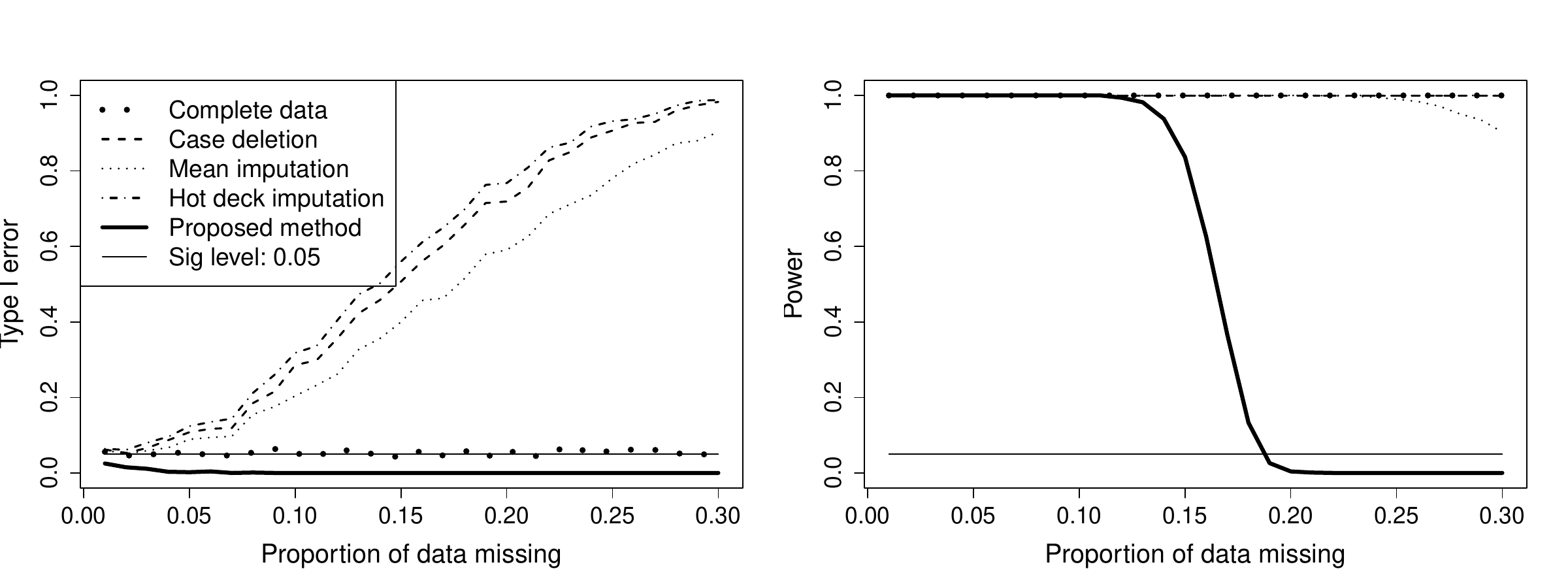}
		\caption{The Type I error and statistical power of the proposed method and 
			the standard Ansari-Bradley test after the missing data is either known or
			has been imputed or ignored as the proportion of missing data increases.
			The data is missing not at random (MNAR), according to the mechanism in 
			Equation~\eqref{eqn:AB:missmechonex}, and \eqref{eqn:AB:missmechoney}.
			(Left) Type I error: $\mathrm{N}(0,1)$ vs $\mathrm{N}(0,1)$; 
			(Right) Power: $\mathrm{N}(0,1)$ vs $\mathrm{N}(0,\sigma^2)$,
			with scale parameter $\sigma = 5$. For both figures, 
			a significance threshold of $\alpha=0.05$ has been used and the total
			sample sizes are $\n=100$, $\m=100$, and $1000$ trials were used.}
		\label{fig:ab:mnar3}
	\end{figure}

	This section considers experiments similar to 
	Figure~\ref{fig:AB:mnar1} when data are missing
	not at random (MNAR), but with different sample
	sizes and scale parameters.
	The results are shown in  
	Figure~\ref{fig:ab:mnar2} and \ref{fig:ab:mnar3}.
	
	Figure~\ref{fig:ab:mnar2} considers different
	sample sizes $\n = \m =500$, while
	Figure~\ref{fig:ab:mnar3} considers the first sample consists of observations
	from $\mathrm{N}(0,1)$, but the second sample consists of observations 
	from $\mathrm{N}(0,\sigma^2)$ with $\sigma = 5$, rather than $3$
	used in Figure~\ref{fig:AB:mnar1}.
	
	\subsection{Scale testing with varying scale value}
	\label{chap4:sec:numeric:mean}
	We now consider two experiments 
	for evaluating the power of the proposed method
	with varying scale parameters $\sigma$ of the distribution 
	generating the sample $\by$. The sample sizes,
	and the proportion of missing data for both $\bx$ and $\by$ are fixed. 
	
	\subsubsection{Missing completely at random (MCAR)}
	
	The first experiment again
	assumes the data are MCAR.
	Observations in the sample $\bx$ are sampled independently from a 
	$\mathrm{N}(0,1)$ distribution, while observations in sample $\by$  are sampled 
	independently from a $\mathrm{N}(0,\sigma^2)$, where
	$\sigma \in \{1.0, 1.1, \ldots, 4.0\}$.
	Sample sizes $\n = \m = 100$,
	and proportion of missing data $\misss = 0.1$ are used.
	
	\begin{figure}
		\centering
		\includegraphics[width=\textwidth]{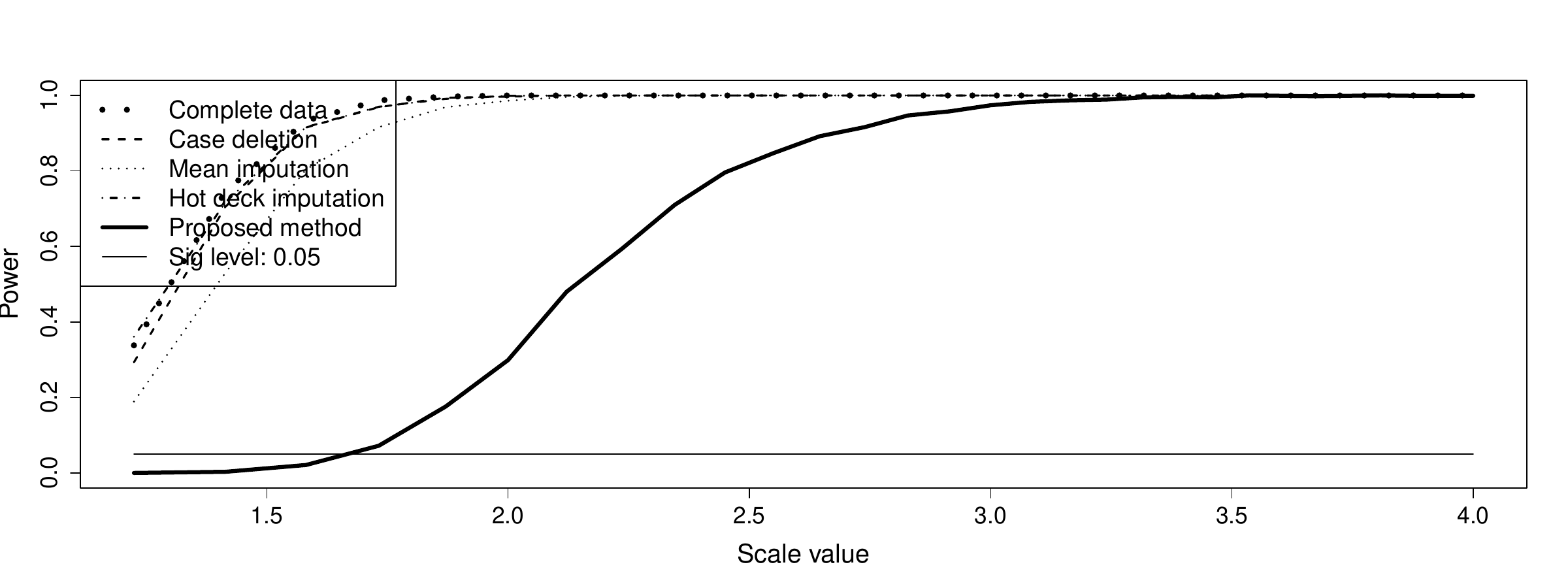}
		\caption{Statistical power of the proposed method and 
			the standard Ansari-Bradley test after the missing data is either known or
			has been imputed or ignored as the sample sizes increase.
			The data is missing completely at random (MCAR).
			Data in samples $\bx$ and $\by$ are generated from 
			$\mathrm{N}(0,1)$, and $\mathrm{N}(0,\sigma^2)$, respectively,
			with $\sigma$ denoting the scale value.
			A significance threshold of $\alpha=0.05$ has been used, 
			the total sample sizes are $\n = \m = 100$,
			the proportion
			of missing data for both samples is $\misss = 0.1$, 
			and $1000$ trials were used.}
		\label{fig:AB:mcar:meanvalueshift1}
	\end{figure}
	
	Figure~\ref{fig:AB:mcar:meanvalueshift1} shows that 
	the proposed method starts to have testing power
	when the scale parameter $\sigma$ is greater than
	1.7. When $\sigma  = 3$, the power of
	the proposed method is close to $1$. The power
	of all other methods increases with $\sigma$, and
	are close to $1$ and when $\Delta \approx 1.7$.
	
	\subsubsection{Missing not at random (MNAR)}
	
	Now we consider an experiment similar to that for Figure~\ref{fig:AB:mcar:meanvalueshift1},
	but when the data are MNAR according to the missingness mechanisms
	specified by Equation~\eqref{eqn:AB:missmechonex}, and \eqref{eqn:AB:missmechoney}.
	
\begin{figure}
		\centering
		\includegraphics[width=\textwidth]{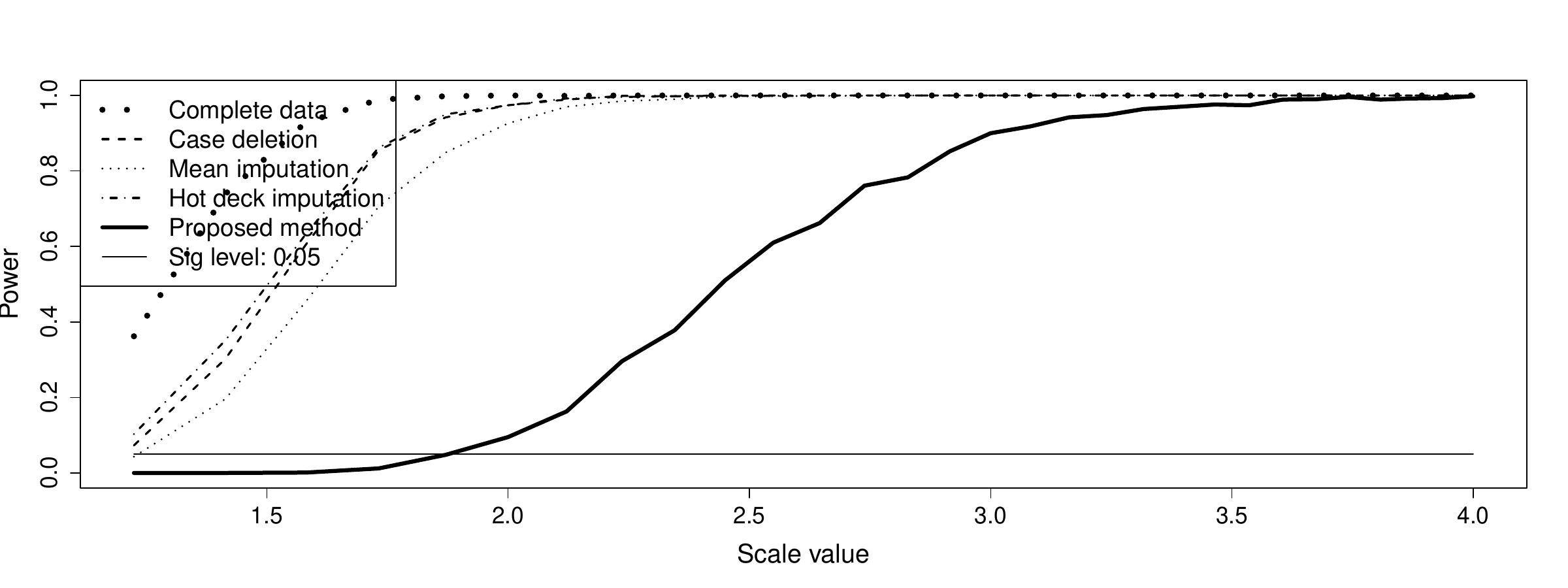}
		\caption{Statistical power of the proposed method and 
			the standard Ansari-Bradley test after the missing data is either known or
			has been imputed or ignored as the sample sizes increase.
			The data is missing not at random (MNAR), according to the mechanism in 
			Equation~\eqref{eqn:AB:missmechonex}, and \eqref{eqn:AB:missmechoney}.
			Data in samples $\bx$ and $\by$ are generated from 
			$\mathrm{N}(0,1)$, and $\mathrm{N}(0,\sigma^2)$, respectively,
			with $\sigma$ denoting the scale value.
			A significance threshold of $\alpha=0.05$ has been used, 
			the total sample sizes are $\n = \m = 100$,
			the proportion
			of missing data for both samples is $\misss = 0.1$, 
			and $1000$ trials were used.}
		\label{fig:AB:mnar:meanvalueshift1}
	\end{figure}

	The results are presented in Figure~\ref{fig:AB:mnar:meanvalueshift1}, 
	which shows similar results 
	to Figure~\ref{fig:AB:mcar:meanvalueshift1}, although in this
	experiment all methods appear to require a larger scale parameter $\sigma$
	for achieving power equal to 1. The power of the proposed 
	method is approximately 1 when the scale parameter $\sigma = 3.5$, 
	while other methods require the scale parameter $\sigma$ to be $2$. 
	
	\subsection{Scale testing for Gamma-distributed data}
	\label{addtionalexp:ab:3}
	
\begin{figure}
		\includegraphics[width=\textwidth]{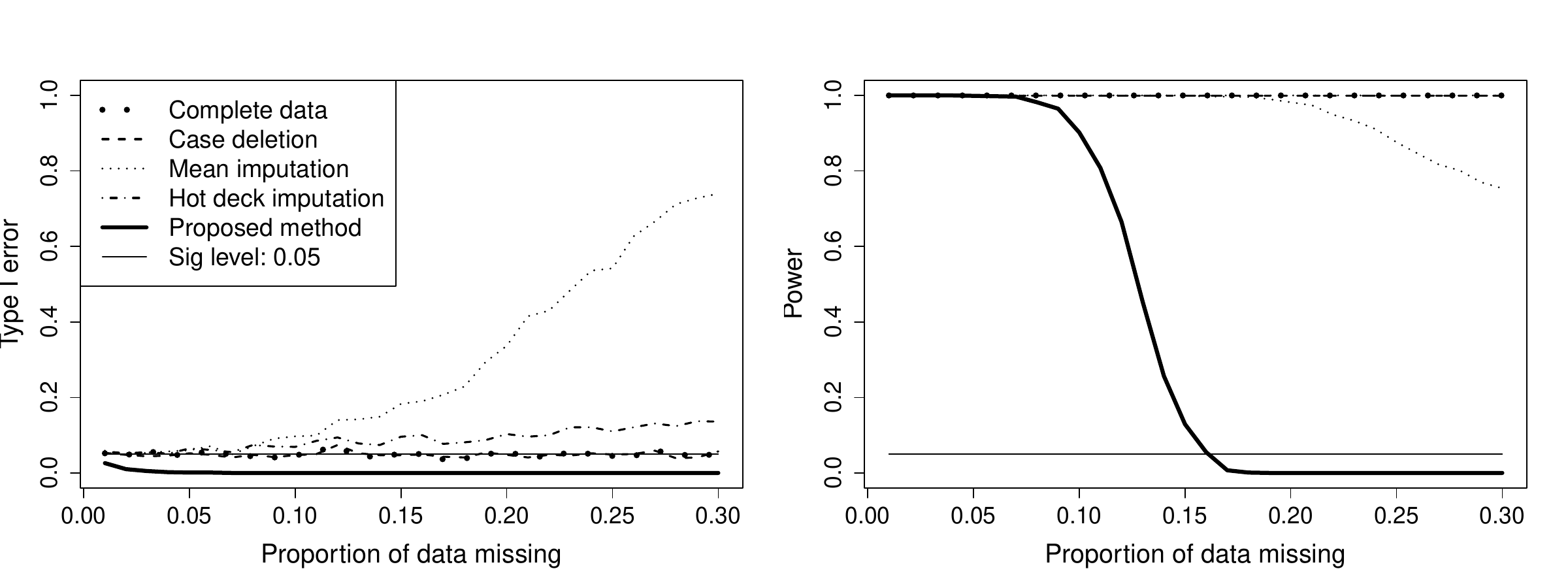}
		\caption{The Type I error and statistical power of the proposed method and 
			the standard Ansari-Bradley test after the missing data is either known or
			has been imputed or ignored as the proportion of missing data increases.
			The data is missing completely at random (MCAR).
			(Left) Type I error: $\mathrm{Gamma}(1,1)$ vs $\mathrm{Gamma}(1,1)$; 
			(Right) Power: $\mathrm{Gamma}(1,1)$ vs $\mathrm{Gamma}(9,0.11)$. For both figures, 
			a significance threshold of $\alpha=0.05$ has been used and the total
			sample sizes are $\n=100$, $\m=100$, and $1000$ trials were used.}
		\label{fig:ab:mcar:gamma}
	\end{figure}

\begin{figure}
		\includegraphics[width=\textwidth]{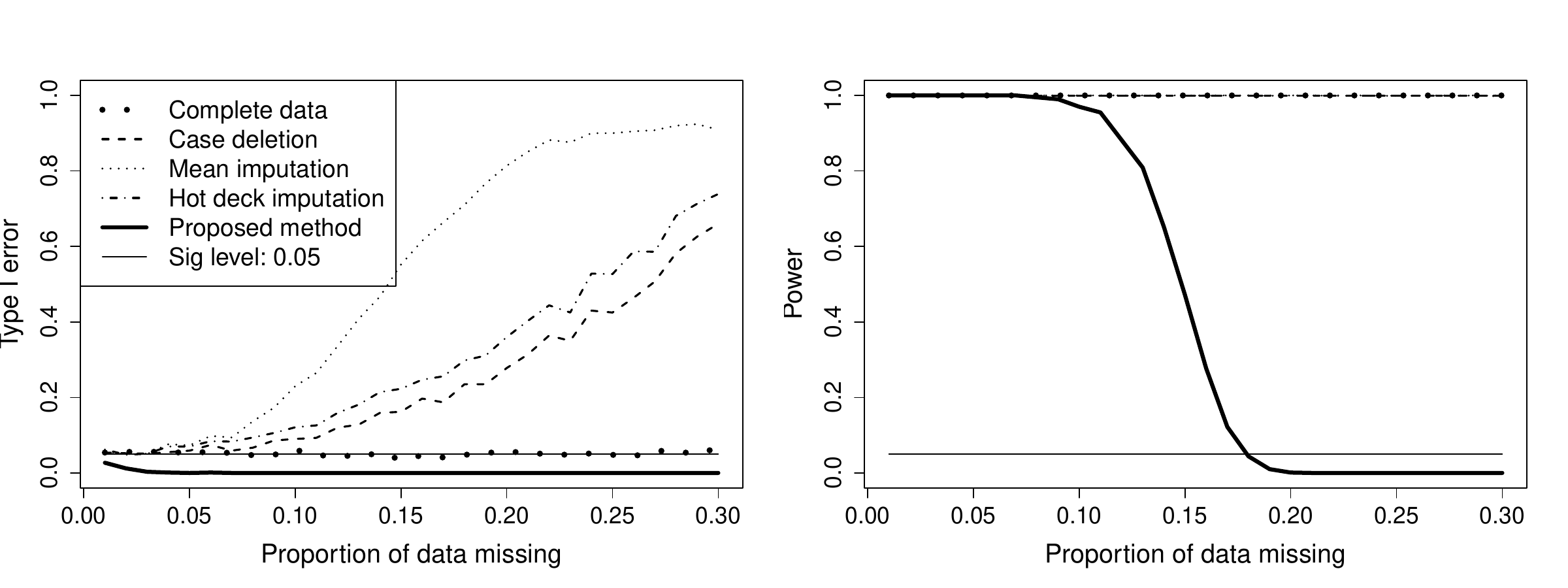}
		\caption{The Type I error and statistical power of the proposed method and 
			the standard Ansari-Bradley test after the missing data is either known or
			has been imputed or ignored as the proportion of missing data increases.
			The data is missing not at random (MNAR), according to the mechanism in 
			Equation~\eqref{eqn:AB:missmechonex}, and \eqref{eqn:AB:missmechoney}.
			(Left) Type I error: $\mathrm{Gamma}(1,1)$ vs $\mathrm{Gamma}(1,1)$; 
			(Right) Power: $\mathrm{Gamma}(1,1)$ vs $\mathrm{Gamma}(9,0.11)$. For both figures, 
			a significance threshold of $\alpha=0.05$ has been used and the total
			sample sizes are $\n=100$, $\m=100$, and $1000$ trials were used.}
		\label{fig:ab:mnar:gamma}
	\end{figure}

	This section presents
	Figure~\ref{fig:ab:mcar:gamma} and Figure~\ref{fig:ab:mnar:gamma},
	which are similar to Figure~\ref{fig:AB:mcar1} and 
	Figure~\ref{fig:AB:mnar1} in the main results,
	but the data are now Gamma-distributed, rather than Normal-distributed.
	
	Similar to Figure~\ref{fig:AB:mcar1},
	Figure~\ref{fig:ab:mcar:gamma} also considers the data to be
	missing completely at random (MCAR). However,
	the data in Figure~\ref{fig:ab:mcar:gamma}
	follow the Gamma distribution, rather than the normal distribution.
	Specifically, the data in sample $\bx$ are generated from Gamma(1,1),
	while the data in sample $\by$ are generated in Gamma(1,1)
	for evaluating the Type I error, and Gamma(9,0.11) for evaluating power.
	The distributions Gamma(1,1)
	and Gamma(9,0.11) are chosen so that the two 
	distributions have roughly the same mean value but
	different variances: Gamma(1,1) has mean 1, and variance 1; 
	Gamma(9,0.11) has mean $0.99$, and variance approximately $0.1$.
	
	\subsection{Scale testing for a different missingness mechanism}
	\label{addtionalexp:ab:4}
	
\begin{figure}
		\includegraphics[width=\textwidth]{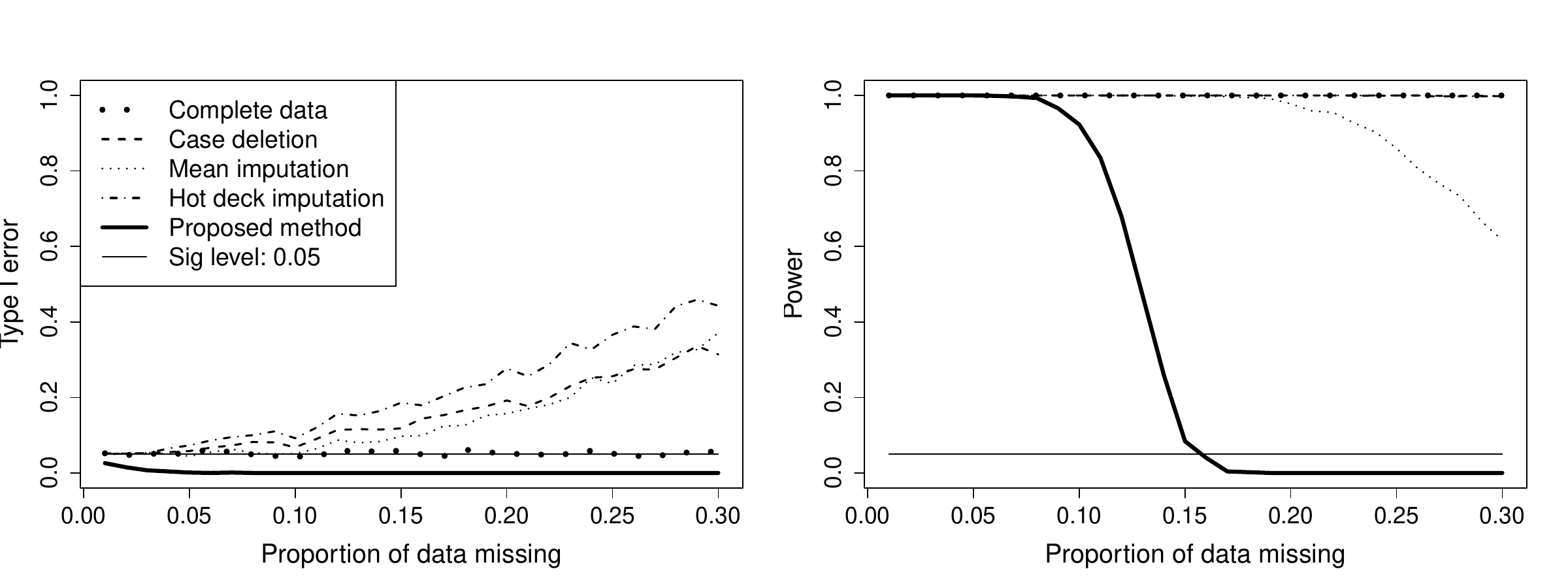}
		\caption{The Type I error and statistical power of the proposed method and 
			the standard Ansari-Bradley test after the missing data is either known or
			has been imputed or ignored as the proportion of missing data increases.
			The data are missing not at random (MNAR). The data in $\bx$ are randomly 
			selected to be missing,
			while the data in $\by$ are missing according to Equation \eqref{eqn:ab:missmechtwoy}.
			(Left) Type I error: $\mathrm{N}(0,1)$ vs $\mathrm{N}(0,1)$; 
			(Right) Power: $\mathrm{N}(0,1)$ vs $\mathrm{N}(0,\sigma^2)$,
			with scale parameter $\sigma = 3$. For both figures, 
			a significance threshold of $\alpha=0.05$ has been used and the total
			sample sizes are $\n=100$, $\m=100$, and $1000$ trials were used.}
		\label{fig:ab:mnar4}
	\end{figure}

	Figure~\ref{fig:ab:mnar4} considers a similar case
	to Figure~\ref{fig:AB:mnar1} when data are missing not
	at random (MNAR), but with a different missingness mechanism.
	For Figure~\ref{fig:ab:mnar4}, the data in $\bx$ are randomly 
	selected to be missing,
	while the data in $\by$ are missing according to
	\begin{align}
		\prob{\textrm{$\yy_i$ is missing}} = \frac{|\yln{\iconstant}|}{\sum_{\jconstant=1}^{\m} |\yln{\jconstant}|}.
		\label{eqn:ab:missmechtwoy}
	\end{align}
	In other words, the probability of a value in $\by$ to be missing is 
	proportional to its absolute value after adding $|\min \by|$.

\end{appendix}

\bibliographystyle{plainnat}
\bibliography{bibliography}       % Bibliography file (usually '*.bib')
\end{document}